\documentclass{article}

\usepackage{amsmath}
\usepackage{amssymb}

\usepackage{ytableau}
\usepackage{multirow}
\usepackage{rotating}
\usepackage{anysize}
\usepackage{simplewick}

\usepackage{tikz}
\usetikzlibrary{arrows}
\usetikzlibrary{positioning}
\usepackage{xparse}

\usepackage{hyperref}
\hypersetup{hidelinks=true}

\newcount\tableauRow
\newcount\tableauCol

\newenvironment{Tableau}[1]{  \tikzpicture[scale=0.4,draw/.append style={thick,black},
                      baseline=(current bounding box.center)]
        \tableauRow=-1.5
    \foreach \Row in {#1} {
       \tableauCol=0.5
       \foreach\k in \Row {
         \draw(\the\tableauCol,\the\tableauRow)rectangle++(1,1);
         \draw(\the\tableauCol,\the\tableauRow)+(0.5,0.5)node{$\k$};
         \global\advance\tableauCol by 1
       }
       \global\advance\tableauRow by -1
    }
}{\endtikzpicture}

\def\cN{{\cal N}}
\def\cO{{\cal O}}
\def\cS{{\cal S}}
\def\umu{{\underline{\smash \mu}}}
\def\ula{{\underline{\smash \lambda}}}

\DeclareMathOperator{\tr}{tr}
\DeclareMathOperator{\str}{str}

\long\def\symbolfootnote[#1]#2{\begingroup\def\thefootnote{\fnsymbol{footnote}}\footnote[#1]{#2}\endgroup}

\marginsize{2.4cm}{2.4cm}{2cm}{2cm}

\begin{document}

\begin{flushright}
DCPT-15/53
\end{flushright}

\vspace{20pt}

\begin{center}

{\Large \bf  Superconformal partial waves }
\vspace{0.3 cm}
{\Large \bf  in Grassmannian field theories}

\vspace{45pt}

{\mbox {\bf  Reza Doobary and  Paul Heslop }}\symbolfootnote[4]{
{\tt  \{ \tt \!\!\!r.c.doobary, paul.heslop\}@durham.ac.uk}
}

\vspace{0.5cm}

  \begin{quote} {\small \em
      Department of Mathematical Sciences\\
      Durham University\\
      South Road, Durham DH1 3LE, United Kingdom }
  \end{quote}

\vspace{60pt}

{\bf Abstract}
\end{center}

\vspace{0.3cm} 

\noindent

\noindent
We derive superconformal partial waves for all scalar four-point functions on a super Grassmannian space $Gr(m|n,2m|2n)$ for all $m,n$. This family of four-point functions includes those of all (arbitrary weight) half BPS operators in both $\cN{=}4$ SYM ($m{=}n{=}2$) and in $\cN{=}2$  superconformal field theories in four dimensions ($m{=}2, n{=}1$) on analytic superspace. It also includes four-point functions of all (arbitrary dimension) scalar fields in non-supersymmetric conformal field theories  ($m{=}2,n{=}0$) on Minkowski space, as well as those of a certain class of representations of the compact $SU(2n)$ coset spaces. As an application we then specialise to $\cN{=}4$ SYM and use these results to perform a detailed superconformal partial wave analysis of the four-point functions of arbitrary weight half BPS operators. We discuss the  non-trivial separation of protected and unprotected sectors for the $\langle 2222\rangle$, $\langle 2233\rangle$ and $\langle 3333\rangle$ cases in an $SU(N)$ gauge theory at finite $N$. The $\langle 2233\rangle$ correlator predicts a non-trivial protected twist four sector for $\langle 3333\rangle$ which we can completely determine using the knowledge that there is precisely one such protected twist four operator for each spin.
 
\setcounter{page}{0}
\thispagestyle{empty}
\newpage

\tableofcontents

\section{Introduction}
\label{sec:introduction}

There has been considerable activity recently in the area of computing four-point functions in conformal field theories, motivated by the conformal bootstrap programme initiated in~\cite{Rattazzi:2008pe}. This programme allows one to obtain non-trivial non-perturbative information about certain quantities, from crossing symmetry and the conformal partial wave expansion.
Independently there has been a great deal of research on the computation of anomalous dimensions and OPE coefficients in conformal field theories -- in particular for $\cN$=4 SYM -- centred around  integrability. The latter programme was given a remarkable boost recently in the work of~\cite{Basso:2015zoa} allowing the computation of non-trivial OPE coefficients non-perturbatively from so-called hexagon functions which are determined from integrability assumptions. It has thus become important to obtain OPE coefficients independently of this in order to test the integrability approach.

Information about OPE coefficients is contained within four-point correlation functions. The method to extract these is via the conformal partial wave expansion.
Dolan and Osborn pioneered the use of
conformal and 
superconformal partial waves for the practical extraction of  data
from known four-point
functions in higher (than two) dimensional theories~\cite{Dolan:2000ut,Dolan:2001tt,Dolan:2243hv}, with further superconformal partial waves in four-dimensions studied in \cite{Fitzpatrick:2014oza}. The main application of this
method so far has been  in $\cal N$=4 SYM whose four-point functions (of half
BPS operators) have been computed both in perturbation theory and at strong coupling in a large number of cases.\footnote{Most work has centred around the four-point function of stress-energy multiplets which has been computed at weak coupling up to
seven loops at the level of the integrand~\cite{corAmpForward} and to  three 
loops analytically~\cite{4ptintegrated12,corAmpBack,Drummond:2013nda}. It is also known at strong coupling via the AdS/CFT 
correspondence~\cite{sugraOld}. Half BPS correlators of (equal) higher charges are known at one- and two-loops~\cite{Arutyunov:2003ad} and at strong coupling~\cite{Arutyunov:2003ae} and recently some mixed charge cases were computed to two-loops~\cite{Chicherin:2014esa}.}
The standard approach has been to solve the superconformal Ward identities via differential equations and then match the superconformal partial waves onto this solution, by summing up all the partial waves of component fields in a multiplet~\cite{Eden:2000bk,Dolan:2001tt,Arutyunov:2002fh,Dolan:2004mu,Nirschl:2004pa}.
More recently superconformal partial waves in $\cN$=4 SYM as well as $\cN=2$ have been reconsidered from the conformal bootstrap perspective~\cite{Beem:2013qxa,Beem:2014zpa,Alday:2014qfa}.

In~\cite{Heslop:2002hp} an alternative approach to solving the Ward identities of arbitrary four-point functions was implemented in $\cN=4$ analytic superspace.
In~\cite{Howe:1995md} a general picture of superspaces as cosets was developed. In particular the study of $\cN=4$ SYM was developed in $\cN=4$ analytic superspace which manifested the full superconformal symmetry in a manner similar to the conformal group in Minkowski space~\cite{Howe:2001je,Heslop:2001zm,Heslop:2243xu}. Using analytic superspace allows one to solve the superconformal Ward identities in a more  direct manner without ever seeing a differential equation. In~\cite{Heslop:2002hp} the four-point functions were written as an expansion in super Schur polynomials. One practical advantage of this approach is that the expansion automatically only ever sees unitary operators, thus there is no issue of disentangling non-unitary operators as in other  approaches  (although as we will see, one still has to understand the real physical problem of disentangling long and short operators). The precise form of the superconformal partial waves in this formalism was however not found at the time.
This paper can be viewed as a continuation of this programme, obtaining the (super)conformal partial waves, first as a sum of Schur polynomials and also then in a summed form and finally using the results to analyse a number of free theory correlation functions of low charge half BPS operators.

We will in fact consider superconformal partial waves in a more general setting
by considering any Grassmannian field theory.\footnote{The idea of considering a generalised Grassmannian field theory was first proposed by Paul Howe.}
By a Grassmannian field theory, we mean any theory with $SU(m,m|2n)$ symmetry given on the complexified space $Gr(m|n,2m|2n)$ of $m|n$-planes in $2m|2n$ dimensions. For $m=2$ this corresponds to an $\cN=2n$ superconformal theory on analytic superspace (which reduces to conformal theory in Minkowski space in the bosonic $n=0$ case). The main case we will pursue in later sections will be $m=n=2$ corresponding to $\cN=4$ SYM. For $m=1$ the results apply to two dimensional superconformal field theories. Finally for $m=0$ this corresponds to a purely internal $SU(2n)$ group written on a coset space (for example for $n=1$ the space would be a 2-sphere).
Coordinates on the Grassmannian take the form
\begin{align*}
  X^{{A}{A}'}=\left(
  \begin{array}{cc}x^{\alpha\dot{\alpha}}&
      \rho^{\alpha  a'}\\
     \bar{\rho}^{a \dot{\alpha}}& y^{aa'}
     \end{array}
\right)
\end{align*}
where $A=(\alpha,a)$ and $A'=(\dot\alpha,a')$, with $\alpha,\dot\alpha=1,\dots m$ and $a, a'=1,\dots n$. In the case $m=2$, $x^{\alpha \dot \alpha}$ is the four-dimensional Minkowski space co-ordinate written in spinor notation.

In this paper we will focus our attention on four-point functions of charged scalars,  $\cO^p$,  on the Grassmannian (meaning they do not transform non-trivially under the two $SL(m|n)$ subgroups which leave the plane invariant). For $\cN$=4 SYM ($m=n=2$) and for  $\cN$=2 superconformal field theories ($m=2, n=1$) these are the half-BPS operators. For conformal theories in four dimensions ($m=2,n=0$) they are Lorentz scalars (with arbitrary dimension $p$)  and in the purely internal case $m=0$ they are representations of $SU(2n)$ defined by rectangular Young tableau of height $n$ and length $p$.

We denote the more general operators which appear in the OPE of two of these special operators by $\cO^{\gamma\ula}$ where $\gamma$ is the charge and $\ula$ is the (in general non-trivial) representation of the isotropy group $GL(m|n)\times GL(m|n)$ (which leaves the plane invariant) under which the operator transforms. A general operator can transform differently under the two copies of $GL(m|n)$ but those appearing in the OPE of scalar operators must transform in the same representation for both subgroups.

Our method for finding the superconformal partial waves is as follows:
\begin{itemize}
\item We start with the well-known bosonic conformal partial waves in four-dimensions~\cite{Dolan:2000ut,Dolan:2243hv}. The contribution of an operator $\cO^{\gamma\ula}$ to a four-point function $\langle \cO^{p_1} \cO^{p_2} \cO^{p_3} \cO^{p_4}\rangle$ is given (up to some propagator factors which we omit here)  
by the conformal partial wave
\begin{align*}
  \text{$GL(4)$:} \qquad 
 F^{\alpha\beta\gamma\underline \lambda}(x_1,x_2) &= \textstyle{\frac{\det\Big(
   x_i^{\lambda_j+2-j}{}_2F_1(\lambda_j{+}1{-}j{+}\alpha ,\lambda_j{+}1{-}j{+}\beta;2\lambda_j{+}2{-}2j{+}\gamma;x_i)\Big)_{1\leq i,j \leq
    2}}{x_1-x_2}}\ ,
\end{align*} 
where
$  \alpha=\tfrac12(\gamma{-}p_{1}{+}p_2) \quad \beta=\tfrac12( \gamma{+}p_{3}{-}p_4)$.
 Here $x_1,x_2$ are the two eigenvalues of the $2\times2$ matrix $(x_{12}x_{24}^{-1}x_{43}x_{31}^{-1})^\alpha{}_\beta$. 
\item We then propose a natural lift of this result to the bosonic $Gr(m,2m)$ Grassmannian field theory  for any integer $m$, namely the contribution of the operator $\cO^{\gamma\ula}$ to any four-point function is
\begin{align*}
\text{$GL(2m)$:} \qquad F^{\alpha\beta\gamma\underline\lambda}(x) &=\textstyle{\frac{\det\Big(
   x_i^{\lambda_j+m-j}{}_2F_1(\lambda_j{+}1{-}j{+}\alpha ,\lambda_j{+}1{-}j{+}\beta;2\lambda_j{+}2{-}2j{+}\gamma;x_i)\Big)_{1\leq i,j \leq
    m}}{\det\Big(
   x_i^{m-j}\Big)_{1\leq i,j \leq
    m}}\ }\ ,
\end{align*}
where similarly, $x_1,x_2,\dots,x_m$ are the eigenvalues of the $m\times m$ matrix $(x_{12}x_{24}^{-1}x_{43}x_{31}^{-1})^\alpha{}_\beta$. We check that this uplift does indeed satisfy the correct Casimir differential equation for the conformal partial wave.

\item We now expand  the above $Gr(m,2m)$ partial wave as a sum over Schur polynomials $s_{\umu}(x)$, where $\umu$ is a representation of $GL(m)$ 
  \begin{align*}
 \text{$GL(2m)$: } \qquad \qquad\qquad   F^{\alpha\beta\gamma\ula}(x) =   \sum_{[\umu]}
R^{\alpha\beta\gamma\underline\lambda}_{\umu} \,
s_{\umu}(x)\ .
  \end{align*}
Note that the numerical coefficients $R^{\alpha\beta\gamma\underline\lambda}_{\umu}$ do not depend on $m$ but only on the Young tableaux of the representation $\ula,\umu$. This is a key point: it must be the case, since on restricting the coordinates to any  $Gr(m{-}1,2m{-}2)$ subgroup both the conformal partial wave and the Schur polynomials reduce to the corresponding $Gr(m{-}1,2m{-}2)$ ones, and since  the numerical coefficients haven't changed under this reduction, they must be independent of $m$.

\item Now we can go directly from here to an expression for the supersymmetric  $Gr(m|n,2m|2n)$  partial waves. Again the key point is that the  coefficients in this expansion will be independent of $m,n$ (by similar reasoning to above) and so we can immediately know that the contribution of the super operator $\cO^{\gamma\ula}$ to any superconformal four-point function is
  \begin{align*}
    \text{$GL(2m|2n)$:} \qquad \qquad \qquad
F^{\alpha\beta\gamma\ula}(x|y) \ &=\   \sum_{[\umu]}
R^{\alpha\beta\gamma\underline\lambda}_{\umu} \,
s_{\umu}(x|y),
\end{align*}
with the $R$ coefficients derived from the $GL(m)$ (and explicitly given later)  and known super Schur polynomials $s_{\umu}(x|y)$. Here $(x|y)=(x_1,\dots,x_m,y_1,\dots y_n)$ are the eigenvalues of the $(m|n)\times (m|n)$ matrix $(X_{12}X_{24}^{-1}X_{43}X_{31}^{-1})^A{}_B$.

Now for finding OPE coefficients we in fact needn't go any further. Indeed one can write any free theory correlator as a sum over super Schur polynomials (using an application of Cauchy's identity) and then comparing with the partial waves expanded in Schur polynomials. Since the Schur polynomials form an independent basis this allows us to equate coefficients on both sides and determine the OPE coefficients. Indeed remarkably one never even needs to know the form of the Schur polynomials themselves in this approach! We do precisely this in a number of cases later in the paper.

\item However for conformal bootstrap applications it is essential to have a summed up form of the partial waves.  Using a beautiful determinantal formula for the super Schur polynomials found by  Moens and van der Jeugt~\cite{moens2003determinantal} as inspiration we then obtain a determinantal formula, summing up the above expansion,  for the superconformal partial waves analogous to the $GL(m)$ one above.

\item As a byproduct we then obtain a formula for the partial waves in the compact $SU(2n)$ case (corresponding to $m=0$). Remarkably this gives an entirely different form for the same numerical coefficients $R^{\alpha\beta\gamma\underline\lambda}_{\umu}$. The equality of these two forms for $R^{\alpha\beta\gamma\underline\lambda}_{\umu}$ produces an infinite number of non-trivial numerical identities. The checking of these remarkable identities provides a strong self-consistency check on our method.

\end{itemize}

Note that we have given a full summary of the final results for the superconformal partial wave expansion both in its expanded and summed up form in section~\ref{sec:summ-superc-part}.

The paper proceeds as follows. In section~\ref{sec:reps} we explain the formalism and notation for fields on Grassmannian spaces. In section~\ref{sec:conf-part-waves-3} we review (super) Schur polynomials and  derive the superconformal partial waves on a general (super)Grassmannian field theory as summarised above.
Both to provide  further checks as well as to obtain new results, in section~\ref{sec:ope-coeff-cn=4} we specialise to the case $m=n=2$ and use our results to initiate a detailed analysis of mixed charge four-point correlators. In particular we  compute the OPE coefficients for a number of low charge cases. In this section all multiplets are considered as being in their naive free theory representations.
In section~\ref{sec:phys-ope-coeff}  we then also consider the problem of multiplet recombination where free-theory short operators can combine to become long operators in the interacting theory and hence develop anomalous dimensions~\cite{Dolan:2242zh,Heslop:2243xu}. In particular, we fully solve this rather intricate problem for the $\langle \tr(W^3)\tr(W^3)\tr(W^3)\tr(W^3)\rangle$ case.
We leave a few more technical points to appendices. In appendix~\ref{sec:proof-conf-part} we give the proof that our simple uplift of the partial waves from $Gr(2,4)$ to $Gr(m,2m)$ is correct, by deriving the Casimir operator which defines  the partial waves and showing that the result satisfies the Casimir eigenvalue equation. In appendix~\ref{sec:further-results-free}
we give some further analysis of some mixed charge correlators which we felt were too detailed to go in the main text. Finally in appendix~\ref{sec:altern-form-glmn} we give an alternative version of the determinantal formula for super Schur polynomials. Our form for the summed up superconformal partial waves reduces to this alternative form rather than the original one.

\hspace{1em}

\noindent During the final writing up stage the preprint~\cite{Bissi:2015qoa} appeared on the arxiv which has partial overlap with the results presented here.

\section{Representations as fields on the (super)Grassmannian}\label{sec:reps}

We will be considering four-point functions in a class of theories
which we call Grassmannian field theories. These are theories whose
configuration space is the super Grassmannian of $(m|n)$-planes
through the origin of a
$(2m|2n)$ complex dimensional vector space. Thus the theories have a $GL(2m|2n)$
symmetry 
(which will be broken down to $SL(2m|2n)$). This symmetry group will
be viewed as the complexification of the group $SU(m,m|2n)$ and the operators we consider will all be unitary representations of this real group. In particular then we view the $SL(m)$ subgroup as non-compact (complexification of $SU(m,m)$) but the $SL(n)$
subgroup to be compact (complexification of $SU(n)$).

This family includes several cases of physical interest (the rest are presumably of only mathematical interest). The case $m=2,n=0$ corresponds to  Minkowski space (well known to be equivalent to the space of
$2$-planes in four dimensions $Gr(2,4)$) and their the symmetry
group is $SU(2,2)$, the conformal group. The case $m=2,n=1$
corresponds to $\cN=2$ analytic superspace~\cite{Galperin:1984av} and the case
$m=2,n=2$ which will be of most interest to us is $\cN=4$ analytic superspace~\cite{Howe:1995md}. In both these
cases the symmetry group, $SU(2,2|2n)$, is the $2n$-extended
superconformal group. 
Furthermore one can consider the cases $m=0$, arbitrary $n$,  which
correspond to the compact spaces $SU(2n)$.

We wish to consider coordinates on $Gr(m|n,2m|2n)$. To do this consider a point in this space (i.e. an $(m|n)$-plane) and consider a basis for this $(m|n)$-plane in the  $(2m|2n)$-dimensional
vector space. This  is equivalent to writing an 
$(m|n)\times(2m|2n)$ matrix (with the rows corresponding to the basis
vectors). Choosing another basis for the same plane is equivalent to multiplication on the left by a
$GL(m|n)$ matrix. We can use this freedom of basis choice to choose unique coordinates
on the Grassmannian as
\begin{align}
  X^{{A}{A}'}=\left(\begin{array}{cc}x^{\alpha\dot{\alpha}}&
      \rho^{\alpha  a'}\\
     \bar{\rho}^{a \dot{\alpha}}& y^{aa'}\end{array}\right),
\end{align}
corresponding to the $(m|n)$-plane specified by the
$(m|n)\times(2m|2n)$ matrix:
\begin{align}
  \left(\begin{array}{cc|cc}
    1_{m\times m}&x       & 0_{n\times m} &\rho\\\hline
    0_{n\times m}&\bar \rho&1_{n\times n}&y      
  \end{array}
\right).
\end{align}
Here the indices $A,A'$ are $(m|n)$-dimensional indices, $\alpha,\dot \alpha$ are
$m$-dimensional and $a,a'$ are $n$-dimensional.
 This superspace is a supersymmetric generalisation of a Grassmannian
 manifold. This Grassmannian can also be thought of as a supercoset,
 and is an example of a much more general construction whereby the
 isotropy group is a parabolic subgroup generated by a parabolic
 subalgebra~\cite{Howe:1995md}.

Representations of $GL(2m|2n)$ are  written as fields (or operators) on this super 
Grassmannian. The operators are specified by the representations of the
two $GL(m|n)$ subgroups which leave the $(m|n)$-plane
invariant. For the operators considered in this paper (i.e. which appear
in the four-point functions we consider here) the representations of
the two $GL(m|n)$ subgroups will always be identical. We
include a further quantum number $\gamma$, which although redundant for
generic representations, is needed to describe short representations
in the supersymmetric case. We thus define our representations through operators on the Grassmannian space $\cO^{\gamma\ula} = \cO^\gamma_{\ula(A) \ula(A')}(X^{BB'})$ where $\ula$ is a Young tableau defining a representation of $GL(m|n)$ via a tensor  product of the fundamental representation, and $\ula(A)$ is a multi-index symmetrised  according to this Young tableau.

It is useful to consider an explicit realisation of the operators. 
We will build all representations
from a very special
representation carrying the trivial representation of the two
$GL(m|n)$ subgroups and with $\gamma=1$. In the case $(m,n)=(2,2)$ this special
representation corresponds to
the $\cN$=4  Maxwell/ Yang-Mills supermultiplet, or for $(m,n)=(2,1)$ it corresponds to
the $\cN$=2  hypermultiplet and for  $(m,n)=(2,0)$ it is a massless
scalar field. When  $m=0$ it corresponds to the representation
of $SU(2n)$ defined by an $n$ row, single column Young tableau (i.e. the representation with dimension $(2n)!/(n!)^2$).
We denote this special representation as a field on the Grassmannian by $W(X)$.

More general operators all then have the 
schematic form 
\begin{align}
\mathcal{O}^{\gamma\ula} \sim \partial^{|\ula|}_{\ula{(A)}\ula{(A')}}  W^\gamma,
\end{align}
where the derivatives $\partial_{AA'}=\partial /\partial X^{AA'}$ can
act on different $W$s. 
We have in mind the case $(m,n)=(2,2)$ of $\cN$=4 SYM where $W$ is the Yang-Mills multiplet and it sits in the adjoint representation of some gauge group.

We define the $GL(m|n)$ representation
$\ula=[\lambda_1,\lambda_2,\dots]$ via Young tableaux where
$\lambda_i$ is the length of row $i$. 
It is also useful to define the heights of column $j$ to be
$\lambda_j^T$ (so $\ula^T$ denotes the conjugate or transpose representation). 
Representations of $GL(m|n)$ are given by all Young tableaux that fit into a thick hook tableau with thickness $m$ horizontally and $n$ vertically.

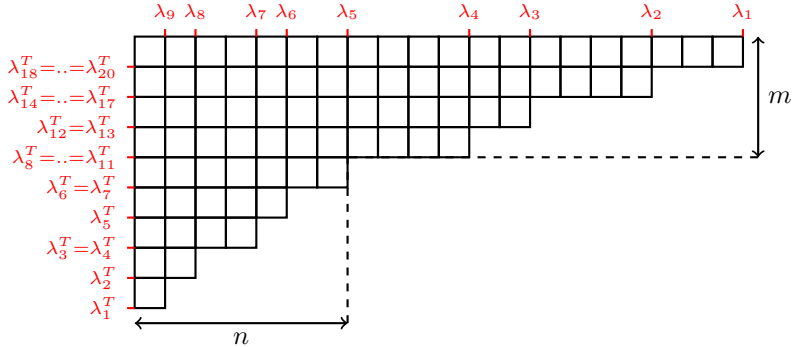
\begin{center}
  \begin{Tableau}
{ {,,,,,,,,,,,,,,,,,,,}, {,,,,,,,,,,,,,,,,},
      {,,,,,,,,,,,,}, {,,,,,,,,,,}, {,,,,,,}, {,,,,}, {,,,}, {,}, {} }
    \draw[dashed] (20.5,-4)--(7,-4); 
    \draw[dashed]
    (7,-9.5)--(7,-4); 
       \draw[thick,<->]
    (20.5,-4)--node[anchor=west]{$m$}(20.5,0) ; 
    \draw[thick,<->]
   (0,-9.5)--node[anchor=north]{$n$}(7,-9.5) ; 
\draw[thick,red]
    (-0,-9)--(-.25,-9)node[anchor=east]{$\scriptstyle \lambda^T_1$} ;
\draw[thick,red]
    (-0,-8)--(-.25,-8)node[anchor=east]{$\scriptstyle \lambda^T_2$}
    ;
\draw[thick,red]
    (-0,-7)--(-.25,-7)node[anchor=east]{$\scriptstyle \lambda^T_3=\lambda^T_4$}
    ;
\draw[thick,red]
    (-0,-6)--(-.25,-6)node[anchor=east]{$\scriptstyle \lambda^T_5$}
    ;
\draw[thick,red]
    (-0,-5)--(-.25,-5)node[anchor=east]{$\scriptstyle \lambda^T_6=\lambda^T_7$} ;\draw[thick,red]
    (-0,-4)--(-.25,-4)node[anchor=east]{$\scriptstyle \lambda^T_8=..=\lambda^T_{11}$} ;\draw[thick,red]
    (-0,-3)--(-.25,-3)node[anchor=east]{$\scriptstyle \lambda^T_{12}=\lambda^T_{13}$} ;\draw[thick,red]
    (-0,-2)--(-.25,-2)node[anchor=east]{$\scriptstyle \lambda^T_{14}=..=\lambda^T_{17}$} ;\draw[thick,red]
    (-0,-1)--(-.25,-1)node[anchor=east]{$\scriptstyle \lambda^T_{18}=..=\lambda^T_{20}$}
    ;
\draw[thick,red]
    (20,0)--(20,0.25)node[anchor=south]{$\scriptstyle\lambda_1$} ;
\draw[thick,red]
    (17,0)--(17,0.25)node[anchor=south]{$\scriptstyle\lambda_2$} ;
\draw[thick,red]
    (13,0)--(13,0.25)node[anchor=south]{$\scriptstyle\lambda_3$} ;
\draw[thick,red]
    (11,0)--(11,0.25)node[anchor=south]{$\scriptstyle\lambda_4$} ;
\draw[thick,red]
    (7,0)--(7,0.25)node[anchor=south]{$\scriptstyle\lambda_5$} ;
\draw[thick,red]
    (5,0)--(5,0.25)node[anchor=south]{$\scriptstyle\lambda_6$} ;
\draw[thick,red]
    (4,0)--(4,0.25)node[anchor=south]{$\scriptstyle\lambda_7$} ;
\draw[thick,red]
    (2,0)--(2,0.25)node[anchor=south]{$\scriptstyle\lambda_8$} ;
\draw[thick,red]
    (1,0)--(1,0.25)node[anchor=south]{$\scriptstyle\lambda_9$} ;
 \end{Tableau}
\end{center}

The operator $\cO^{\gamma\ula}$ defines a representation of $GL(2m|2n)$ and
thus of $SL(2m|2n)$  and in turn then of the real form $SU(m,m|2n)$.
Representations of $SU(m,m|2n)$ are more familiarly given via  Dynkin labels for the compact $SU(2n)$ subgroup $m_1,\dots m_{2n-1}$, then Dynkin labels for the two (left and right) $SL(m)$ groups $j^L_{1},\dots j^L_{m{-}1},j^R_{1}, \dots j^R_{m{-}1}$ (in the physical case with  $m=2$ this is just (twice) the left and right spin) and  finally giving the dilatation weight $\Delta$ (weight under $x\rightarrow \lambda x$ as usual). The translation between the labels of the operator then $\cO^{\gamma\ula}$ and the corresponding representation is given by
\begin{align}\label{eq:2612}
  m_i &= m_{n-1-i} = \lambda^T_{n-i}-\lambda_{n-i+1}^T \text{ for
        $1\leq i\leq n{-}1$},\notag\\
  m_n&= \gamma-2\lambda^T_1, \notag \\
  j_i= j^L_{i}&=j^R_{i} = \hat \lambda_{m-i}-\hat \lambda_{m-i+1}  \text{ for
        $1\leq i\leq m{-}1$},\notag\\
  \Delta &= 
  \frac m2 \gamma +\sum_{i-1}^m j_i  
  \ ,
\end{align}
where we defined
\begin{align}
  \hat \lambda_i :=\left\{
  \begin{array}{ll}
    \lambda_i -n & \text{if } \lambda_i\geq n\\
    0 & \text{if } \lambda_i< n
  \end{array}
\right.
\end{align}
This translation can be obtained by considering the highest weight state (HWS) in a standard way (see~\cite{Heslop:2243xu}). 
In particular, the special representation $W$ has $m_n=1, \Delta= m/2 $ and all other quantum numbers vanishing.

We can now consider  the degeneracy in our description of operators $\cO^{\gamma\ula}$ mentioned above. A generic $GL(m|n)$ Young Tableau can be uniquely determined by $m+n$ numbers (e.g. the first $m$ row lengths, $\lambda_1, \dots , \lambda_m$ and the first $n$ column heights, $\lambda^T_1, \dots, \lambda_n^T$). Together with $\gamma$ then $\cO^{\gamma\ula}$ has $m+n+1$ quantum numbers. On the other hand the corresponding $SL(m|n)$ representations require only $n+m$ quantum numbers ($m_1, \dots , m_n, j_1, \dots j_{m-1}, \Delta$). Thus there must be some degeneracy in~\eqref{eq:2612}. Indeed we see that the relations~\eqref{eq:2612} are invariant under the following shift:
\begin{align}&\text{(if } \notag \lambda_m\geq n+1)&&\text{(if }  \lambda^T_n\geq m+1)\notag \\
  \lambda_i &\rightarrow \lambda_i -1, \text{ for }1\leq i \leq m
 & \lambda^T_i &\rightarrow \lambda^T_i -1, \text{ for }1\leq i \leq n \notag
  \\ \lambda^T_i &\rightarrow \lambda^T_i +1, \text{ for }1\leq i \leq n  &\lambda_i &\rightarrow \lambda_i+1, \text{ for }1\leq i \leq m \notag \\
  \gamma&\rightarrow \gamma + 2 &   \gamma&\rightarrow \gamma - 2\ .\label{eq:41}
\end{align}
This corresponds to deleting a full (height $m$) column from the
horizontal part of the ``hook'' and adding a full (length $n$)
column to the vertical part (or vice versa). The condition
$\lambda_m\geq n+1$ is simply the condition that there exists a full
(height $m$) column to delete, and similarly the condition
$\lambda^T_n\geq m+1$ states  that there exists a full (length $n$)
row to delete. Such Young tableau necessarily correspond to long
(typical) representations of $GL(m|n)$. The transformation~\eqref{eq:41} relates
representations that are equivalent under $SL(m|n)$ but not under
$GL(m|n)$. The modification of $\gamma$ then ensures the corresponding
induced $SL(2m|2n)$ representation is unchanged. Note that the above
transformations are also valid as they stand in the two bosonic cases $m=0$
or $n=0$. For $n=0$ the condition  $\lambda_n^T \geq m+1$  does not
make sense and is interpreted as always being satisfied for any Young
tableau. Then the transformation adds columns to the Young tableau in
favour of reducing $\gamma$. One possibility is to use this freedom to ensure that $\gamma=0$. This then corresponds precisely to the form
chosen in~\cite{Dolan:2243hv}. Similarly in the case $n=0$ we can ensure that
$\gamma=0$. However for short supersymmetric representations we can not
remove $\gamma$ entirely. Furthermore, if we perform this transformation to change $\gamma$, we no longer have 
the direct connection between $\gamma$ and the number of basic fields
$W$.~\footnote{A simple example of this in conformal field theory is
  provided by considering  the two operators $W^n$ and $\square W^{n-2}$
  in Minkowski space  where $W$ is a scalar field and the derivatives
  in $\square = \partial_{\alpha \dot \alpha} \partial^{\alpha \dot
    \alpha}$ can act anywhere appropriately to make a conformal
  primary (in fact one needs sums of such terms but we are being
  schematic here).  These two operators have the same dimension and spin and
  thus transform under the same representation of the conformal
  group. In our notation the first operator is given  as $\cO^{n[0]}$, the second as
  $\cO^{n{-}2\, [1,1]}$.}  
Indeed a simple way of removing the ambiguity would be to insist that we always have $\lambda^T_{n}\leq m$ (or equivalently $\lambda_{m+1}<n$) and if this is not the case then we use the above transformation to make it so.

We finish this section by giving three tables with  the translation between our description of
representations and the usual one in three cases of interest: the
bosonic conformal group, $\cN$=2 and $\cN$=4 SYM.

$$
\begin{array}{|c||c|c|}
  \multicolumn{3}{c}{\text{Translation between 4$d$ conformal reps and fields $\cO^{\gamma\ula}$}}\\ \hline
  GL(2) \text{ rep } \ula&  \text{dimension} & \text{spin} \\\hline
  [\lambda_1,\lambda_2]  & \gamma+\lambda_1+\lambda_2&\lambda_1-\lambda_2 \\ \hline
\end{array}
$$

$$
\begin{array}{|c||c|c|c|c|}
  \multicolumn{5}{c}{\text{Translation between $\cN=2$  superconformal reps and superfields $\cO^{\gamma\ula}$}}\\ \hline
  GL(2|1) \text{ rep }\ula& \text{dimension} & \text{spin}& SU(2) \text{ rep}&\text{multiplet type} \\\hline
[0]&\gamma&0&\gamma&\text{half BPS}\\\hline
  [\lambda]\ (\lambda\geq 1)& \gamma{+}\lambda{-}1&\lambda{-}1&\gamma-2&\text{semi-short} \\ \hline
{ [\lambda_1,\lambda_2,1^\mu]\  (\lambda_2\geq 1)} & \gamma{+}\lambda_1{+}\lambda_2{-}2&\lambda_1{-}\lambda_2&\gamma{-}2\mu-4&\text{long} \\ \hline
\end{array}
$$

$$
\begin{array}{|c||c|c|c|c|}
  \multicolumn{5}{c}{\text{Translation between $\cN=4$  superconformal reps and superfields $\cO^{\gamma\ula}$}}\\ \hline
  GL(2|2) \text{ rep }\ula& \text{dimension} & \text{spin}& SU(4) \text{ rep}&\text{multiplet type} \\\hline
[0]&\gamma&0&[0,\gamma,0]&\text{half BPS}\\\hline
  [\lambda,1^{\mu}]\ (\lambda\geq 2)& \gamma{+}\lambda{-}2&\lambda{-}2&[\mu,\gamma{-}2\mu{-}2,\mu]&\text{semi-short} \\  
  \left[1^\mu\right] & \gamma&0&[\mu,\gamma{-}2\mu,\mu]&\text{quarter BPS} \\ \hline
{ [\lambda_1,\lambda_2,2^{\mu_2},1^{\mu_1}]\  (\lambda_2\geq 2)} & \gamma{+}\lambda_1{+}\lambda_2{-}4&\lambda_1{-}\lambda_2&[\mu_1{-}\mu_2,\gamma{-}2\mu_1-4,\mu_1{-}\mu_2]&\text{long} \\ \hline
\end{array}
$$

\section{Conformal partial waves in (super)Grassmannian field theories }
\label{sec:conf-part-waves-3}
In this section we  consider four-point functions of scalar operators of arbitrary weight on the Grassmannian and  in particular obtain the (super) conformal partial wave associated with any operator occurring in the OPE of two of them. We will obtain explicit formulae for the partial waves, both as an expansion in Schur polynomials with given coefficients, and in a summed up form.

\subsection{The OPE and its relation to an expansion  in Schur polynomials}
\label{sec:ope-grmn-2m2n}

We here examine the connection between the OPE and conformal partial waves
of four-point functions in a general $Gr(m|n,2m|2n)$ field theory. We
take the OPE of two scalar operators, $\cO^{p_1},\cO^{p_2}$ with arbitrary integer weight $p_1,p_2$. In the $\cN=4$ context this corresponds to taking two half BPS operators with dimension $p_i$ and lying in the $SU(4)$ reps with Dynkin labels $[0,p_i,0]$.

The OPE takes the general form~\cite{Heslop:2001gp}
\begin{align}
  \label{eq:10}
  \cO^{p_1}(X_1) \cO^{p_2}(X_2) &= 
 \sum_{\cO}
  C_{p_1p_2}^{\cO} \,   g_{12}^{\frac{p_1+p_2-\gamma}2}  C^{\gamma,\underline \lambda;\underline
  A\underline A'}(X_{12},\partial_2)
  \cO^{\gamma\ula}_{\underline A \underline A'}(X_2),\nonumber\\
& \gamma=|p_{21}|,|p_{21}|+2,\dots, p_1+p_2\, ,
\end{align}
where we define $p_{ij}=p_i-p_j$ and where
\begin{align}
  g_{ij}=\text{sdet}(X_i-X_j)^{-1}
\end{align}
which becomes the (super)propagator in the physical cases where $m=2$. Here the sum  is over all superconformal primary operators  in the theory.  The object $C^{\gamma,\underline \lambda;\underline
  A\underline A'}(X_{12},\partial_2)$ is a formal expansion in powers of $X_{12}^{AA'}$ and derivatives $(\partial/\partial X_2)_{AA'}$ which act on the primary operator (thus producing descendant operators).
It takes the form
\begin{align}\label{eq:14}
  C^{\gamma,\underline \lambda;\underline
  A\underline A'}(X_{12},\partial_2) \cO^{\gamma\ula}_{\underline A \underline A'}(X_2)= \sum_{\umu \geq \underline \lambda} C_{\umu}^{\gamma\underline \lambda} \big(X_{12}^{|\umu|}\big)^{\underline B \underline B'} \big[\partial_2^{|\umu|-|\underline \lambda|}\cO^{\gamma\underline \lambda}\big]_{\underline
  B\underline B'},
\end{align}
where the sum is over all Young tableaux $\umu$ containing $\underline \lambda$, with $|\umu|=\sum_i \mu_i$ the number of boxes in the Young tableau $\umu$. There are $|\umu|$ powers of $X_{12}$ and both primed and unprimed indices are symmetrised into the representation $\umu$ according to the usual  Young tableau rules. This appropriately symmetrised multi-index is denoted $\underline B$ and $\underline B'$. Similarly in the descendant operator there are a total of $|\umu|$ primed and unprimed downstairs indices coming from both $\cO$ and the derivatives. These too are to be both symmetrised into the rep $\umu$ as indicated by the multi-index $\underline B,\underline B'$. Finally one should contract the $\underline B$ and $\underline B'$ indices

The first term in this expansion is always normalised to one
\begin{align}\label{eq:15}
  C_{\underline \lambda}^{\gamma\underline \lambda} =1,
\end{align}
but the remaining coefficients are unknown in general (although they are fixed by symmetry).

To obtain the contribution of operators to the four-point
function, insert the OPE into the four-point function twice (once at
points 1,2 and once at points 3,4) and use the two-point functions
(fixed by symmetry)
\begin{align}
  \langle \cO^{\gamma\ula}_{\underline A\underline A'}(X_2)
  \tilde \cO^{\gamma\ula}_{\underline B\underline B'}(X_4)\rangle = C_{\cO
    \tilde\cO} \, g_{24}^{\gamma}
  (X_{24}^{-|\ula|})_{\underline A' \underline B}  (X_{24}^{-|\ula|})_{\underline B' \underline A},
\end{align}
to obtain 
\begin{align}
  \label{eq:11}
&\langle  \cO^{p_1}(X_1) \cO^{p_2}(X_2)   \cO^{p_3}(X_3) \cO^{p_4}(X_4)
\rangle \nonumber \\&= 
\sum_{\cO,\tilde \cO}
  C_{p_1p_2}^{\cO}   C_{p_3p_4}^{\tilde \cO}  C_{\cO \tilde \cO}   \,
 g_{12}^{\frac{p_1+p_2-\gamma}2} g_{34}^{\frac{p_3+p_4-\gamma}2}   C^{\gamma,\underline A\underline A'}(X_{12},\partial_2)
   C^{\gamma,\underline B\underline B'}(X_{34},\partial_4)
 g_{24}^{\gamma}
  (X_{24}^{-|\ula|})_{\underline A' \underline B}
  (X_{24}^{-|\ula|})_{\underline B' \underline A}\ .
\end{align}
Here for $C_{\cO \tilde \cO}$ to be non-zero, the representations of
$\cO$ and $\tilde \cO$ must be the same.
In particular 
$\gamma$ takes on values appearing both in the range for the OPE $\cO^{p_1}(X_1)
\cO^{p_2}(X_2)$, ($|p_{12}| \leq \gamma\leq p_1+p_2$) as well as for the OPE $\cO^{p_3}(X_3)
\cO^{p_4}(X_4)$, ($|p_{34}| \leq \gamma\leq p_1+p_2$). If we assume
(without loss of generality) that $p_1+p_2
\leq p_3+p_4$ then there are two inequivalent
cases to consider
\begin{align}
  \label{eq:17}
&  \text{Case 1:} & |p_{12}|&\geq |p_{34}|\qquad \Rightarrow &
|p_{12}|\leq \gamma\leq p_1+p_2\nonumber \\
&  \text{Case 2:} & |p_{12}|&\leq |p_{34}|\qquad \Rightarrow & |p_{34}|\leq \gamma\leq p_1+p_2
\end{align}
Note that in case $2$, for a non-zero four-point function we clearly
need $p_1+p_2 - |p_{34}|$
to be positive and even. In $\cN=4$ SYM, the minimal cases with $p_1+p_2 -
|p_{34}|$ and
$p_1+p_2 -
|p_{34}|=2$ correspond to  the so-called extremal and next-to-extremal
cases and 
are protected~\cite{Eden:2000gg,Eden:1999kw}.

The conformal partial wave expansion given in~\eqref{eq:11} hides the conformal symmetry
of the four-point function. It is however possible to re-expand the
conformal partial wave in a way that makes the superconformal symmetry
manifest in terms of Schur polynomials.
\begin{align}
  \label{eq:11b}
&\langle  \cO^{p_1}(X_1) \cO^{p_2}(X_2)   \cO^{p_3}(X_3) \cO^{p_4}(X_4)
\rangle  \nonumber \\&= 
\sum_{\gamma,\ula}
 A^{p_1p_2p_3p_4}_{\gamma\ula}  \,
g_{12}^{\frac{p_1+p_2}2} g_{34}^{\frac{p_3+p_4}2} 
 \left(\frac{g_{24}}{g_{14}}\right)^{\frac12{p_{21}}}
 \left(\frac{g_{14}}{g_{13}}\right)^{\frac12{p_{43}}}
\left(\frac{g_{13}g_{24}}{g_{12}g_{34}}\right)^{\frac12\gamma}
\, F^{\alpha\beta \gamma\underline\lambda}(Z),
                       \nonumber\\[5pt]
&\qquad \qquad \alpha=\tfrac12(\gamma-p_{12}) \quad \beta=\tfrac12( \gamma +p_{34})\ ,
\end{align}
where
\begin{align}
&A^{p_1p_2p_3p_4}_{\gamma\ula}= \sum_{\cO^{\gamma\ula},\tilde \cO^{\gamma\ula}}
  C_{p_1p_2}^{\cO}   C_{p_3p_4}^{\tilde \cO}  C_{\cO \tilde \cO}\end{align}
and 
where  the conformal partial wave is given as a sum over Schur polynomials $s_{\mu}(Z)= Z^{\mu(A)}{}_{\mu(A)}$ (traces over irreps as described in the next section) 
\begin{align}
  \label{eq:12}
  F^{\alpha\beta\gamma\underline\lambda}(Z) = \sum_{\umu} R^{\alpha\beta \gamma\underline
    \lambda}_{\umu} Z^{\mu(A)}{}_{\mu(A)}\ ,
\end{align}
of the $GL(m|n)$ cross-ratio matrix $Z$
\begin{align}
  \label{eq:13}
  Z=X_{12}X_{24}^{-1}X_{43}X_{31}^{-1}\ ,
\end{align}
for some numerical coefficients $R^{\alpha\beta \gamma\underline
    \lambda}_{\umu}$ with
\begin{align}
  \label{eq:18}
  R^{\alpha\beta \gamma\underline
    \lambda}_{\underline \lambda}=1\ .
\end{align}

Here we have restricted ourselves to two cases without loss of generality
\begin{align}
&\text{Case 1:}\qquad  \big(p_1+p_2\leq p_3+p_4,\ p_1\geq p_2,\ p_3\geq p_4,\
  p_{12} \geq p_{34}\big) \nonumber \\&\qquad \qquad \qquad
                                        \alpha=\big(0,1, \dots p_2\big)\qquad \beta=\big(\tfrac12(p_{12}+p_{34}\big),\tfrac12(p_{12}+p_{34})+1,\dots, \tfrac12(p_{1}+p_2+p_{34}) \big)\nonumber \\
 & \qquad \qquad \qquad \gamma=\big(p_{12},p_{12}+2,\dots, p_1+p_2\big)  \nonumber\\[5pt]
&\text{Case 2:}\qquad  \big(p_1+p_2\leq p_3+p_4,\ p_2\geq p_1,\ p_4\geq p_3,\
  p_{21} \leq p_{43}\big) \nonumber \\&\qquad \qquad \qquad
    \alpha=\big(\tfrac12(p_{21}+p_{43}) ,\tfrac12(p_{21}+p_{43})+1, \dots p_2\big)\qquad \beta=\big(0,1,\dots,
    \tfrac12(p_{1}+p_2+p_{34}) \big)\nonumber \\
& \qquad \qquad \qquad \gamma=\big(p_{43},p_{43}+2,\dots, p_1+p_2\big)
\label{eq:42}
\end{align}
Note that in~\eqref{eq:11b} we have fixed the symmetry in swapping
points $1,2$ and $3,4$ differently in the two cases. This allows
a universal form for the prefactor. We can always choose an ordering
of operators consistent with the conformal partial wave expansion which fits into one of
the two cases above.

It is one of the main purposes of this paper  to derive a formula for the numerical  coefficients in~\eqref{eq:12},  $R^{\alpha\beta \gamma\underline
  \lambda}_{\umu}$.  Furthermore we would like to sum up the conformal partial wave expansion.

Crucially the coefficients $R^{\alpha\beta \gamma\underline \lambda}_{\umu}$ {\em only depend on $\alpha,\beta,\gamma$
  and the   Young tableaux $\umu,\underline\lambda$ but are independent of the group}. 
This fact can be seen by considering the limit of the $GL(2m|2n)$ Grassmannian field theory to either $GL(2(m{-}1)|2n)$ or $GL(2m|2(n{-}1))$. In this limit the partial waves $F^{pab\underline\lambda}(Z)$ simply become the equivalent partial waves for the reduced group (or vanish if the corresponding representation $\underline \lambda$ does not exist for the reduced isotropy group $GL(m{-}1|n)$ or $GL(m|n{-}1)$ respectively). Similarly the Schur polynomials  $Z^{\mu(A)}{}_{\mu(A)}$ become the equivalent Schur polynomial for the reduced $Z$ (or vanish). We thus conclude that the coefficients of the Schur polynomials in the partial wave must reduce directly, and hence be independent of $m,n$.

Let us derive explicitly the first term in the expansion as a sum over Schur polynomials~\eqref{eq:11b} starting from  the  form~\eqref{eq:11b}.
The first term in~\eqref{eq:11}  is obtained by inserting the first term in the expansion~\eqref{eq:14} together with~\eqref{eq:15} into~\eqref{eq:11b} to obtain  
\begin{align}
  &\langle  \cO^{p_1}(X_1) \cO^{p_2}(X_2)   \cO^{p_3}(X_3) \cO^{p_4}(X_4)
\rangle \nonumber\\&= 
\sum_{\cO,\tilde \cO}
  C_{p_1p_2}^{\cO}   C_{p_3p_4}^{\tilde \cO}  C_{\cO \tilde \cO}   \,
 g_{12}^{\frac{p_1+p_2-\gamma}2} g_{34}^{\frac{p_3+p_4-\gamma}2}   (X_{12}^{|\underline \lambda|})^{\underline A\underline A'}
  (X_{34}^{|\underline \lambda|})^{\underline B\underline B'} 
 g_{24}^{\gamma}
  (X_{24}^{-1})_{\underline A' \underline B}
  (X_{24}^{-1})_{\underline B' \underline A} +  O(X_{12},X_{34})\nonumber\\&= 
\sum_{\cO,\tilde \cO}
  C_{p_1p_2}^{\cO}   C_{p_3p_4}^{\tilde \cO}  C_{\cO \tilde \cO}   \,
 g_{12}^{\frac{p_1+p_2-\gamma}2} g_{34}^{\frac{p_3+p_4-\gamma}2}  g_{24}^{\gamma}
  (X_{12}X_{24}^{-1}X_{34}X_{24}^{-1})^{\ula(A)}{}_{\ula(A)}\ +\   O(X_{12},X_{34})\ .
\end{align}
The object $(X_{12}X_{24}^{-1}X_{34}X_{24}^{-1})^{\underline A}{}_{\underline A}$ is the trace over the representation $\ula$ of $Z=X_{12}X_{24}^{-1}X_{34}X_{24}^{-1}$ and is hence equal to the Schur polynomial $s_{\ula}(x|y)$ (as we shall see shortly).

\subsection{Free field theory OPE and Wick's theorem}
\label{sec:free-field-theory}

The discussion of the OPE in section~\ref{sec:ope-grmn-2m2n} is completely general and essentially only uses symmetry. However in a free quantum field theory we can be much more explicit and give precise expressions for the operators under consideration. 

As described in~\cite{Arutyunov:2240ku} the easiest way to derive the
OPE in a free field theory context is to simply use Wick's theorem. The time
ordered product of two operators $\cO_{p_1}(X_1) \cO_{p_2}(X_2)$ is
equal to the normal ordered product, together with the sum over
contractions multiplied by appropriate powers of propagators. In this
context, we get that (for $p_1 \leq p_2$)
\begin{align}
  \label{eq:8}
\cO_{p_1}(X_1) \cO_{p_2}(X_2) &= :\cO_{p_1}(X_1) \cO_{p_2}(X_2): \ +\
\sum_{p=0}^{p_1-1}  g_{12}^{p_1-p} \cO_{p_2-p_1+2p}(X_1,X_2)\, ,
\end{align}
where for example $\cO_{p_1+p_2-2}$ is the result of a single contraction\footnote{Here, so this can be applied to $\cN=4$ SYM we are including the
possibility of some
colour  structure in the definition of our operators. So $ \cO_{p_1}:=\tr(W^{p_1})$ is a
single trace gauge invariant operator. Then $:\cO_{p_1}(X_1)
\cO_{p_2}(X_2):$ is a double trace bilocal operator. We can
of course ignore the gauge structure if we wish to consider a more
abstract context (as we will do shortly) or equivalently simply consider the gauge group to be $U(1)$. } 
\begin{align}
  \label{eq:9}
 \contraction{\cO_{p_1+p_2-2}(X_1,X_2)= :\tr(W^{p_1-1}}{W}{)(X_1) \tr(}{W}
\cO_{p_2-p_1+2p}(X_1,X_2)= \tr(W^{p_1-1} W)(X_1) \tr(W W^{p_2-1})(X_2):\ ,
\end{align}
whereas $\cO_{p_1-p_2-4}$ will involve two contractions etc. Here the contractions simply give a Kronecker delta in the corresponding adjoint gauge index. 

Now one Taylor expands the RHS and rearranges into primaries and
descendants to obtain~\eqref{eq:10}  but with explicit expressions for the operators which appear.

So if $\gamma=p_1+p_2$, the operators are double trace operators from the  product (in general with derivatives) of $\cO_{p_1}$ and $\cO_{p_2}$.  If however  $\gamma=p_1+p_2-2$, then in the $U(N)$ theory the single Wick contraction will glue together  the two traces to form a single trace. Similarly for the  $SU(N)$ theory in the large $N$ limit. For finite $N$ in the $SU(N)$ theory however there will be a $1/N$ correction (from writing the Kronecker delta's in adjoint indices back in terms of fundamental gauge indices via $T^a_{ij}T^a_{kl} = \delta_{il}\delta_{jk}-1/N \delta_{ij}\delta_{kl}$)
giving back a double trace operator.

\subsection{Schur polynomials of $GL(m|n)$}

\subsubsection{$GL(m)$ characters (Schur polynomials)}

Given a partition $\underline \lambda =[\lambda_1,\lambda_2,\dots ,\lambda_m]$ with $\lambda_1\geq \lambda_2\geq\dots \geq \lambda_m$, the corresponding Schur polynomial is the symmetric polynomial of $m$ variables $x_i, i=1\dots m$,  given by 
\begin{align}
s_{\underline\lambda}(x) =  \frac{\det \Big(x_i^{\lambda_j+m-j} \Big)_{1\leq i,j \leq
    m}}{\det\Big(x_i^{m-j}\Big)_{1\leq i,j \leq
    m}}\ .\label{eq:3}
\end{align}
The Schur polynomial is the character of the corresponding $GL(m)$
representation described by a Young tableau with row lengths
$\lambda_i$.
In particular, the Schur polynomial is the trace over the representation
$R_{\underline\lambda}$ of an element $Z \in GL(m)$ written as a
function of the $m$ eigenvalues $x_i$ of $Z$,
\begin{align}
  \label{eq:4}
s_{\underline \lambda}(x) = \tr\big(R_{\underline \lambda}(Z)\big)\ . 
\end{align}

A  $GL(m)$ Schur polynomial containing a full, length $m$, column
is equal to the Schur polynomial with that column deleted, multiplied
by the product of all $x$'s:
\begin{align}
  \label{eq:38}
  s_{[\underline\lambda+1]}(x) = (\textstyle{\prod_{i=1}^m x_i })\times 
  s_{[\underline\lambda]}(x) 
\end{align}
where $[\ula+1]:=[\lambda_1+1,\lambda_2+2,\dots]$.

For example for $GL(2)$ the
fundamental  representation
has character
 $\tr(Z)=x_1+x_2$ in agreement with
the formula above for $\underline\lambda=[1]$.
As another example, again for $GL(2)$, consider
$\underline\lambda=[1,1]$ corresponding to the antisymmetric rep. The
trace over the representation gives
\begin{align}
  \tr\big(R_{\ytableausetup{mathmode,
      boxsize=.2em}\ydiagram{1,1}}(Z)\big)=Z^{[i}_iZ^{j]}_j=1/2\big(\tr(Z)^2-\tr(Z^2)\big)=x_1x_2\label{eq:1}
\end{align}
and the Schur polynomial formula~\eqref{eq:3} gives the same result $s_{[1,1]}(x)=x_1x_2$.

\subsubsection{$GL(m|n)$ characters (super Schur polynomials)}

In just the same way we define the super-Schur polynomial as the
characters of the supergroup $GL(m|n)$ just as
in~\eqref{eq:4} but this time using the supertrace
\begin{align}
  \label{eq:4s}
s_{\underline \lambda}(x|y) = \str\big(R_{\underline \lambda}(Z)\big)\ ,
\end{align}
where we define the eigenvalues of $g \in GL(m|n)$ to be $x_i\, y_j\,
i=1\dots m,\, j=1\dots n.$ Thus for example for the fundamental
representation the character is  simply
the supertrace of $g$ so $s_{(1)}(x|y)=\str(Z)=\sum_i x_i - \sum_j
y_j$ with the minus sign due to the nature of the supertrace.

In 2003 Moens and Van der Jeugt wrote down a
remarkable 
 determinantal formula for the super Schur
 polynomials~\cite{moens2003determinantal}. This formula is the
 analogue of the determinantal formula~\eqref{eq:3} for the standard
 Schur polynomials and takes the form of a $(n+k-1)\times (n+k-1)$
 determinant\footnote{The minus signs here agree with those
  of~\cite{moens2003determinantal} after sending $y_j\rightarrow
  -y_j$ (bringing a $(-1)^{n(n-1)/2}$ from $D$) and swapping  the
  columns so that $R$ appears in the top left block.}
 \begin{align}
   \label{eq:5}
  s_{\underline \lambda}(x|y)=
(-1)^{(n-1)(m + (k - 1) + n /2)}\, D^{-1}
 \det \left(
     \begin{array}{cc}
X_{\underline\lambda}&R\\
         0&  Y_{\underline\lambda^T}
         \end{array}
\right)\ ,
 \end{align}
where
\begin{align}
  \label{eq:6}
X_{\underline \lambda}&=\Big(x_i^{\lambda_j+m-n-j} \Big)_{\substack{1\leq i
  \leq m\\1\leq j
  \leq k-1}}&
R&=\left(\frac1{x_i-y_j}\right)_{1\leq i \leq m,\   1\leq j \leq n}
\nonumber\\[5pt]
Y_{\underline \lambda^T}&=\Big((-y_j)^{\lambda^T_i+n-m-i} \Big)_{\substack{1\leq i
  \leq k'-1\\1\leq j
  \leq n}}
& 
  D&=\frac{\prod_{1\leq i<j \leq m} (x_i-x_j)\prod_{1\leq i<j \leq n}
  (y_i-y_j)}{\prod_{1\leq i\leq m,\,1\leq j \leq n} (x_i-y_j)  }
\ .
\end{align}
and 
\begin{align}
  \label{eq:21}
k=\text{min}\{j: \lambda_j+m-n-j<0\} \qquad k'=\text{min}\{i:
\lambda^T_i+n-m-i<0\}\ .
\end{align}
In~\cite{moens2003determinantal}, the number $k$ was called the ``atypicality'' of the
representation and in fact, as we shall see shortly 
\begin{align}
  k'=k-m+n\label{eq:7}  \ .
\end{align}

Here $\underline \lambda^T$
is the conjugate partition to $\underline \lambda$ (so $\lambda^T_i$
is the length of column $i$).
This formula is only valid if the Young Tableau has an allowed shape
consistent with $GL(m|n)$ i.e. $\lambda_{m+1}\leq n$. If this is not
the case the Schur polynomial vanishes (although the above formula
will not give this automatically).

The restriction on the number of columns of $X_{\underline \lambda}$ to $k-1$ is explained by
considering the power appearing in $X_{\underline \lambda}$ and
comparing with the definition of $k$~\eqref{eq:21}. Clearly the number
of columns of 
$X_{\underline \lambda}$ is defined to be as large as possible without
having negative powers of $x_i$. The same is true for the restriction
on the number of rows of $Y_{\underline \lambda^T}$ to be less than or
equal to $k'-1$.  
It is useful to consider this pictorially. Here we consider an example
of a $GL(m|n)$ rep (with $m=7,n=10$).
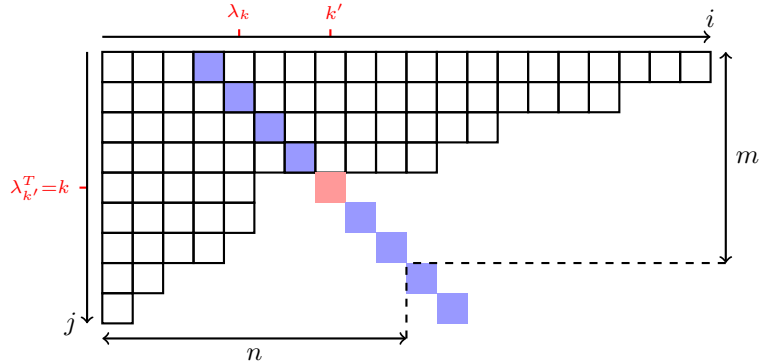
\begin{center}
  \begin{Tableau}{ {,,,,,,,,,,,,,,,,,,,}, {,,,,,,,,,,,,,,,,},
      {,,,,,,,,,,,,}, {,,,,,,,,,,}, {,,,,}, {,,,,}, {,,,}, {,}, {} }
\filldraw[draw=black, fill=blue!40!white] (3,-1) rectangle (4,0);
\filldraw[draw=black, fill=blue!40!white] (4,-2) rectangle (5,-1);
\filldraw[draw=black, fill=blue!40!white] (5,-3) rectangle (6,-2);
\filldraw[draw=black, fill=blue!40!white] (6,-4) rectangle (7,-3);
\fill[red!40!white ] (7,-5) rectangle (8,-4);
\fill[blue!40!white] (8,-6) rectangle (9,-5);
\fill[blue!40!white] (9,-7) rectangle (10,-6); 
\fill[blue!40!white] (10,-8) rectangle (11,-7);
\fill[blue!40!white] (11,-9) rectangle (12,-8);   
    \draw[dashed] (20.5,-7)--(10,-7); \draw[dashed]
    (10,-9.5)--(10,-7); \draw[thick,<->]
    (20.5,-7)--node[anchor=west]{$m$}(20.5,0) ; 
    \draw[thick,<->]
    (0,-9.5)--node[anchor=north]{$n$}(10,-9.5) ; 
    \draw[thick,red]
    (-0.5,-4.5)--(-.75,-4.5)node[anchor=east]{$\scriptstyle \lambda^T_{k'}=k$} ;
        \draw[thick,red]
    (4.5,0.5)--(4.5,0.75)node[anchor=south]{$\scriptstyle\lambda_k$} ;
            \draw[thick,red]
    (7.5,0.5)--(7.5,0.75)node[anchor=south]{$\scriptstyle
      \ k'$} ;
            \draw[thick,black,->]
    (0,.5)--(20,.5) node[anchor=south]{$i$} ;
    \draw[thick,black,->]
    (-0.5,0)--(-0.5,-9) node[anchor=east]{$j$} ;
 \end{Tableau}
\end{center}
Any non-zero $GL(m|n)$ Young Tableau is restricted to
fit into a hook shape of height $m$ and width $n$ as illustrated by the
dashed lines. This is equivalent to the statement that
$\lambda_{m+1}\leq n$ for a non-zero representation. We label the row number as $i$ and the column number
with $j$. Then consider boxes with $i-n=j-m$ (shaded boxes in the
diagram). The
atypicality of the representation, $k$,  is the row number (and $k'$
the column number) of the shaded box lying just below (or just to the
right) of the Young Tableau
(the pink box in the diagram). 

The power of $x_i$ in the matrix $X_{\underline \lambda}$,
$\lambda_j+m-n-j$ is represented by the number of boxes to the right
of the shaded box in row $j$. Clearly this number becomes negative if
$j \geq k$ and thus the matrix must be  restricted to $j\leq k-1$ if
we wish to  avoid
negative powers. Similarly the power of $y_j$ in the matrix $Y_{\underline \lambda^T}$,
$\lambda^T_i+n-m-i$ is represented by the number of boxes below the
shaded box in column $i$ (one should think of the shaded boxes as
continuing above the Young tableau in the example). This number becomes negative if
$i \geq k'$ and thus this matrix must be  restricted to $i\leq k'-1$.
From the diagram it is also clear that~\eqref{eq:7} $k'=k-m+n$.

Let us give an explicit example. Consider $GL(2|3)$ and
$\underline\lambda=(3,2,2,1)$. We have $\underline\lambda^T=(4,3,2)$
and $(k,k')=(2,3)$ so the formula for the Schur polynomial~\eqref{eq:5} and the associated shaded Young
tableau are
\begin{align}
  \label{eq:16}
  s_{\underline\lambda}(x|y)= D^{-1}
\det
\left(
\begin{array}{cccc}
 \frac{1}{x_1-y_1} & \frac{1}{x_1-y_2} & \frac{1}{x_1-y_3} & x_1 \\
 \frac{1}{x_2-y_1} & \frac{1}{x_2-y_2} & \frac{1}{x_2-y_3} & x_2 \\
 y_1^4 & y_2^4 & y_3^4 & 0 \\
 y_1^2 & y_2^2 & y_3^2 & 0 \\
\end{array}
\right)
\qquad \qquad \qquad 
  \begin{Tableau}{{,,},{,},{,}, {\ }}
\fill[blue!40!white] (0,0) rectangle (1,1);
\draw[thick] (0,0)--(1,0);
\filldraw[draw=black, fill=blue!40!white] (1,-1) rectangle (2,0);
\fill[color=red!40!white] (2,-2) rectangle (3,-1);
\draw[thick] (2,-2)--(2,-1)--(3,-1);
\fill[blue!40!white] (3,-3) rectangle (4,-2);
\fill[blue!40!white] (4,-4) rectangle (5,-3);
    \draw[dashed] (5,-2)--(3,-2); 
    \draw[dashed]
    (3,-2)--(3,-4); 
                         \end{Tableau}\ .
\end{align}
Here we see explicitly that the row lengths to the right of the shaded diagonal give the $x$
exponents (here just a single row of length 1) and the column lengths
to the left of the diagonal give the $y$ exponents (here they are $2$
and $4$).

In appendix~\ref{sec:altern-form-glmn} we give an alternative form for  the super Schur polynomials. The alternative form reduces straightforwardly to the form here, but has a closer relation to the super conformal partial waves.

\subsubsection{Long (typical) reps and multiplet shortening for Schur polynomials}

In supergroups, representations occur as ``typical'' or ``atypical''
representations. Typical representations are Long representations,
essentially having the maximal odd dimension allowed, whereas
``atypical'' representations are short. Typical representations of
$GL(m|n)$ are ones for which the atypicality $k=m+1$ (implying
$k'=n+1$ from~\eqref{eq:7}) and so the first $m$ rows and first $n$
columns are fully occupied and $\lambda_m\geq n, (\lambda^T)_n\geq m$.
Thus their Young Tableau can be described by the arbitrarily long horizontal Young
tableau $\ula_x$ to the right of the $m\times n$ block, and the 
arbitrarily high vertical Young tableau $\ula_y$ 
attached to the bottom of the $m\times n$ block. 
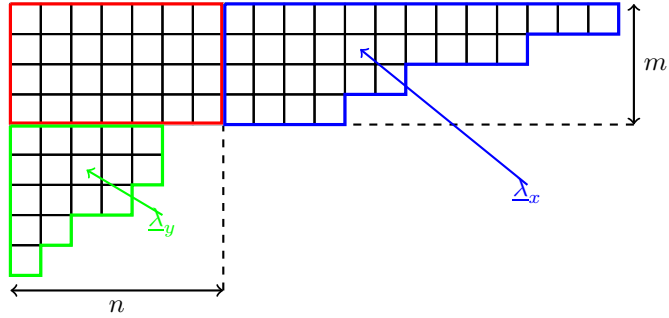
\begin{center}
  \begin{Tableau}{ {,,,,,,,,,,,,,,,,,,,}, {,,,,,,,,,,,,,,,,},
      {,,,,,,,,,,,,}, {,,,,,,,,,,}, {,,,,}, {,,,,}, {,,,}, {,}, {} }
    \draw[dashed] (20.5,-4)--(7,-4); 
    \draw[dashed]
    (7,-9.5)--(7,-4); \draw[thick,<->]
    (20.5,-4)--node[anchor=west]{$m$}(20.5,0) ; 
    \draw[thick,<->]
    (0,-9.5)--node[anchor=north]{$n$}(7,-9.5) ; 
    \draw[thick,white]
    (0,0)--(7,0)--(7,-4)--(0,-4)--(0,0);
    \draw[very thick,red]
    (0,0)--(6.95,0)--(6.95,-3.95)--(0,-3.95)--(0,0); 
  \draw[very thick,blue]
    (7.05,0)--(20,0)--(20,-1)--(17,-1)--(17,-2)--(13,-2)--(13,-3)--(11,-3)--(11,-4)--(7.05,-4)--(7.05,0);
  \draw[very thick,green]
    (0,-4.05)--(5,-4.05)--(5,-6)--(4,-6)--(4,-7)--(2,-7)--(2,-8)--(1,-8)--(1,-9)--(0,-9)--(0,-4.05);
    \draw[thick,blue,->]
    (17,-6)node[anchor=north]{$\ula_x$}--(11.5,-1.5); 
    \draw[thick,green,->]
    (5,-7)node[anchor=north]{$\ula_y$}--(2.5,-5.5); 
 \end{Tableau}
\end{center}
In this example the $m\times n$ block is bounded in red. If one
deleted this block you would be left with two  Young tableaux one we
call $\ula_x$ and the other $\ula_y$. So the full Young tableau
is given in terms of $\ula_x$ and $\ula_y$ as
\begin{align}
  \label{eq:32}
  \ula= [\ula_x+n,\ula_y] 
\end{align}
where by $\ula_x+n$ we simply mean add $n$ to each row.

Typical representations are very simple and this is reflected in
their Schur polynomials which factorise:
\begin{align}
  \label{eq:33}
 \ula \text{ typical} \quad \Rightarrow \quad  s_{\ula}(x|y)= s_{\ula_x}(x)s_{\ula_y^T}(-y) \times \prod_{1\leq i\leq m,\,1\leq j \leq n} (x_i-y_j).
\end{align}
where $s_{\ula_y^T}(-y)$ is the ordinary bosonic $SU(n)$ Schur polynomial in the variables $-y_i$ of the conjugate representation to $\ula_y$.

This can be easily verified from determinantal form of the super Schur
polynomial~\eqref{eq:5} since when $k=m+1,k'=n+1$, the matrix  splits into an $m\times m$
block and an $n\times n$ block with a zero in the lower $n\times m$
block. Thus the determinant factorises into the determinant of
$X_\ula$ and $Y_{\ula}$.  

Furthermore, if we consider this factorisation together with~\eqref{eq:38}, this then implies
that if $\ula_x$ contains a  full 
($m$ row) column then we can delete this column in favour of adding a
full (length $n$) row, up to multiplication by a  factor:
\begin{align}
  \label{eq:39}
  s_{\ula}(x|y)= \frac{\prod_{i=1}^m x_i}{\prod_{j=1}^n (-y_j)}\times
  s_{\ula'}(x|y) \qquad \quad [\ula]= [\ula_x{+}n,\ula_y],\  [\ula']= [\ula_x{-}1{+}n,n,\ula_y]
\end{align}

What is less obvious is that the sum of certain atypical
representations with $k=m, k'=n$ can sum to a
factorised form. Specifically, let $\ula_x$ be an  $SL(m)$ (i.e. $m-1$
row) Young
tableau  and similarly let $\ula_y^T$ be a $SL(n)$ (i.e. $n-1$
row) Young
tableau. Then consider the three $GL(m|n)$ Young tableaux
$\ula, \ula_1,\ula_2$ with $\ula$ the typical  representation defined
in~\eqref{eq:32} and $\ula_1,\ula_2$ the two short Young tableaux
\begin{align}\label{eq:37}
  \ula_1&=[\ula_x+(n{-}1),n{-}1,\ula_y]\\
  \ula_2&=[\ula_x+n,\ula_y]\\
  \ula&=[\ula_x+n,n,\ula_y]\ .
\end{align}
Then the sum of the appropriately weighted 
$GL(m|n)$ Schur polynomials factorise:
\begin{align}
  \label{eq:35}
 \left(\textstyle{\prod_{j=1}^n x_j}\right) \times s_{\ula_1}(x|y) +
 \left(\textstyle{\prod_{i=1}^m(-y_i)}\right) \times  s_{\ula_2}(x|y) = s_{\ula}(x|y)
\end{align}

In $\cN=4$ SYM this phenomenon corresponds to long multiplets
decomposing into short multiplets at the unitary bound. We illustrate
this in the following diagram

\begin{center}
$\left({\prod_{j=1}^n x_j}\right)\ \times$
\scalebox{0.6}{
\begin{Tableau}{ {,,,,,,,,,,,,,,,,}, {,,,,,,,,,,,,,},
      {,,,,,,,,,}, {,,,,,}, {,,,,}, {,,,,}, {,,,}, {,}, {} }
    \draw[dashed] (17.5,-4)--(7,-4); 
    \draw[dashed]
    (7,-9.5)--(7,-4); \draw[thick,<->]
    (17.5,-4)--node[anchor=west]{$m$}(17.5,0) ; 
    \draw[thick,<->]
    (0,-9.5)--node[anchor=north]{$n$}(7,-9.5) ; 
                  \draw[very thick,blue]
    (6,0)--(17,0)--(17,-1)--(14,-1)--(14,-2)--(10,-2)--(10,-3)--(8,-3)--(6,-3)--(6,0);
  \draw[very thick,green]
    (0,-4)--(5,-4)--(5,-6)--(4,-6)--(4,-7)--(2,-7)--(2,-8)--(1,-8)--(1,-9)--(0,-9)--(0,-4);
    \draw[thick,blue,->]
    (17,-6)node[anchor=north]{$[\ula_x]$}--(11.5,-1.5); 
    \draw[thick,green,->]
    (5,-7)node[anchor=north]{$[\ula_y]$}--(2.5,-5.5); 

 \end{Tableau}
}
$+\ \left({\prod_{i=1}^m(-y_i)}\right)\ \times$
\scalebox{0.6}{
\begin{Tableau}{ {,,,,,,,,,,,,,,,,,}, {,,,,,,,,,,,,,,},
      {,,,,,,,,,,}, {,,,,}, {,,,,}, {,,,}, {,}, {} }
    \draw[dashed] (18.5,-4)--(7,-4); 
    \draw[dashed]
    (7,-8.5)--(7,-4); \draw[thick,<->]
    (18.5,-4)--node[anchor=west]{$m$}(18.5,0) ; 
    \draw[thick,<->]
    (0,-8.5)--node[anchor=north]{$n$}(7,-8.5) ; 
  \draw[very thick,blue]
    (7,0)--(18,0)--(18,-1)--(15,-1)--(15,-2)--(11,-2)--(11,-3)--(9,-3)--(7,-3)--(7,0);
  \draw[very thick,green]
    (0,-3)--(5,-3)--(5,-5)--(4,-5)--(4,-6)--(2,-6)--(2,-7)--(1,-7)--(1,-8)--(0,-8)--(0,-3);
    \draw[thick,blue,->]
    (17,-6)node[anchor=north]{$[\ula_x]$}--(11.5,-1.5); 
    \draw[thick,green,->]
    (5,-7)node[anchor=north]{$[\ula_y]$}--(2.5,-5.5); 
 \end{Tableau}
}
\end{center}
\begin{center}
 $ =$
\scalebox{0.7}{
\begin{Tableau}{ {,,,,,,,,,,,,,,,,,}, {,,,,,,,,,,,,,,},
      {,,,,,,,,,,}, {,,,,,,}, {,,,,}, {,,,,}, {,,,}, {,}, {} }
    \draw[dashed] (18.5,-4)--(7,-4); 
    \draw[dashed]
    (7,-9.5)--(7,-4); \draw[thick,<->]
    (18.5,-4)--node[anchor=west]{$m$}(18.5,0) ; 
    \draw[thick,<->]
    (0,-9.5)--node[anchor=north]{$n$}(7,-9.5) ; 
  \draw[very thick,blue]
    (7,0)--(18,0)--(18,-1)--(15,-1)--(15,-2)--(11,-2)--(11,-3)--(9,-3)--(7,-3)--(7,0);
  \draw[very thick,green]
    (0,-4)--(5,-4)--(5,-6)--(4,-6)--(4,-7)--(2,-7)--(2,-8)--(1,-8)--(1,-9)--(0,-9)--(0,-4);
    \draw[thick,blue,->]
    (17,-6)node[anchor=north]{$[\ula_x]$}--(11.5,-1.5); 
    \draw[thick,green,->]
    (5,-7)node[anchor=north]{$[\ula_y]$}--(2.5,-5.5); 
 \end{Tableau}
}
\end{center}

This equality can 
be proved from the
determinantal formula for Schur polynomials~\eqref{eq:5} and we just give a very brief sketch of how the proof goes here.  The matrices corresponding to the ``nearly long'' cases $\ula_1,\ula_2$  are ``nearly block triangular'' and thus the determinant takes the form of a sum of products of minors multiplied by components of $R$, $1/(x_i-y_j)$. The minors being summed over are very similar in each case $\ula_1$ and $\ula_2$. The non-trivial part of the sum on the LHS of~\eqref{eq:35} reduces then to $x_i/(x_i-y_j)-y_j/(x_i-y_j)=1$. We then end up with a sum of products of minors and one can match that with the RHS via the standard formula for determinants.

We should also point out here that long (typical) supersymmetric representations can have non-integer quantum numbers. This can be incorporated into this Young tableau setting by introducing  ``quasi-tensors'' as in~\cite{Heslop:2243xu}.

\subsection{Conformal  Partial waves}
\label{sec:conf-part-waves}

\subsubsection{$GL(m)$ conformal partial waves}

The four-dimensional conformal partial waves are well known
from~\cite{Dolan:2000ut}. In the Grassmannian $GL(m|n)$ set up that we are considering
here, they correspond to $m=2,n=0$ and are given by
\begin{align}
  \label{eq:22}
  F^{\alpha\beta\gamma\underline \lambda}(x_1,x_2) &= \frac{x_1^{\lambda_1+1}
    x_2^{\lambda_2}{}_2F_1(\lambda_1{+}\alpha,\lambda_1{+}\beta;2\lambda_1{+}\gamma;x_1){}_2F_1(\lambda_2{+}\alpha{-}1,\lambda_2{+}\beta{-}1;2\lambda_2{+}\gamma{-}2;x_2)
  \ -\ x_1 \leftrightarrow x_2 }{x_1-x_2}\,
\end{align}
where from~\eqref{eq:11b}
\begin{align}
   \alpha=\tfrac12(\gamma-p_{12}) \quad \beta=\tfrac12( \gamma +p_{34})\ .
\end{align}
Note that here,  and for $GL(m)$ groups in general, there is a redundancy in this description, since
\begin{align}
  F^{\alpha\beta\gamma[\underline\lambda]}(x)  = (x_1\dots x_m)^{-\delta} F^{(\alpha{-}\delta)(\beta{-}\delta)(\gamma{-}2\delta) [\underline\lambda+\delta] }(x)  \label{eq:24}
      \end{align}
      where $[\underline\lambda+\delta]:=[\lambda_1+\delta,\lambda_2+\delta,\dots]$.
This can be seen from its definition~\eqref{eq:11b}, together with the redundancy in the definition of the operators as discussed in~\eqref{eq:41}. It can also be seen directly to be the case for $GL(2)$ from~\eqref{eq:22}. 
This redundancy can be used for example to set $\gamma=0$.
Nevertheless we keep it in here for easier comparison to the
supersymmetric case where it is not
redundant (at least for short representations).

First note that~\eqref{eq:24} can be rewritten in the suggestive
determinantal form
\begin{align}\label{eq:26}
 F^{\alpha\beta\gamma\underline \lambda}(x_1,x_2) &= \frac{\det\Big(
   x_i^{\lambda_j+2-j}{}_2F_1(\lambda_j{+}1{-}j{+}\alpha ,\lambda_j{+}1{-}j{+}\beta;2\lambda_j{+}2{-}2j{+}\gamma;x_i)\Big)_{1\leq i,j \leq
    2}}{x_1-x_2}\ .
\end{align}
This form has a close correspondence with the formula for Schur polynomials
in~\eqref{eq:3}. Indeed it is manifestly a sum of
Schur polynomials, as in~\eqref{eq:12} and, in particular one can see very directly that the
first term in 
the OPE expansion (obtained by setting all the hypergeometric functions to
one) {\em is} the corresponding Schur polynomial. 

This form also then suggests to consider a simple generalisation to arbitrary
$GL(m)$ groups, namely
\begin{align}\label{eq:25}
 F^{\alpha\beta\gamma\underline\lambda}(x) &=\frac{\det\Big(
   x_i^{\lambda_j+m-j}{}_2F_1(\lambda_j{+}1{-}j{+}\alpha ,\lambda_j{+}1{-}j{+}\beta;2\lambda_j{+}2{-}2j{+}\gamma;x_i)\Big)_{1\leq i,j \leq
    m}}{\det\Big(
   x_i^{m-j}\Big)_{1\leq i,j \leq
    m}}\ .
\end{align}

Remarkably we find that this natural  generalisation is indeed the correct
answer as we show in appendix~\ref{sec:proof-conf-part}. Furthermore it allows us to derive the superconformal partial waves in an arbitrary $GL(m|n)$ theory.

First we expand out the $GL(m)$ partial waves into Schur polynomials,
expanding out the hypergeometric functions:
\begin{align}
 x_i^{\lambda_j+m-j} {}_2F_1(\lambda_j{+}1{-}j{+}\alpha
 ,\lambda_j{+}1{-}j{+}\beta;2\lambda_j{+}2{-}2j{+}\gamma;x_i)=
\sum_{\mu_j=0}^\infty
\frac{(\lambda_j{+}1
{-}j{+}\alpha)^{(\mu_j{-}\lambda_j)}
(\lambda_j{+}1{-}j{+}\beta)^{(\mu_j{-}\lambda_j)}  } {(\mu_j{-}\lambda_j)!(2\lambda_j{+}2{-}2j{+}\gamma)^{(\mu_j{-}\lambda_j)}   }  x_i^{\mu_j+m-j}
\end{align}
where $a^{(n)}=a(a+1)\dots(a+n-1)$ is the rising factorial or
Pochhammer symbol. 
Plugging this expansion into the determinant~\eqref{eq:25} 
we obtain
\begin{align}\label{eq:29b}
F^{\alpha\beta\gamma\ula}(x) \ &=\   \sum_{\mu_1=0}^\infty\dots\sum_{\mu_m=0}^\infty
r^{\alpha\beta\gamma\underline\lambda}_{\mu_1\dots \mu_m} \, \frac{\det\Big(
   x_i^{\mu_j+m-j}\Big)_{1\leq i,j \leq
    m}}{\det\Big(
   x_i^{m-j}\Big)_{1\leq i,j \leq
    m}}\nonumber\\
&=\   \sum_{[\umu]}
R^{\alpha\beta\gamma\underline\lambda}_{\umu} \,
s_{\umu}(x),
\end{align}
where
\begin{align}
  \label{eq:27}
r^{\alpha\beta\gamma\underline\lambda}_{\mu_1\dots \mu_m}\ &=\ \prod_{j=1}^m\frac{(\lambda_j +1
-j+\alpha)^{(\mu_j{-}\lambda_j)}
(\lambda_j +1 -j+\beta)^{(\mu_j-\lambda_j)}  }
{(\mu_j-\lambda_j)!(2\lambda_j +2 -2j+\gamma)^{(\mu_j-\lambda_j)}   },
\nonumber\\
R^{\alpha\beta\gamma\underline\lambda}_\umu\ &=
 \ \sum_{\sigma\in  S_m}
  (-1)^{|\sigma|}\,r^{\alpha\beta\gamma\underline\lambda}_{w_\sigma(\mu_1,\dots,\mu_m)},
\end{align}
and where
\begin{align}
  \label{eq:28}
  w_\sigma(\mu_1,\dots,\mu_m)=
  (\mu_{\sigma_1}+1-\sigma_1,\mu_{\sigma_2}+2-\sigma_2,\dots ,\mu_{\sigma_m}+m-\sigma_m),
\end{align}
is an affine Weyl reflection. The first line of~\eqref{eq:29b} is
obtained by simply inserting the expansion of the hypergeometric functions and
factoring out the coefficients from the determinant. In the second line
we first recognise the ratio of determinants as a Schur
polynomial~\eqref{eq:3} and we reorder the sum so that it runs
over ordered $\mu_j$'s,   $\mu_1 \geq \mu_2 \geq ...\geq \mu_m$. We do
this by performing an affine Weyl reflection whenever they are in the
wrong order. For the Schur polynomial this just corresponds to swapping
columns of the matrix in the numerator and hence brings a minus sign
for each swap. As an example of this is the $\gamma=6$ conformal partial wave, with $\alpha=\beta=3$. We need to consider $S_{3}$ in which case there are $6$ generators of the affine Weyl group.

\begin{align}&\notag F^{336 \underline{\lambda}}=\sum_{\sigma\in S_3}\sum_{\underline{\mu}\geq \underline{\lambda}}r^{336\underline{\lambda}}_{w_{\sigma}(\mu_1,\mu_2,\mu_3)}s_{\underline{\mu}}(x)\\& = \sum_{\underline{\mu}\geq \underline{\lambda}}\left[r^{336\underline{\lambda}}_{\mu_{1},\mu_{2},\mu_{3}}-r^{336\underline{\lambda}}_{\mu_{2}-1,\mu_{1}+1,\mu_{3}}-r^{336\underline{\lambda}}_{\mu_{3}-2,\mu_{2},\mu_{1}+2}-r^{336\underline{\lambda}}_{\mu_{1},\mu_{3}-1,\mu_{2}+1}+r^{336\underline{\lambda}}_{\mu_{3}-2,\mu_{1}+1,\mu_{2}+1}+r^{336\underline{\lambda}}_{\mu_{2}-1,\mu_{3}-1,\mu_{1}+2}\right]s_{\underline{\mu}}(x).\end{align}
Here the sum over $\umu \geq \underline \lambda$ is over all Young tableau $\umu$ which fully contain the Young tableau $\underline \lambda$. Notice that  the  factorial in the denominator of $r^{\alpha\beta\gamma}_{\mu_1 \dots \mu_m}$ diverges as the argument of the factorial becomes negative and thus we do not need to be too careful about the summation boundary.

\subsubsection{$GL(m|n)$ conformal partial waves}
\label{sec:glmn-conf-part}

The coefficients of the Schur
polynomials in any $GL(m|n)$ partial wave expansion are universal, which implies that
they do not depend on the group but only on the representations (Young
tableau). This means that having obtained the $GL(m)$ partial waves
for any $m$,  we can immediately write down the $GL(m|n)$ partial
waves as an explicit expansion over super Schur polynomials! Namely, we
have for any group $GL(m|n)$ (including $m=0$ or $n=0$) 
\begin{align}\label{eq:29}
F^{\alpha\beta\gamma\ula}(x|y) \ &=\   \sum_{[\umu]}
R^{\alpha\beta\gamma\underline\lambda}_{\umu} \,
s_{\umu}(x|y),
\end{align}
where $R^{\alpha\beta\gamma\underline\lambda}_{\umu}$ are {\em exactly the
same} numerical coefficients as
defined in~\eqref{eq:27} and $s_{\umu}(x|y)$ are the $GL(m|n)$ Schur
polynomials defined in~\eqref{eq:5}.
Indeed in the practical computation of  OPE coefficients -- as we will do for $\cN$=4 SYM in section~\ref{sec:ope-coeff-cn=4} -- this form of the partial wave is the most useful one. It turns out that we can expand the free theory correlator in Schur polynomials, and equate with the above expansion of the partial wave in Schur polynomials and simply equate the coefficient of each Schur polynomial on both sides.

However we also have in mind possible conformal bootstrap applications, and for these we will need to sum up the expansion. It is the purpose of this section to seek   a simple formula summing up
this 
$GL(m|n)$ partial wave.

It turns out that such a simple formula can be obtained. Just as
the summed up $GL(m)$ partial wave had a close relation with the
corresponding Schur polynomial, the summed up $GL(m|n)$ Schur
polynomial has a close relationship with an alternative form of the
$GL(m|n)$ Schur polynomial derived  in appendix~\ref{sec:altern-form-glmn} and defined in~\eqref{eq:5b}.
In particular we find
 \begin{align}\label{eq:2}
F^{\alpha\beta\gamma\ula}(x|y) \ &=\  (-1)^{\frac12(2 m + 2 p + n) (n - 1)}
D^{-1}
  \det \left(
     \begin{array}{cc}
F^X_{\underline\lambda}&R\\
       K_{\underline \lambda}   &   F^Y
         \end{array}
\right)\ ,
 \end{align}
where here we define
\begin{align}
p=\min\left\{\alpha,\beta\right\}  
\end{align}
and  $D,R$  are just as defined previously for the super Schur
polynomial, in~\eqref{eq:6}, $K_{\underline \lambda}$ is as defined for the alternative form of the Schur polynomials in~\eqref{eq:6b} and
$F^X_{\underline\lambda}$ and $F^Y$ are matrices
of hypergeometric functions
\begin{align}
  \label{eq:6c}
F^X_{\underline \lambda}&=\Big([x_i^{\lambda_j +m - n  - j}
  {}_2F_1( \lambda_j + 1 - j+\alpha, 
  \lambda_j + 1 - j+\beta;  2 \lambda_j + 2 - 2 j+\gamma; 
  x_i)] \Big)_{  \substack{
1\leq i
  \leq m\\
1\leq j\leq p
}
}
\nonumber\\[5pt]
F^Y&=\Big((y_j)^{i - 1}
  {}_2F_1(i + m - n-\alpha, i + m - n-\beta; 
  2 i + 2 (m - n)-\gamma; y_j)  \Big)_{\substack{1\leq i\leq p+n-m\\1\leq j
  \leq n}}\ .
\end{align}
 Here we again define the square brackets to mean ``the regular part
 at $x=0$'' i.e. with the principal part subtracted off. In the current context the function is a hypergeometric function in $x$ (which has a
non-singular expansion around $x=0$) multiplied by a power of $x$
which can be negative in which case
\begin{align}\label{eq:23}
[x^{-\ell} {}_2F_1(a,b;c;x)]&:=x^{-\ell} {}_2F_1(a,b;c;x) - 
\sum_{k=0}^{\ell-1} \frac{a^{(k)}b^{(k)}}{k! \,c^{(k)}} x^{k-\ell}
\nonumber \\
 &=  \sum_{k=0}^{\infty} \frac{a^{(k+\ell)}b^{(k+\ell)}}{(k+\ell)! \,
   c^{(k+\ell)}} x^{k}\ .
\end{align}

Note that we have not been able to prove this formula, indeed as we shall see shortly, even in the case $m=0$ it relies on an infinite number of remarkable, non-trivial numerical identities. Nevertheless we have checked it in sufficiently many cases to be confident of its veracity.

\subsubsection{Long reps and multiplet shortening for the conformal partial waves}
\label{sec:long-reps-multiplet}
The superconformal partial waves for long (typical) operators factorise just as for the Schur polynomials~\eqref{eq:33}, and the superconformal partial waves also satisfy multiplet shortening formulae analogous to~\eqref{eq:35}.
So for a long (or typical) representation we have that the conformal partial wave for a  long representation factorises into an $x$ partial wave and a $y$ partial wave
\begin{align}\label{eq:53}
  \begin{array}{c}
   \ula = [\ula_x+n,\ula_y] \qquad  \text{(long $GL(m|n)$ rep)} \\ \Downarrow \\
   F^{\alpha\beta\gamma\underline \lambda}(x|y)= F^{(\alpha+n)(\beta+n)(\gamma+2n)\ula_x}(x|0)\times F^{(\alpha-m)(\beta-m)(\gamma-2m)\ula_y}(0|y) \times \prod_{\substack{ 1\leq i\leq m,\\ 1\leq j \leq n}} (x_i-y_j)
 \end{array}
\end{align}
where $\ula_x,\ula_y$ are defined in~\eqref{eq:32} and the figure above.

This further implies relations between the partial waves of long reps,
when $\ula_x$ has a full column, just
as for the Schur polynomials~\eqref{eq:39}:
\begin{align}
  \label{eq:40}
  F^{\alpha\beta\gamma\underline \lambda}(x|y)= \frac{\prod_{i=1}^m x_i}{\prod_{j=1}^n (-y_j)}\times
  F^{(\alpha+1)(\beta+1)(\gamma+2)\underline \lambda'}(x|y) \qquad
  \quad \ula= [\ula_x{+}n,\ula_y],\  \ula'=
  [\ula_x{-}1{+}n,n,\ula_y]  \ .
\end{align}

Similarly for reps of the form $\ula_1,\ula_2,\ula$ defined as in~\eqref{eq:37},  we have analogous multiplet shortening formulae to~\eqref{eq:35}
\begin{align}\label{eq:recombo1}
   &\left(\textstyle{\prod_{i=1}^m x_i}\right) \times F^{\alpha\beta\gamma\ula_1}(x|y) +
 \left(\textstyle{\prod_{j=1}^n(-y_j)}\right) \times
 F^{(\alpha-1)(\beta-1)(\gamma-2)\ula_2}(x|y) \nonumber\\
=&  F^{(\alpha+n-1)(\beta+n-1)(\gamma+2n-2)\ula_x}(x)\times F^{(\alpha-m)(\beta-m)(\gamma-2m)\ula_y}(0|-y) \times \prod_{\substack{ 1\leq i\leq m,\\ 1\leq j \leq n}} (x_i-y_j)\ .
\end{align}

The proofs of these identities follow from considering the determinantal formula in a similar way  (albeit more involved) to that of the Schur polynomial case described below \eqref{eq:35}.

We note here also that as is well known in $\cal N$=4 SYM, the long operators can gain non-integer anomalous dimensions. The easiest way to incorporate this into the formalism is to simply define the long superconformal partial wave via the factorised form~\eqref{eq:35} and then continue the appropriate parameters to real values.

\subsubsection{$GL(0|n)$ partial waves and remarkable numerical identities}
\label{sec:gl0n-partial-waves}

The formula for the partial waves~\eqref{eq:2} is valid for all $m,n$. It was obtained from the case $n=0$, but should now also be valid for the other extreme case, when $m=0$ where it becomes
 \begin{align}\label{eq:2b}
F^{\alpha\beta\gamma\ula}(0|y) \ &=\  (-1)^{\frac12( \gamma + n) (n - 1)}
D^{-1}
  \det \left(
     \begin{array}{cc}
       K_{\underline \lambda}   &   F^Y
         \end{array}
\right)\ ,
 \end{align}
where $K_{\underline \lambda} $ is a $(p+n)\times p$ matrix and $F^Y$ is a $(p+n)\times n$ matrix (recalling that $p=\min(\alpha,\beta))$. 
 However in this case  the formula can be simplified: the $p$ columns of $K_\ula$ together with the unique corresponding row containing a non-zero entry can be deleted  from the matrix without changing the determinant and we are left with a formula for the $GL(0|n)$ partial waves:
\begin{align}
  \label{eq:31}
  F^{\alpha\beta\gamma\underline\lambda}(0|y) &=\frac{\det\Big(
   y_j^{\lambda^T_i+n-i}{}_2F_1(\lambda^T_i{+}1{-}i{-}\alpha ,\lambda^T_i{+}1{-}i{-}\beta;2\lambda^T_i{+}2{-}2i{-}\gamma;y_j)\Big)_{1\leq i,j \leq
    n}}{\det\Big(
   y_j^{n-i}\Big)_{1\leq i,j \leq
    m}}\ .
\end{align}
As for the $GL(m|0)$ case there is a redundancy in the description here. If the Young tableau contains  a complete (length $n$) row then we can delete it via
\begin{align}
  F^{\alpha\beta\gamma[n,\underline\lambda]}(0|y)  = (y_1\dots y_n) F^{(\alpha{-}1)(\beta{-}1)(\gamma{-}2) \underline\lambda }(0|y)
      \end{align}
Recall that although this is an ordinary bosonic group, the Young tableau are the transpose of the Young tableau discussed previously, i.e. they have length $n$ and infinite height (rather than the usual height $n$, infinite length).

Also recall that $m=0$ corresponds to the group $SU(n)$ (whereas $n=0$ is $SU(2,2)$) and so this is giving us the contribution of a representation  of $SU(2n)$ in the tensor product  of two representations, to a four-point function of four representations.

Note the close similarity with the $GL(m|0)$ case~\eqref{eq:25}.  Essentially the only difference is the sign with which the parameters $\alpha,\beta,\gamma$ appear as arguments of the hypergeometric function. This sign is crucial as it ensures that the arguments are all negative and so the hypergeometric functions become finite polynomials. The case $n=2$ corresponds to the group $SU(4)$ and was found previously in the $\cN=4$ context by~\cite{Dolan:2243hv} in terms of Legendre polynomials. The relation between the two forms arises through the identity given in~\href{http://functions.wolfram.com/Polynomials/LegendreP/26/01/02/0003/}{http://functions.wolfram.com/Polynomials/LegendreP/26/01/02/0003/}.

But now recall that writing the partial waves  as an expansion in Schur polynomials, the coefficients  are independent of the symmetry group. Expanding out the hypergeometric functions in~(\ref{eq:31}), we thus find an alternative formula for the coefficients, namely  
        \begin{align}\label{eq:49}
          R^{\alpha\beta\gamma,\ula}_{\umu} = \sum_{\sigma \in S_p} (-1)^{|\sigma|}\,\hat r^{\alpha\beta\gamma\ula}_{w_\sigma(\mu^T_1,\mu^T_2,\dots)},
        \end{align}
where
\begin{align}\hat r^{\alpha\beta\gamma,\ula}_{\umu}=\prod_{i=1}^n \frac{ \left(\alpha-\mu^T_i+i-1\right)_{\mu^T_i-\lambda^T_i}\left(\beta-\mu^T_i+i-1\right)_{\mu^T_i-\lambda^T_i} }{\left(\mu^T_i-\lambda^T_i\right)!  \left(\gamma-2 \mu^T_i+2i-2\right)_{\mu^T_i-\lambda^T_i} },\end{align}
and the Weyl transformation acts as in (\ref{eq:28}). Here $x_{n}$ is the {\em falling} Pochhammer symbol
\begin{align}
  x_{n}:= x(x-1)(x-2)\dots (x-n+1)\ .
\end{align}

But now we seem to have two completely different expressions for the coefficients $R$, \eqref{eq:27} and \eqref{eq:49}: 
\begin{align}\label{eq:52}
R_{\umu}^{\alpha\beta\gamma\ula}=\sum_{\sigma \in S_p} (-1)^{|\sigma|}\,\hat r^{\alpha\beta\gamma,\ula}_{w_\sigma(\mu^T_1,\mu^T_2,\dots)}=\sum_{\sigma \in S_q} (-1)^{|\sigma|}\,r^{\alpha\beta\gamma,\ula}_{w_\sigma(\mu_1,\mu_2,\dots)}.\end{align}
(where $p$ is the number of rows of $\umu$ and $q$ the number of columns.)

 We consider a couple of simple examples of this identity. In both cases, let us fix as before $\alpha=\beta=\frac12\gamma=3$. Let us consider in both cases $\ula=[0]$, and consider $\umu=[3,3,3]$ so that here $\umu^{T}=\umu$.  We perform the sums such that the terms are ordered according to following generators of the affine Weyl group; $(e), (12), (13), (23),(123)$ and $(132)$ of $S_3$. Then we obtain the following two expressions
\begin{align}&\notag \sum_{\sigma \in S_3} (-1)^{|\sigma|}\, r^{336[0]}_{w_\sigma(3,3,3)}=\frac{5}{14}-\frac{15}{49}-\frac{1}{5}-\frac{9}{28}+\frac{3}{14}+\frac{9}{35}=\frac{1}{980}, \\& \sum_{\sigma \in S_3} (-1)^{|\sigma|}\,\hat r^{336[0]}_{w_\sigma(3,3,3)}=\frac{1}{84}-\frac{1}{140}-\frac{1}{588}-\frac{3}{392}+\frac{1}{392}+\frac{3}{980}=\frac{1}{980}\ .\end{align}
One notices that each term associated to a particular affine Weyl group generator are rather different, yet remarkably all the terms of the  entire sum all contributes to give the same number. As a further example we may consider again $\ula=[0]$ with $\umu=[3,1]$ and $\umu^{T}=[2,1,1]$, we find
\begin{align}&\notag \sum_{\sigma \in S_3} (-1)^{|\sigma|}\, r^{336[0]}_{w_\sigma(3,1,0)}=\frac{25 }{14}-\frac{5 }{7}=\frac{15}{14}, \\& \sum_{\sigma \in S_3} (-1)^{|\sigma|}\,\hat r^{336[1]}_{w_\sigma(2,1,1)}=3 -\frac{5 }{7}-\frac{4 }{3}+\frac{5 }{42}=\frac{15}{14},\end{align}
where in the first line only the generator $(e)$ and $(12)$ contribute all other terms being zero, whilst in the second line the non-zero terms come from the generators $(e), (12), (23)$ and $(132)$.
It would be very interesting to prove and gain further insight into the identity~\eqref{eq:52}.

\subsection{Summary of the superconformal partial wave result}
\label{sec:summ-superc-part}
We here summarise the result in one place for easy access. We have found that the contribution of an operator $\cO^{\gamma \ula}$ to a four-point function $\langle p_1p_2p_3p_4\rangle$ is given by~\eqref{eq:11b} 
  \begin{align}
&\langle  \cO^{p_1}(X_1) \cO^{p_2}(X_2)   \cO^{p_3}(X_3) \cO^{p_4}(X_4)
\rangle  \nonumber \\&= 
\sum_{\gamma,\ula}
 A^{p_1p_2p_3p_4}_{\gamma\ula}  \,
g_{12}^{\frac{p_1+p_2}2} g_{34}^{\frac{p_3+p_4}2} 
 \left(\frac{g_{24}}{g_{14}}\right)^{\frac12{p_{21}}}
 \left(\frac{g_{14}}{g_{13}}\right)^{\frac12{p_{43}}}
\left(\frac{g_{13}g_{24}}{g_{12}g_{34}}\right)^{\frac12\gamma}
\, F^{\alpha\beta \gamma\underline\lambda}(Z),
                       \nonumber\\[5pt]
&\qquad \qquad \alpha=\tfrac12(\gamma-p_{12}) \quad \beta=\tfrac12( \gamma +p_{34})\ ,
  \end{align}
  where, in terms of OPE coefficients,
\begin{align}
&A^{p_1p_2p_3p_4}_{\gamma\ula}= \sum_{\cO^{\gamma\ula},\tilde \cO^{\gamma\ula}}
  C_{p_1p_2}^{\cO}   C_{p_3p_4}^{\tilde \cO}  C_{\cO \tilde \cO}\end{align}
Here we have that~(\ref{eq:29},\ref{eq:27})
\begin{align}
F^{\alpha\beta\gamma\ula}(x|y) \ &=\   \sum_{[\umu]}
R^{\alpha\beta\gamma\underline\lambda}_{\umu} \,
                                   s_{\umu}(x|y),
  \nonumber\\
R^{\alpha\beta\gamma\underline\lambda}_\umu\ &=
 \ \sum_{\sigma\in  S_m}
  (-1)^{|\sigma|}\,r^{\alpha\beta\gamma\underline\lambda}_{w_\sigma(\mu_1,\dots,\mu_m)},\nonumber \\
r^{\alpha\beta\gamma\underline\lambda}_{\mu_1\dots \mu_m}\ &=\ \prod_{j=1}^m\frac{(\lambda_j +1
-j+\alpha)^{(\mu_j{-}\lambda_j)}
(\lambda_j +1 -j+\beta)^{(\mu_j-\lambda_j)}  }
{(\mu_j-\lambda_j)!(2\lambda_j +2 -2j+\gamma)^{(\mu_j-\lambda_j)}   }\ ,
\end{align}
and $s_{\umu}(x|y)$ are the super Schur polynomials. Since one can immediately write down the free correlator as a sum of Schur polynomials, this form is enough to obtain free OPE coefficients (even without knowing the explicit form of the Schur polynomials themselves) as will do explicitly in the next section.

If one is interested in the summed up version of the conformal partial waves then instead we have
 \begin{align}
F^{\alpha\beta\gamma\ula}(x|y) \ &=\  (-1)^{\frac12(2 m + 2 p + n) (n - 1)}
D^{-1}
  \det \left(
     \begin{array}{cc}
F^X_{\underline\lambda}&R\\
       K_{\underline \lambda}   &   F^Y
         \end{array}
\right)\ ,
 \end{align}
where
\begin{align}
  p&=\min\{\alpha,\beta\}\nonumber \\
F^X_{\underline \lambda}&=\Big([x_i^{\lambda_j +m - n  - j}
  {}_2F_1( \lambda_j + 1 - j+\alpha, 
  \lambda_j + 1 - j+\beta;  2 \lambda_j + 2 - 2 j+\gamma; 
  x_i)] \Big)_{  \substack{
1\leq i
  \leq m\\
1\leq j\leq p
}
}
\nonumber\\[5pt]
F^Y&=\Big((y_j)^{i - 1}
  {}_2F_1(i + m - n-\alpha, i + m - n-\beta; 
  2 i + 2 (m - n)-\gamma; y_j)  \Big)_{\substack{1\leq i\leq p+n-m\\1\leq j
  \leq n}}
\nonumber\\[5pt]
  K_\ula&=\Big( -\delta_{i;-(\lambda_j+m-n-j)}\Big)_{\substack{1\leq i\leq p+n-m\\ 1\leq j  \leq p}}
  \nonumber\\[5pt]
  R&=\left(\frac1{x_i-y_j}\right)_{1\leq i \leq m,\   1\leq j \leq n} \nonumber\\[5pt]
   D&=\frac{\prod_{1\leq i<j \leq m} (x_i-x_j)\prod_{1\leq i<j \leq n}
  (y_i-y_j)}{\prod_{1\leq i\leq m,\,1\leq j \leq n} (x_i-y_j)  }
\end{align}

Note all the above formulae are straightforward to implement in a computer algebra programme.

\subsubsection{Summary for $\cN=4$}
The above formula is for a general superconformal field theory with
symmetry group $SU(m,m|2n)$. If one is interested in $\cN$=4 SYM
simply put $m=n=2$ in the above formulae. Using simple properties of
the determinant, the results can be rewritten in terms of two
functions, a one variable (in each of $x$ and $y$) function,
$f(x,y)$, and a two-variable function $f(x_1,x_2,y_1,y_2)$. The full
correlator is written in terms of these  simply as
\begin{align}
  \label{eq:30}
  F^{\alpha\beta\gamma\underline \lambda}(x|y) = \delta_{\ula; 0}
  +D^{-1}
\left[\left(
  \frac{f(x_2,y_2)}{x_1-y_1} \ - \ y_1 \leftrightarrow y_2\right) \ -\ x_1
  \leftrightarrow x_2 \right] \ +\   \ D^{-1} f(x_1,x_2,y_1,y_2).
\end{align}
where here
\begin{align}
  \label{eq:36}
  D^{-1}=\frac{(x_1-y_1)(x_1-y_2)(x_2-y_1)(x_2-y_2)}{(x_1-x_2)(y_1-y_2)}\ .
\end{align}
The functions are given explicitly as 
\begin{align}
  \label{eq:34}
  &{\bf{\lambda_2>1 \ (\text{\bf {long}}):}} \notag\\& f(x,y)=0 \notag \\
                                    & f(x_1,x_2,y_1,y_2)= (-1)^{\lambda' _1+\lambda' _2} \left(F_{\lambda _1}^{\alpha \beta \gamma }\left(x_1\right) F_{\lambda _2-1}^{\alpha \beta \gamma
   }\left(x_2\right)- x_1\leftrightarrow x_2\right)
   \left(G_{\lambda' _1}^{\alpha \beta \gamma }\left(y_1\right)
                                      G_{\lambda' _2-1}^{\alpha \beta
                                      \gamma }\left(y_2\right)-y_1
                                      \leftrightarrow
                                      y_2\right)\notag\\[15pt]
  &{\bf{\lambda_2=0,1 \ (\text{\bf{semi-short / quarter BPS}}):}} \notag\\& f(x,y)=(-1)^{\lambda' _1} F_{\lambda _1}^{\alpha \beta \gamma }(x) G_{\lambda' _1}^{\alpha \beta \gamma }(y) \notag \\
                                    & f(x_1,x_2,y_1,y_2)= 
\sum _{j=\lambda' _1+1}^p (-1)^{\lambda' _1} \left(F_{1-j}^{\alpha \beta \gamma }\left(x_2\right) F_{\lambda _1}^{\alpha \beta
   \gamma }\left(x_1\right)- (x_1 \leftrightarrow x_2)\right) 
   \left(G_j^{\alpha \beta \gamma }\left(y_2\right) G_{\lambda'
                                      _1}^{\alpha \beta \gamma
                                      }\left(y_1\right)- (y_1 \leftrightarrow y_2)\right)\notag\\
&\qquad \qquad \qquad \qquad +\sum _{j=2}^{\lambda' _1} (-1)^{\lambda'
   _1} \left(F_{2-j}^{\alpha \beta \gamma }\left(x_2\right) F_{\lambda
  _1}^{\alpha \beta \gamma }\left(x_1\right)- (x_1 \leftrightarrow x_2) \right) \left(G_{j-1}^{\alpha \beta \gamma }\left(y_2\right) G_{\lambda' _1}^{\alpha \beta \gamma }\left(y_1\right)-(y_1 \leftrightarrow y_2)\right)\notag\\[15pt]
  &{\bf{\ula=0 \ (\text{\bf{half BPS}}):}} \notag\\& f(x,y)=-\sum _{i=1}^p F_{1-i}^{\alpha \beta \gamma }(x) G_i^{\alpha \beta \gamma }(y)\notag\\& f(x_1,x_2,y_1,y_2)=\sum_{1\leq i<j\leq p}  \left(F_{1-i}^{\alpha \beta \gamma }\left(x_2\right) F_{1-j}^{\alpha \beta \gamma }\left(x_1\right)-F_{1-i}^{\alpha
   \beta \gamma }\left(x_1\right) F_{1-j}^{\alpha \beta \gamma }\left(x_2\right)\right) \left(G_i^{\alpha \beta \gamma }\left(y_1\right)
   G_j^{\alpha \beta \gamma }\left(y_2\right)-G_i^{\alpha \beta \gamma }\left(y_2\right) G_j^{\alpha \beta \gamma }\left(y_1\right)\right)
\end{align}
where we have defined the functions
\begin{align}
  \label{eq:54}
  F^{\alpha\beta\gamma}_\lambda(x)& :=
                              [x^{\lambda-1}{}_2F_1(\lambda+\alpha,\lambda+\beta;2\lambda+\gamma;x)]\notag\\
  G^{\alpha\beta\gamma}_{\lambda'}(y)&:=
  y^{\lambda'-1}{}_2F_1(\lambda'-\alpha,\lambda'-\beta;2\lambda'-\gamma;y)
\end{align}
where we recall that the square brackets indicate we must take the
regular part of the function.

The combination of short reps into long reps described for a general
supergroup in section~\ref{sec:long-reps-multiplet} can here be seen
from the vanishing of the sum of the corresponding one-variable
functions.
A semi-short operator defined by $\lambda_1,\lambda_1',\gamma$
combines with another defined by quantum numbers
$\lambda_1-1,\lambda_1'+1,\gamma+2$. The corresponding one-variable
functions cancel via the identity
\begin{align}
 (-1)^{\lambda' _1} F_{\lambda _1}^{\alpha
  \beta \gamma }(x) G_{\lambda' _1}^{\alpha \beta \gamma
  }(y)+(-1)^{\lambda' _1+1} \left(\frac xy\right) F_{\lambda _1-1}^{(\alpha+1)
  (\beta+1) (\gamma+2) }(x) G_{\lambda'_1+1}^{(\alpha+1) (\beta+1)
  (\gamma+1)  }(y)\ =\ 0\ .
\label{eq:55}
  \end{align}

\section{OPE coefficients in $\cN$=4 SYM}
\label{sec:ope-coeff-cn=4}
For this section we specialise to $\cN$=4 SYM. We thus take the
partial waves of the previous section and set
$(m,n)=(2,2)$. We wish to perform a superconformal partial wave
expansion on free theory correlation functions in order to illustrate
and confirm the partial waves of the previous section, and obtain new
results in this theory. 

A general free theory correlation function of four arbitrary charge
half-BPS operators is given by a sum of products of propagators
\begin{align}
g_{ij}=\det{(X_{j}-X_j)}^{-1} = \frac{y_{ij}^2}{x_{ij}^2} + O(\rho \bar
  \rho)\ .
\end{align}
Any free theory correlation function can be written, by observing that
\begin{align}
\text{sdet}\left(1-Z\right) =
  \left(\frac{g_{14}g_{23}}{g_{13}g_{24}}\right)^{-1}\ ,
\end{align}
in the general form:
\begin{align} 
&\langle  p_1p_2p_3p_4
\rangle    = 
g_{12}^{\frac{p_1+p_2}2} g_{34}^{\frac{p_3+p_4}2} 
 \left(\frac{g_{24}}{g_{14}}\right)^{\frac12{p_{21}}}
 \left(\frac{g_{14}}{g_{13}}\right)^{\frac12{p_{43}}}\sum_{\gamma}\left(\frac{g_{13}g_{24}}{g_{12}g_{34}}\right)^{\frac12\gamma}
\times \sum_{i=0}^{\left\lfloor\frac12\gamma\right\rfloor}\ a_{\gamma i}\ 
\text{sdet}\left(1-Z\right)^{-i}
\end{align}
where $p_{ij}=p_i-p_j$ and where $a_{\gamma \, i }$ are colour factors
which can be computed using Wick contractions. The restrictions on
$\gamma$ are the same as in~(\ref{eq:42}).

On the other hand we wish to compare this with the conformal partial
wave expansion~\eqref{eq:11b} 
\begin{align} 
&\langle  p_1p_2p_3p_4
\rangle  \nonumber \\&= 
\sum_{\cO,\tilde \cO}
  C_{p_1p_2}^{\cO}   C_{p_3p_4}^{\tilde \cO}  C_{\cO \tilde \cO}   \,
g_{12}^{\frac{p_1+p_2}2} g_{34}^{\frac{p_3+p_4}2} 
 \left(\frac{g_{24}}{g_{14}}\right)^{\frac12{p_{21}}}
 \left(\frac{g_{14}}{g_{13}}\right)^{\frac12{p_{43}}}
\left(\frac{g_{13}g_{24}}{g_{12}g_{34}}\right)^{\frac12\gamma}
\, F^{\alpha\beta \gamma\underline\lambda}(Z).\end{align}

The exercise is then to equate
\begin{align}\label{eq:43}
\sum_{i=0}^{\left\lfloor\frac12\gamma\right\rfloor}a_{\gamma i}\, 
\text{sdet}\left(1-Z\right)^{-i}=\sum_{[\ula]} \, A_{\gamma\underline{\lambda}}F^{\alpha\beta\gamma\underline\lambda}(Z)\end{align}
in order to find the OPE coefficients $A_{\gamma\underline{\lambda}}= C_{p_1p_2}^{\cO}
C_{p_3p_4}^{\tilde \cO}  C_{\cO \tilde \cO}$. 

The simplest way to do this is to 
use the Cauchy identity to rewrite
the RHS of~\eqref{eq:43} as an infinite sum over the super Schur
polynomials. This then allows for a direct comparison with the superconformal partial wave (SCPW)  expansion (which we also view as a sum over Schur polynomials) and
thus allows us to solve for the OPE coefficients. Remarkably, this
means we never in fact need to know the form of the Schur polynomials
themselves, both sides are given as expansions in Schur polynomials
and since we know these are independent this allows us to equate the
coefficients of each Schur polynomial.

\subsection{The Cauchy Identity}\label{sec:cauchy}

The Cauchy identity provides a way to write functions of
$\text{sdet}(1-Z)^{-q}$ for some $q$ a an expansion in super Schur polynomials.  Cauchy's identity states that (see for example appendix A of~\cite{FULTONHARRIS}):
\begin{align}\frac{1}{\prod_{i,j}(1-x_{i}z_{j})}=\sum_{\underline{\lambda}}s_{\underline{\lambda}}(x)s_{\underline{\lambda}}(z),\end{align}
where $\underline{\lambda}$ is some Young tableau. If we set
the $z_j$'s to 1 we gain the following formula relevant to the bosonic case:
\begin{align}   \det(1-Z)^{-p} = \frac{1}{\prod_{i}(1-x_{i})^p}=\sum_{\lambda}s_{\underline{\lambda}}(x)d^{GL(p)}_{\underline{\lambda}},\end{align}
where $d^{GL(p)}_{\underline{\lambda}}$ is the dimension of
some Young tableau  $\underline{\lambda}$ in $GL(p)$. In
particular this means we can never see Young tableaux with more than $p$ rows.

In the supersymmetric
case, this formula generalises naturally to
\begin{align}\label{eq:c}\prod_{i}\left(\frac{1-y_{i}}{1-x_{i}}\right)^p=\sum_{\underline{\lambda}}s_{\underline{\lambda}}(x|y)d^{GL(p)}_{\underline{\lambda}}.\end{align}
The standard Hook dimension formula gives
\begin{align}\label{eq:dim}d^{GL(p)}_{\underline{\lambda}}=\frac{\prod_{i=1}^{p}(p-i+1)^{(\lambda_{i})}}{\prod_{i=j}^{p}\prod_{j=1}^{p}(\lambda_{j}-\lambda_{i}+(i-j+1))^{(\lambda_{i}-\lambda_{i+1})}},\end{align}
where $x^{(n)}$ is the ascending Pochhammer symbol. Implicitly, this formula has a label for $p+1$ which we must switch off, namely $\lambda_{p+1}=0$.

For example for $p=1$, in $\cN=4$ SYM, one finds that
\begin{align}\text{sdet}(1-Z)^{-1}=\frac{(1-y_1)(1-y_2)}{(1-x_1)(1-x_2)}=\sum_{\lambda=0}^{\infty}s_{[\lambda,0,\dots]}(x|y).\end{align}
whereas for $p=2$, we get
\begin{align}&\notag
               \text{sdet}(1-Z)^{-2}=\frac{(1-y_1)^2(1-y_2)^2}{(1-x_1)^2(1-x_2)^2}=\sum_{\lambda_1\geq
               \lambda_2\geq 0}^{\infty}(\lambda _1-\lambda
               _2+1)s_{[\lambda_1,\lambda_2,0,..]}(x|y)\ .\end{align}

Using the above results it is now straightforward to obtain the OPE
coefficients in the free theory. In the next section we give a number
of low weight
examples of this. Note that at this stage we are not considering
the fact that in the interacting theory certain short multiplets can
combine together to become long. We will consider this in the following subsection.

Let us outline a basic example for precisely how this works. In the example of $\left\langle 1111\right\rangle$ which we study in the next subsection, we will encounter the function $f_{2}(A,A)$ which we want to compare with a linear combination of superconformal partial wave expansions of the form $F^{112[\lambda]}$ (corresponding to twist two operators). So using the Cauchy identity  we equate
\begin{align}f_{2}(A,A)=A(1+\text{sdet}(1-Z)^{-1})=2A s_{[0]}(x|y)+A\sum_{i\geq 1} s_{[\lambda]}(x|y)=\sum_{\lambda\geq 0}A_{2[\lambda]}F^{112[\lambda]}\end{align}
We can expand the rightmost-side explicitly using~\eqref{eq:29} giving 
\begin{align}&\notag 2A s_{[0]}(x|y)+A\sum_{i\geq 1} s_{[\lambda]}(x|y)=A_{2[0]}\underbrace{\left(s_{[0]}(x|y)+\frac12 s_{[1]}(x|y)+\frac{1}{3}s_{[2]}(x|y)+\dots\right)}_{F^{112[0]}}\\&+A_{2[1]}\underbrace{\left(s_{[1]}(x|y)+\frac12 s_{[2]}(x|y)+\frac{9}{10}s_{[3]}(x|y)+\dots\right)}_{F^{112[1]}}+A_{2[2]}\underbrace{\left(s_{[2]}(x|y)+\frac{3}{2} s_{[3]}(x|y)+\frac{12}{7}s_{[4]}(x|y)+\dots\right)}_{F^{112[2]}}+\dots\end{align}
One can already see that $A_{2[1]}=0$. Comparing the coefficients of $s_{[0]}(x|y)$ requires that $A_{2[0]}=2A$. A consequence of this is that this automatically sets coefficient of $s_{[1]}(x|y)$ to $A$ on the RHS, which yields an overall equality if we set $A_{2[1]}=0$. We may continue to the next order to find $A_{2[2]}$ and there onwards to find the rest of the coefficients. With enough terms, one can spot a pattern and write a general formula. As we will see in the next subsection, it turns out that the only non-zero OPE coefficients in this case are $\lambda\in \mathbb{Z}_{\text{even}}$, corresponding to even spin operators. All results are found in this way. Note that as mentioned previously, one never even needs to know the explicit form of the Schur polynomials for this.

\subsection{Results: Free theory OPE coefficients (before recombination)}

The purpose of this section is to display the OPE coefficients before
taking into account any recombination in the interacting theory. We do
this for the list of the correlation functions $\left\langle
  1111\right\rangle$, $\left\langle 1122\right\rangle$, $\left\langle
  2222\right\rangle$, $\left\langle 2233\right\rangle$, $\left\langle
  3333\right\rangle$, $\left\langle 2433\right\rangle$ and
$\left\langle 3544\right\rangle$. Clearly the first two correlators
can only exist in the U$(N)$ gauge theory (since $\tr(W^1)=0$ for $SU(N)$) whilst the others may exist in either U$(N)$ or SU$(N)$. 

For notational convenience we 
 have defined
\begin{align}f_{\gamma}\left(a_{\gamma0},a_{\gamma1},\dots,a_{\gamma\left\lfloor\frac12\gamma\right\rfloor}\right):=\sum_{i=0}^{\left\lfloor\frac12\gamma\right\rfloor}a_{\gamma i}\text{sdet}\left(1-Z\right)^{-i}\end{align}
where $a_{\gamma i}$ are the associated colour factors.

We consider all half BPS operators, both single- and multi-trace at finite $N$. We denote $A_\gamma=\tr(W^\gamma)$ so the multi-trace operator $\tr(W^2)^2$ is denoted $(A_2)^2$ etc.

\subsection*{\underline{$\mathbf{\left\langle 1111\right\rangle}$}}

This correlator many only exist in the U$(N)$ gauge theory and is given by
\begin{align}&\notag \left\langle 1111\right\rangle=A \left(g_{14} g_{23}+g_{13} g_{24}+g_{12} g_{34}\right)= g_{12}g_{34}\left(f_{0}(A)+\left(\frac{g_{13}g_{24}}{g_{12}g_{34}}\right)f_{2}(A,A)\right)\end{align}
The colour factor is given by
\begin{align}A=N^2\end{align}
In comparing with the SCPW expansion, one finds that
\begin{align}\label{eq:1111scpw}&\notag \left\langle 1111\right\rangle=g_{12}g_{34}\left(A+\left(\frac{g_{13}g_{24}}{g_{12}g_{34}}\right)\sum_{\lambda\geq 0}A_{2[\lambda]}F^{112[\lambda]}\right)\\&\text{with }A_{2[\lambda]}=\frac{2A(\lambda!)^2}{(2\lambda)!}\text{ for $\lambda\in\mathbb{Z}_{\text{even}}$ and zero otherwise.}\end{align}

\subsection*{\underline{$\mathbf{\left\langle 1122\right\rangle}$}}

\begin{align} \left\langle 1122\right\rangle=A g_{12} g_{34}^2+B \left(g_{14} g_{23} g_{34}+g_{13} g_{24} g_{34}\right)=g_{12}g_{34}^2\left(f_{0}(A)+\left(\frac{g_{13}g_{24}}{g_{12}g_{34}}\right)f_{2}(B,B)\right)\end{align}
The colour factors for U$(N)$ for the various types of correlators may be tabulated as
\begin{align}
\begin{tabular}{|l|l|l|}
\hline
Correlator type                                  & A      & B      \\ \hline
$\left\langle A_1A_1A_2A_2\right\rangle$         & $2N^3$ & $4N$   \\
$\left\langle A_1A_1(A_1)^2 A_2\right\rangle$    & $2N^2$ & $4N^2$ \\
$\left\langle A_1A_1(A_1)^2(A_1)^2\right\rangle$ & $2N^3$ & $4N^3$ \\ \hline
\end{tabular}\end{align}

Since $p_{12}=p_{34}=0$ (which means we use the same set of SCPW's), we see that this result is structurally identical to the (\ref{eq:1111scpw}), but for the change
\begin{align}A_{2[\lambda]}=\frac{2B(\lambda!)^2}{(2\lambda)!},\end{align}
which is simply a change in the colour factors.

\subsection*{\underline{$\mathbf{\left\langle 1133\right\rangle}$}}

\begin{align}&\notag \left\langle 1122\right\rangle=A g_{12} g_{34}^3+B \left(g_{14} g_{23} g_{34}^2+g_{13} g_{24} g_{34}^2\right)=g_{12}g_{34}^3\left(f_{0}(A)+\left(\frac{g_{13}g_{24}}{g_{12}g_{34}}\right)f_{2}(B,B)\right)\end{align}
The U$(N)$ colour factors for the various types of correlators is given by
\begin{align}
\begin{tabular}{|l|l|l|}
\hline
Correlator type                                  & A      & B      \\ \hline
$\left\langle A_1A_1A_3A_3\right\rangle$         & $3N^2(1+N^2)$ & $18N^2$   \\
$\left\langle A_1A_1(A_1A_2) A_3\right\rangle$    & $6N^3$ & $6N(2+N^2)$ \\
$\left\langle A_1A_1(A_1A_2)(A_1A_2)\right\rangle$ & $2N^2(2+N^2)$ & $2N^2(8+N^2)$ \\ 
$\left\langle A_1A_1(A_1A_2)(A_1)^3\right\rangle$ & $6N^3$ & $18N^3$ \\
$\left\langle A_1A_1(A_1)^3(A_3)\right\rangle$ & $6N^2$ & $18N^2$ \\
$\left\langle A_1A_1(A_1)^3(A_1)^3\right\rangle$ & $6N^4$ & $18N^4$ \\   \hline
\end{tabular}\end{align}
The result of the SCPW expansion is identical to the $\left\langle 1122\right\rangle$ previously shown but for the precise colour factors.

\subsection*{\underline{$\mathbf{\left\langle 2222\right\rangle}$}}

This is the first case where we have a correlator which may exist in either the U$(N)$ or SU$(N)$ guage theory. The correlator is given by
\begin{align}&\notag \left\langle 2222\right\rangle=A(g_{12}^2g_{34}^2+g_{13}^2g_{24}^2+g_{14}^2g_{23}^2)+B(g_{12}g_{23}g_{34}g_{41}+g_{13}g_{32}g_{21}g_{14}+g_{13}g_{34}g_{42}g_{21})\\&=g_{12}^2g_{34}^2\left(f_{0}(A)+\left(\frac{g_{13}g_{24}}{g_{12}g_{34}}\right)f_{2}(B,B)+\left(\frac{g_{13}g_{24}}{g_{12}g_{34}}\right)^2f_{4}(A,B,A)\right)\end{align}
For the SU$(N)$ theory, there is only one possible colour structure where the operator is $A_{2}$, and we have
\begin{align}
\begin{tabular}{|l|l|l|}
\hline
Correlator type $SU(N)$                                 & A      & B      \\ \hline
$\left\langle A_2A_2A_2A_2\right\rangle$         & $4(N^2-1)^2$ & $16(N^2-1)$  \\   \hline
\end{tabular}\end{align}
On the other hand there are a few variations in the U$(N)$ theory, which are given by
\begin{align}
\begin{tabular}{|l|l|l|}
\hline
Correlator type $U(N)$                                 & A      & B      \\ \hline
$\left\langle A_2A_2A_2A_2\right\rangle$         & $4N^4$ & $16N^2$   \\
$\left\langle (A_1)^2A_2A_2A_2\right\rangle$    & $4N^3$ & $16N$ \\
$\left\langle (A_1)^2(A_1)^2A_2A_2\right\rangle$ & $4N^4$ & $16N^2$ \\ 
$\left\langle (A_1)^2(A_1)^2(A_1)^2A_2\right\rangle$ & $4N^3$ & $16N^3$ \\
$\left\langle (A_1)^2(A_1)^2(A_1)^2(A_1)^2\right\rangle$ & $4N^4$ & $16N^4$ \\
$\left\langle (A_1)^2A_2(A_1)^2A_2\right\rangle$ & $4N^2$ & $16N^2$ \\   \hline
\end{tabular}\end{align}
Comparing to an SCPW expansion yields
\begin{align}\label{eq:2222} \left\langle 2222\right\rangle=g_{12}^2g_{34}^2\left(A+\left(\frac{g_{13}g_{24}}{g_{12}g_{34}}\right)\sum_{\lambda\geq 0}A_{2[\lambda]}F^{112[\lambda]}+\left(\frac{g_{13}g_{24}}{g_{12}g_{34}}\right)^2\sum_{\lambda_1\geq \lambda_2\geq 0}A_{4[\lambda_1,\lambda_2]}F^{224[\lambda_1,\lambda_2]}\right),\end{align}
where the coefficients are given by
\begin{align}\label{eq:2222coeffs}&\notag A_{2[\lambda]}=\frac{2 B (\lambda!)^2}{(2 \lambda)!}\text{ for }\lambda\in \mathbb{Z}_{\text{even}}\text{ zero otherwise},\\&\notag A_{4[\lambda_{1},\lambda_{2}]}=\frac{\lambda _1! \left(\lambda _1+1\right)! \left(\lambda _2!\right){}^2 \left(A \left(\lambda _1-\lambda _2+1\right) \left(\lambda _1+\lambda _2+2\right)+B (-1)^{\lambda_2}\right)}{\left(2 \lambda _2\right)! \left(2\lambda _1+1\right)!}\\&\text{    for $\lambda_{1}-\lambda_{2}\in \mathbb{Z}_\text{even}\geq 0$,  $\lambda_{2}\in \mathbb{Z}\geq 0$ and zero otherwise.} \end{align}

\subsection*{\underline{$\mathbf{\left\langle 2233\right\rangle}$}}

One may write the free theory correlator as
\begin{align}&\notag \left\langle 2233\right\rangle=A g_{12}^2 g_{34}^3+B \left(g_{14}^2 g_{34} g_{23}^2+g_{13}^2 g_{24}^2 g_{34}\right)+C \left(g_{12} g_{14} g_{23} g_{34}^2+g_{12} g_{13} g_{24}
   g_{34}^2\right)\\&\notag+D g_{13} g_{14} g_{23} g_{24} g_{34},\\&=g_{12}^2g_{34}^3\left(f_{0}(A)+\left(\frac{g_{13}g_{24}}{g_{12}g_{34}}\right)f_{2}(C,C)+\left(\frac{g_{13}g_{24}}{g_{12}g_{34}}\right)^2f_{4}(B,D,B)\right)\end{align}
The colour factors for $\text{SU}(N)$ can only come from one correlator:
\begin{align}\label{eq:2233col} 
\begin{tabular}{|l|l|l|l|l|}
\hline
Correlator type   $SU(N)$                               & A      & B      & C    & D    \\ \hline
$\left\langle A_2A_2A_3A_3\right\rangle$         & $\frac{6 (N^2-1)^2(N^2-4)}{N}$&       $0$&     $\frac{36(N^2-1)(N^2-4)}{N}$&       $\frac{72(N^2-1)(N^2-4)}{N}$    \\     \hline
\end{tabular}\end{align}

For the $\text{U}(N)$ theory we have $18$ possible ways of partitioning the $p_i$'s into local operators:

\begin{align}
\begin{tabular}{|l|l|l|l|l|}
\hline
Correlator type   $U(N)$                               & A      & B      & C    & D    \\ \hline
$\left\langle A_2A_2A_3A_3\right\rangle$         & $6N^3(1+N^2)$ & $36N^3$ & $36N(1+N^2)$ & $72N(1+N^2)$  \\
$\left\langle (A_1)^2A_2A_3A_3\right\rangle$    & $6N^2(1+N^2)$ & $36N^2$ & $72N^2$ & $72N(1+N^2)$ \\
$\left\langle (A_1)^2(A_1)^2A_3A_3\right\rangle$ & $6N^3(1+N^2)$ & $36N$ & $72N^3$ & $144N$ \\ 
$\left\langle A_2A_2(A_1A_2)A_3\right\rangle$ & $12N^4$ & $12N^2(2+N^2)$ & $72N^2$ & $144N^2$ \\
$\left\langle A_2A_2(A_1A_2)(A_1A_2)\right\rangle$ & $4N^3(2+N)$ & $4N(2+N^2)^2$ & $24N(2+N^2)$ & $48N(2+N^2)$ \\
$\left\langle A_2A_2(A_1)^3A_3\right\rangle$ & $12N^3$ & $36N^3$ & $72N$ & $144N$ \\ 
$\left\langle A_2A_2(A_1)^3(A_1)^3\right\rangle$ & $12N^5$ & $36N^3$ & $72N^3$ & $144N^3$ \\
$\left\langle A_2A_2(A_1)^3(A_1A_2)\right\rangle$ & $12N^4$ & $12N^2(2+N^2)$ & $72N^2$ & $144N^2$ \\ 
$\left\langle (A_1)^2A_2(A_1)^3A_3\right\rangle$ & $12N^2$ & $36N^2$ & $72N^2$ & $144N^2$ \\ 
$\left\langle (A_1)^2A_2(A_1A_2)A_3\right\rangle$ & $12N^3$ & $12N(2+N^2)$ & $24N(2+N^2)$ & $48N(2+N^2)$ \\
$\left\langle (A_1)^2A_2(A_1)^3(A_1A_2)\right\rangle$ & $12N^3$ & $36N^3$ & $72N^3$ & $144N^3$ \\  
$\left\langle (A_1)^2A_2(A_1)^3(A_1)^3\right\rangle$ & $12N^4$ & $36N^4$ & $72N^4$ & $144N^4$ \\
$\left\langle (A_1)^2A_2(A_1A_2)(A_1A_2)\right\rangle$ & $4N^2(2+N)$ & $12N^2(2+N^2)$ & $8N^2(8+N^2)$ & $16N^2(8+N^2)$ \\  
$\left\langle (A_1)^2(A_1)^2(A_1)^3A_3\right\rangle$ & $12N^4$ & $36N^4$ & $72N^4$ & $144N^4$ \\      
$\left\langle (A_1)^2(A_1)^2(A_1A_2)A_3\right\rangle$ & $12N^4$ & $36N^2$ & $24N^2(2+N^2)$ & $144N^2$ \\ 
$\left\langle (A_1)^2(A_1)^2(A_1)^3(A_1A_2)\right\rangle$ & $12N^4$ & $36N^4$ & $72N^4$ & $144N^4$ \\    
$\left\langle (A_1)^2(A_1)^2(A_1A_2)(A_1A_2)\right\rangle$ & $4N^3(2+N^2)$ & $36N^3$ & $8N^3(8+N^2)$ & $144N^3$ \\
$\left\langle (A_1)^2(A_1)^2(A_1)^3(A_1)^3\right\rangle$ & $12N^5$ & $36N^5$ & $72N^5$ & $144N^5$ \\     \hline
\end{tabular}\end{align}

We see that this result here is structurally identical to the $\left\langle 2222\right\rangle$ case, the only difference is as in previous cases the precise difference in the colour factors. Namely, the result is identical to (\ref{eq:2222}), but instead we have
\begin{align}&\notag A_{2[\lambda]}=\frac{2 C (\lambda!)^2}{(2 \lambda)!}\text{ for }\lambda\in \mathbb{Z}_{\text{even}}\text{ zero otherwise},\\&\notag A_{4[\lambda_{1},\lambda_{2}]}=\frac{\lambda _1! \left(\lambda _1+1\right)! \left(\lambda _2!\right){}^2 \left(B \left(\lambda _1-\lambda _2+1\right) \left(\lambda _1+\lambda _2+2\right)+D (-1)^{\lambda_2}\right)}{\left(2 \lambda _2\right)! \left(2\lambda _1+1\right)!}\\&\text{    for $\lambda_{1}-\lambda_{2}\in \mathbb{Z}_\text{even}\geq 0$,  $\lambda_{2}\in \mathbb{Z}\geq 0$ and zero otherwise.} \end{align}

\subsection*{\underline{$\mathbf{\left\langle 3333\right\rangle}$}}
	
The free theory correlator is given by
\begin{align}&\notag\left\langle 3333\right\rangle =A \left(g_{14}^3 g_{23}^3+g_{13}^3 g_{24}^3+g_{12}^3 g_{34}^3\right)+B (g_{13} g_{14}^2 g_{24} g_{23}^2+g_{12} g_{14}^2 g_{34}
   g_{23}^2\\&\notag +g_{13}^2 g_{14} g_{24}^2 g_{23}+g_{12}^2 g_{14} g_{34}^2 g_{23}+g_{12}^2 g_{13} g_{24} g_{34}^2+g_{12} g_{13}^2 g_{24}^2
   g_{34})+C g_{12} g_{13} g_{14} g_{23} g_{24} g_{34},\\&=g_{12}^3g_{34}^3\left(f_{0}(A)+\left(\frac{g_{13}g_{24}}{g_{12}g_{34}}\right)f_{2}(B,B)+\left(\frac{g_{13}g_{24}}{g_{12}g_{34}}\right)^2f_{4}(B,C,B)+\left(\frac{g_{13}g_{24}}{g_{12}g_{34}}\right)^3f_{6}(A,B,B,A)\right)\end{align}
There is only one SU$(N)$ correlator which has colour factors
\begin{align}
\begin{tabular}{|l|l|l|l|}
\hline
Correlator type                                  & A      & B      & C    \\ \hline
$\left\langle A_3A_3A_3A_3\right\rangle$         & $\frac{9(N^2{-}4)^2(N^2{-}1)^2}{N^2}$ & $\frac{81 (N^2{-}4)^2(N^2{-}1)}{N^2}$ & $\frac{162(N^2{-}4)(N^2{-}1) \left(N^2{-}12\right)}{N^2}$    \\ \hline
\end{tabular}\end{align}

For the $\text{U}(N)$ theory we have $17$ possible ways of partitioning the $p_i$'s into local operators:
\begin{align}
\begin{tabular}{|l|l|l|l|}
\hline
Correlator type                                  & A      & B      & C    \\ \hline
$\left\langle A_3A_3A_3A_3\right\rangle$         & $9N^2(1+N^2)^2$ & $81N^2(3+N^2)$ & $162N^2(7+N^2)$  \\
$\left\langle (A_1)^3A_3A_3A_3\right\rangle$         & $18N^2(1+N^2)$ & $108N^2(2+N)$ & $1296N^2$  \\
$\left\langle (A_1)^3(A_1)^3A_3A_3\right\rangle$        & $18N^4(1+N^2)$ & $324N^2$ & $1296N^2$  \\ 
$\left\langle (A_1)^3(A_1)^3(A_1)^3A_3\right\rangle$        & $36N^4$ & $324N^4$ & $1296N^4$  \\ 
$\left\langle (A_1)^3A_3(A_1)^3A_3\right\rangle$        & $36N^2$ & $324N^2$ & $1296N^2$  \\ 
$\left\langle (A_1A_2)A_3A_3A_3\right\rangle$        & $18N^3(1+N^2)$ & $108N(2+N^3)$ & $1296N$  \\ 
$\left\langle (A_1A_2)(A_1A_2)A_3A_3\right\rangle$        & $6N^2(1+N^2)(2+N^2)$ & $36N^2(8+N^2)$ & $72N^2(17+N^2)$  \\
$\left\langle (A_1A_2)(A_1A_2)(A_1A_2)A_3\right\rangle$        & $12N^3(2+N^2)$ & $12N(12+14N^2+N^4)$ & $48N(14_13N^2)$   \\ 
$\left\langle (A_1A_2)A_3(A_1A_2)A_3\right\rangle$        & $36N^4$ & $36N^2(8+N^2)$ & $72N^2(17+N^2)$   \\ 
$\left\langle (A_1)^3(A_1A_2)(A_1A_2)(A_1A_2)\right\rangle$        & $12N^3(2+N^2)$ & $36N^3(8+N^2)$ & $48N^3(26+N^2)$   \\ 
$\left\langle (A_1)^3(A_1)^3(A_1A_2)(A_1A_2)\right\rangle$        & $12N^4(2+N^2)$ & $324N^4$ & $1296N^4$   \\  
$\left\langle (A_1)^3(A_1)^3(A_1)^3(A_1A_2)\right\rangle$        & $36N^5$ & $324N^5$ & $1296N^5$   \\  
$\left\langle (A_1)^3(A_1)^3(A_1)^3(A_1)^3\right\rangle$        & $36N^6$ & $324N^6$ & $1296N^6$   \\  
$\left\langle (A_1)^3(A_1A_2)(A_1)^3(A_1A_2)\right\rangle$        & $36N^4$ & $324N^4$ & $1296N^4$   \\ 
$\left\langle (A_1)^3(A_1A_2)A_3(A_1)^3\right\rangle$        & $36N^3$ & $108N^3(2+N^2)$ & $1296N^3$   \\ 
$\left\langle (A_1)^3(A_1A_2)A_3(A_1A_2)\right\rangle$        & $36N^4$ & $108N^2(2+N^2)$ & $144N^2(8+N^2)$   \\ 
$\left\langle (A_1A_2)(A_1A_2)(A_1A_2)(A_1A_2)\right\rangle$        & $4N^2(2+N^2)^2$ & $4N^2(60+20N^2+N^4)$ & $48N^2(22+5N^2)$   \\ \hline
\end{tabular}\end{align}
Upon comparing to an SCPW expansion we get
\begin{align}&\notag \left\langle 3333\right\rangle=g_{12}^3g_{34}^3\Bigg(A+\left(\frac{g_{13}g_{24}}{g_{12}g_{34}}\right)\sum_{\lambda\geq 0}A_{2[\lambda]}F^{112[\lambda]}+\left(\frac{g_{13}g_{24}}{g_{12}g_{34}}\right)^2\sum_{\lambda_1\geq \lambda_2\geq 0}A_{4[\lambda_1,\lambda_2]}F^{224[\lambda_1,\lambda_2]}\\&+\sum_{\lambda_1\geq \lambda_2\geq \lambda_3\geq 0}A_{6[\lambda_1,\lambda_2,\lambda_3]}F^{336[\lambda_1,\lambda_2,\lambda_3]}\Bigg),\end{align}
Similarly to previous examples we see structures repeating again. Namely, the $\gamma=2$ is identical to (\ref{eq:2222coeffs}) and $\gamma=4$ sector is structurally identical to (\ref{eq:2222coeffs}) but for the change of colour factor $A\rightarrow B$ and $B\rightarrow C$. We also get a $\gamma=6$ sector where the OPE coefficients are
\begin{align}\label{eq:3333cof}&\notag A_{6[\lambda_1,\lambda_2]}=m_{\lambda_1,\lambda_2} \frac{1}{2} \Big(A\left(\lambda _1+2\right) \left(\lambda _1+3\right) \left(\lambda _1-\lambda _2+1\right) \left(\lambda _2+1\right) \left(\lambda _2+2\right) \left(\lambda
   _1+\lambda _2+4\right)\\&\notag+4 B
    \left(\left((-1)^{\lambda _2}+1\right) \lambda _1 \left(\lambda _1+5\right)+8 (-1)^{\lambda _2}+\left((-1)^{\lambda _2}-1\right) \lambda _2 \left(\lambda
   _2+3\right)+4\right)\Big)\\&\notag \text{for $\lambda_{1}-\lambda_{2}\in \mathbb{Z}_{\text{even}}\geq 0, \lambda_{2}\geq 0$ and zero otherwise,}
	\\&\notag A_{6[\lambda_{1},\lambda_{2},1]}=m_{\lambda_1,\lambda_2}\frac{1}{4} \Big(A \left(\lambda _1+1\right) \left(\lambda _1+4\right) \left(\lambda _1-\lambda _2+1\right) \lambda _2 \left(\lambda _2+3\right) \left(\lambda _1+\lambda _2+4\right)\\&\notag +4 B\left((-1)^{\lambda _2}-1\right) \left(\lambda _1-\lambda _2+1\right) \left(\lambda _1+\lambda _2+4\right)\Big)\\&\notag \text{for $\lambda_{1}-\lambda_{2}\in \mathbb{Z}_{\text{odd}}\geq 1, \lambda_{2}\geq 1$ and zero otherwise,}
	\\&\notag A_{6[\lambda_1,\lambda_2,2]}=m_{\lambda_1,\lambda_2}\frac{1}{12} \Big(A \lambda _1 \left(\lambda _1+5\right) \left(\lambda _1-\lambda _2+1\right) \left(\lambda _2-1\right) \left(\lambda _2+4\right) \left(\lambda _1+\lambda _2+4\right)\\&\notag+4 B\left(\left((-1)^{\lambda _2}+1\right) \lambda _1 \left(\lambda _1+5\right)+\left((-1)^{\lambda _2}-1\right) \left(\lambda _2-1\right) \left(\lambda
   _2+4\right)\right)\Big)\\& \text{for $\lambda_{1}-\lambda_{2}\in \mathbb{Z}_{\text{even}}\geq 0, \lambda_{2}\geq 2$ and zero otherwise,}\end{align}
where 
\begin{align}m_{\lambda_{1},\lambda_{2}}=\frac{\left(\lambda _1+2\right)!{}^2 \left(\lambda _2+1\right)!{}^2}{\left(2\lambda_2+2\right)! \left(2\lambda_1+4\right)!}.\end{align}

We give two  further cases  in appendix~\ref{sec:further-results-free}, namely $\langle4233\rangle$ and $\langle5344\rangle$.

\subsection{Consistency checks for the above OPE coefficients}
\label{sec:cons-checks-above}
It is possible to perform non-trivial consistency checks for the above results if we have some information concerning the number of operators in each representation.

To see where these consistency checks come from, consider writing the OPE coefficients as follows, 
\begin{align} A^{p_1p_2p_3p_4}_{\gamma\ula}   =   \left\langle C_{p_1 p_2},C_{p_3 p_4}\right\rangle:=\sum_{\cO^{\gamma\ula},\tilde \cO^{\gamma\ula}}C_{p_1 p_2}^{\cO}C^{\tilde{\cO}}_{p_3 p_4}C_{\cO\tilde{\cO}}\ .\end{align}
Namely, we can consider the inner product of the structure constants of the three-point function with a metric defined by the two point function.
Here we sum over all operators in the same representation ($\gamma\ula$) 
and we may regard $C_{p_ip_j}$ as being a vector with dimension equal to the number of operators in this representation. If we choose a basis for the operators where we have diagonalised the two-point functions, then we have  simply $C_{\cO\tilde{\cO}}\sim \delta_{\cO\tilde{\cO}}$ and this becomes the standard scalar product.

Various results follow from this. Firstly, notice that
\begin{align}\label{eq:angle}\text{cos}^2(\theta)=\frac{\left\langle C_{p_1p_2},C_{p_3p_4}\right\rangle^2}{\left\langle C_{p_1p_2},C_{p_1p_2}\right\rangle\left\langle C_{p_3p_4},C_{p_3p_4}\right\rangle},\end{align}
where $\theta$ is the angle between the two vectors $C^{\cO}_{p_1p_2}$ and $C^{\cO}_{p_3p_4}$, and so it follows that 
\begin{align}\label{eq:47}
0\leq\frac{\left(A^{p_1p_2p_3p_4}\right)^2}{A^{p_1p_2 p_1p_2} A^{p_3p_4p_3p_4}}\leq 1\end{align}
for {\em all} OPE coefficients.~\footnote{For long operators, this need only be true after taking into account the equivalence relation~\eqref{eq:41}.}  

Furthermore, if there is only one  operator $\cO$ in the representation in question, then the vector space has dimension $1$ and we must get 1.

Indeed if we know how many operators there are in a particular representation, $b$,  (so we know the dimension of the relevant inner product space) then we know that any  Gram determinant of dimension $b+1$ must vanish. So 
\begin{align}
  \label{eq:45}
  \det   \left( A^{p_ip_jp_kp_l}\right)_{\substack{(p_i, p_j) \in S\\(p_k, p_l) \in S}},
\end{align}
where $S$ is any set of pairs $(p_i,p_j)$ such that $|S|=b+1$.

So for the previously mentioned case where the number of operators is one we let $S=\{(p_1,p_2),(p_3,p_4)\}$ and then
\begin{align}
  \label{eq:46}
  \text{Gram}=\det \left(
  \begin{array}{cc}
    A^{p_1p_2p_1p_2}& A^{p_1p_2p_3p_4}\\
A^{p_1p_2p_3p_4}&A^{p_3p_4p_3p_4}
  \end{array}
\right) = A^{p_1p_2p_1p_2}A^{p_3p_4p_3p_4} - (A^{p_1p_2p_3p_4})^2 = 0 ,
\end{align}
which is equivalent to equation~\eqref{eq:47} being equal to one.
For the case where we have two operators we have 
\begin{align}
  \label{eq:46b}
  \text{Gram}=\det \left(
  \begin{array}{ccc}
    A^{p_1p_2p_1p_2}& A^{p_1p_2p_3p_4}& A^{p_1p_2p_5p_6}\\
A^{p_1p_2p_3p_4}&A^{p_3p_4p_3p_4}& A^{p_3p_4p_5p_6}\\
A^{p_1p_2p_5p_6}&A^{p_3p_4p_5p_6}& A^{p_5p_6p_5p_6}
  \end{array}
\right) =  0 \ .
\end{align}

Let us check these conditions in a few cases. Firstly, consider the case with only one operator. This is the case for all twist two operators $\cO^{2[\lambda]}$ in the $SU(N)$ theory. Looking back at the results above one can straightforwardly check that indeed
\begin{align}
A^{2222}_{2[\lambda]}A^{3333}_{2[\lambda]}-(A^{2233}_{2[\lambda]})^2 = \left(\frac{2  (\lambda!)^2}{(2 \lambda)!}\right)^2 \left[16(N^2{-}1)\times \frac{81 (N^2{-}4)^2(N^2{-}1)}{N^2}  {-}  \left( \frac{36(N^2{-}1)(N^2{-}4)}{N}\right)^2 \right] =0\ .\label{eq:48}
\end{align}

Similarly in the $U(N)$ case there are two twist 2 operators $\cO^{2[\lambda]}$ for each spin $\lambda$ (a single-trace and a double-trace one). Thus  the following $3\times3$ Gram determinant should vanish
\begin{align}
\det \left(
  \begin{array}{ccc}
    A_{2[\lambda]}^{1111}& A_{2[\lambda]}^{1122}& A_{2[\lambda]}^{1133}\\
A_{2[\lambda]}^{1122}&A_{2[\lambda]}^{2222}& A_{2[\lambda]}^{2233}\\
A_{2[\lambda]}^{1133}&A_{2[\lambda]}^{2222}& A_{2[\lambda]}^{3333}
  \end{array}
\right)=0\label{eq:50}
\end{align}
which can be readily seen to be the case using the results above.

As can be seen there will are many such consistency checks which can be performed. They require knowing the number of operators of each representation which can be read off from~\cite{Bianchi:2006ti}. Furthermore in the next section we will show how  similar considerations give information about the disentangling of protected and unprotected operators. Indeed we can use this  to completely disentangle the protected and unprotected sectors in the $\langle3333\rangle$ correlator.

\section{Physical OPE coefficients: recombination in $SU(N)$}
\label{sec:phys-ope-coeff}
It is well known that free theory supermultiplets in $\cal N$= 4 SYM combine together to form long supermultiplets, which are then free to develop an anomalous dimension. In order to separate out the OPE coefficients into free and interacting pieces, it is useful to be able to disentangle the genuine short multiplets from those which become part of long multiplets. This is also a crucial element of the conformal bootstrap programme, since there one needs to know  the contribution to the free correlator of all protected operators~\cite{Alday:2014qfa}.

It is impossible to uniquely disentangle this information from the free theory alone, one requires some information from the interacting theory. At least in some situations however, knowledge of mixed charge correlators, together with simply the knowledge of the number of long/short operators (the precise form of them is however not required) allows us to uniquely disentangle the protected and unprotected sectors. The number of short and long operators can be obtained by an examination of the classical interacting theory~\cite{Heslop:2001dr,Bianchi:2006ti}.  We will give an example of this in the current section, and we will obtain  the precise separation of the  free $SU(N)$ correlator $\langle 3333 \rangle$ into protected and unprotected sectors by making use of the $\langle2233\rangle$ and $\langle2222\rangle$ correlators.

In order to gain the correct answer, we make repetitive use of the reducibility equation at the unitary bound~\eqref{eq:recombo1} which in $\cN=4$ SYM reads 
\begin{align}F_{\text{long}}^{\alpha\beta\gamma[\lambda+1,1^{\nu+1}]}:=\text{lim}_{\rho\rightarrow 1}F^{\alpha\beta\gamma[\lambda+\rho,\rho,1^\nu]}=\left(\frac{g_{13}g_{24}}{g_{12}g_{34}}\right)^{-1}F^{\alpha-1\hspace{1mm}\beta-1\hspace{1mm}\gamma-2[\lambda+2,1^{\nu}]}+F^{\alpha\beta\gamma[\lambda+1,1^{\nu+1}]},\end{align}
where the LHS is understood for arbitrary real $\rho$ via an analytic continuation of the results for the long representations $\rho=2,3,4,\dots$.
It is thus convenient to introduce the  notation $F_{\text{long}}^{\alpha\beta\gamma[\lambda+1,1^{\nu+1}]}$ to take care of this situation.

There then remains the question as to how to decide which operators become long without doing explicit computations.

In this subsection we present the physical OPE coefficients of gauge group $\text{SU}(N)$, in particular for $\left\langle 2222\right\rangle$, $\left\langle 2233\right\rangle$ and $\left\langle 3333\right\rangle$. Let us begin with the $\left\langle 2222\right\rangle$ case.

\subsection*{\underline{$\mathbf{\left\langle2222\right\rangle}$}}
Stating the result again, we had 
\begin{align}\left\langle 2222\right\rangle=g_{12}^2g_{34}^2\left(A+\left(\frac{g_{13}g_{24}}{g_{12}g_{34}}\right)\sum_{\lambda\geq 0}A_{2[\lambda]}F^{112[\lambda]}+\left(\frac{g_{13}g_{24}}{g_{12}g_{34}}\right)^2\sum_{\lambda_1\geq \lambda_2\geq 0}A_{4[\lambda_1,\lambda_2]}F^{224[\lambda_1,\lambda_2]}\right),\end{align}
where the coefficients are given by (\ref{eq:2222coeffs}), but for convenience we repeat them
\begin{align}&\notag A_{2[\lambda]}=\frac{2 B (\lambda!)^2}{(2 \lambda)!}\text{ for }\lambda\in \mathbb{Z}_{\text{even}}\text{ zero otherwise},\\&\notag A_{4[\lambda_{1},\lambda_{2}]}=\frac{\lambda _1! \left(\lambda _1+1\right)! \left(\lambda _2!\right){}^2 \left(A \left(\lambda _1-\lambda _2+1\right) \left(\lambda _1+\lambda _2+2\right)+B (-1)^{\lambda_2}\right)}{\left(2 \lambda _2\right)! \left(2\lambda _1+1\right)!}\\&\text{    for $\lambda_{1}-\lambda_{2}\in \mathbb{Z}_\text{even}\geq 0$,  $\lambda_{2}\in \mathbb{Z}\geq 0$ and zero otherwise.} \end{align}
with
\begin{align}
\begin{tabular}{|l|l|l|l|l|}
\hline
Correlator type   $SU(N)$                               & A      & B       \\ \hline
$\left\langle A_2A_2A_2A_2\right\rangle$         & $4(N^2-1)^2$&       $16(N^2-1)$   \\     \hline
\end{tabular}\end{align}

We recognise the term $F^{112[2]}$ as being the Konishi operator. Famously, the Konishi operator gains an anomalous dimension in the interacting theory, hence it should be long whilst as it stands it is short. By looking at the structure of the Wick contractions, one also observes that the semi-short operators that follow, namely $F^{112[\lambda\geq 4]}$ are all long in the interacting theory and have the form $\tr(W_{AB}(\partial)^{\lambda}\bar{W}^{AB})$ \cite{Heslop:2001dr}. The operator corresponding to  $F^{112[0]}$, on the other hand, corresponds to the stress-tensor multiplet, and is the only $\gamma=2$ protected operator. It will remain short in the interacting theory.

In order to manifest these points one may make use of the reducibility equation
\begin{align}\label{eq:51}
F^{112[\lambda]}=\left(\frac{g_{13}g_{24}}{g_{12}g_{34}}\right)\left(F^{224[\lambda-1,1]}_{\text{long}}-F^{224[\lambda-1,1]}\right).\end{align}
In which we get
\begin{align}&\notag\left\langle 2222\right\rangle=g_{12}^2g_{34}^2\Bigg(A+\left(\frac{g_{13}g_{24}}{g_{12}g_{34}}\right)2BF^{112[0]}+\left(\frac{g_{13}g_{24}}{g_{12}g_{34}}\right)^2\Bigg(\sum_{\lambda \geq 0}^{\infty} A_{4[\lambda]}F^{224[\lambda]}+\sum_{\lambda \geq 1}^{\infty}A'_{4[\lambda,1]}F^{224[\lambda,1]}\\&+\sum_{\lambda_{1}\geq \lambda_{2}\geq 2}^{\infty}A_{4[\lambda_{1},\lambda_{2}]}F^{224[\lambda_{1},\lambda_{2}]}+\sum_{\lambda\geq 1}^{\infty}A_{2[\lambda+1]}F^{224[\lambda,1]}_{\text{long}}\Bigg)\Bigg),\end{align}
where\begin{align}A'_{4[\lambda,1]}=A_{4[\lambda,1]}-A_{2[\lambda+1]}.\end{align}

Here the second line consists of unprotected operators, whereas the first line corresponds to genuine short operators.

So we have used qualitative knowledge (essentially that all twist two operators become long) to disentangle the protected and unprotected sectors.  This result is consistent with~\cite{Dolan:2001tt}.

\subsection*{\underline{$\mathbf{\left\langle2233\right\rangle}$}}

As we discussed above, the structural form  of $\left\langle 2233\right\rangle$ is the same as that of $\left\langle 2222\right\rangle$. The reason for this is that we are computing the overlap of the $22$ OPE with the $33$ OPE, which in fact contains all the sectors of the $22$ OPE.
With coefficients given by
For convenience we repeat them
\begin{align}&\notag A_{2[\lambda]}=\frac{2 C (\lambda!)^2}{(2 \lambda)!}\text{ for }\lambda\in \mathbb{Z}_{\text{even}}\text{ zero otherwise},\\&\notag A_{4[\lambda_{1},\lambda_{2}]}=\frac{\lambda _1! \left(\lambda _1+1\right)! \left(\lambda _2!\right){}^2 \left(B \left(\lambda _1-\lambda _2+1\right) \left(\lambda _1+\lambda _2+2\right)+D (-1)^{\lambda_2}\right)}{\left(2 \lambda _2\right)! \left(2\lambda _1+1\right)!}\\&\text{    for $\lambda_{1}-\lambda_{2}\in \mathbb{Z}_\text{even}\geq 0$,  $\lambda_{2}\in \mathbb{Z}\geq 0$ and zero otherwise.} \end{align}
with~\eqref{eq:2233col}
\begin{align}
\begin{tabular}{|l|l|l|l|l|}
\hline
Correlator type   $SU(N)$                               & A      & B      & C    & D    \\ \hline
$\left\langle A_2A_2A_3A_3\right\rangle$         & $\frac{6 (N^2-1)^2(N^2-4)}{N}$&       $0$&     $\frac{36(N^2-1)(N^2-4)}{N}$&       $\frac{72(N^2-1)(N^2-4)}{N}$    \\     \hline
\end{tabular}\end{align}

The multiplet recombination is then identical to the $\langle2222\rangle$ case: essentially remove all $F^{112[\ula]}$ except for the half BPS case $F^{112[0]}$ in favour of long operators.

The result of performing this is:
\begin{align}&\notag\left\langle 2233\right\rangle=g_{12}^2g_{34}^3\Bigg(A+\left(\frac{g_{13}g_{24}}{g_{12}g_{34}}\right)2CF^{112[0]}+\left(\frac{g_{13}g_{24}}{g_{12}g_{34}}\right)^2\Bigg(\sum_{\lambda \geq 0}^{\infty}A_{4[\lambda]}F^{224[\lambda]}+\sum_{\lambda \geq 1}^{\infty}A'_{4[\lambda,1]}F^{224[\lambda,1]}\\&+\sum_{\lambda_{1}\geq \lambda_{2}\geq 2}^{\infty}A_{4[\lambda_{1},\lambda_{2}]}F^{224[\lambda_{1},\lambda_{2}]}+\sum_{\lambda\geq 1}^{\infty}A_{2[\lambda+1]}F^{224[\lambda,1]}_{\text{long}}\Bigg)\Bigg),\end{align}
where\begin{align}A'_{4[\lambda,1]}=A_{4[\lambda,1]}-A_{2[\lambda+1]},\end{align}
and again the first line consists of protected operators and the second line unprotected ops.

Interestingly, the coefficient $A'_{4[1,1]}$ of $F^{224[1,1]}$, namely $\frac{1}{6}(4B-2C-D)$  is subleading in the planar limit, whereas for the $\langle2222\rangle$ case it is not.
This can be understood as follows. The coefficient $A'_{4[1,1]}$ is related to the OPE coefficient of the genuine twist four quarter BPS operator. In the large N limit this is a double trace operator (see~\cite{Heslop:2001dr,D'Hoker:2243vf}). As described in section~\ref{sec:free-field-theory} the twist four operators arising from the $\cO_2 \cO_2$ OPE are double trace operators whereas the twist four operators arising from the $\cO_3 \cO_3$ OPE on the other hand involve a Wick contraction, which in the large N limit reduces to a single trace operator.

Also note that the presence of non-zero coefficients $A_{4[\lambda]}$ and $A'_{4[\lambda,1]}$ imply that the OPE coefficient $C_{33}^{\cO^{\text{twist 4}}}$ where $\cO^{\text{twist 4}}$ are the protected twist four operators,  can not be zero. This in turn has some unexpected  implications for the twist four part of the protected sector of the $\langle3333\rangle$ correlator as we shall see.

\subsection*{\underline{$\mathbf{\left\langle3333\right\rangle}$}}

Now we come to a more non-trivial case, the $\left\langle3333\right\rangle$ correlator which contains operators up to twist 6.  

Firstly we restate the result before recombination from the previous section. The OPE coefficients here are as in (\ref{eq:2222coeffs}) and (\ref{eq:3333cof}) where for the $A_{4[\underline{\lambda}]}$ coefficient of the former, we must do the change $A\rightarrow B$ and $B\rightarrow C$. 
\begin{align}&\notag \left\langle 3333\right\rangle=g_{12}^3g_{34}^3\Bigg(A+\left(\frac{g_{13}g_{24}}{g_{12}g_{34}}\right)\sum_{\lambda\geq 0}A_{2[\lambda]}F^{112[\lambda]}+\left(\frac{g_{13}g_{24}}{g_{12}g_{34}}\right)^2\sum_{\lambda_1\geq \lambda_2\geq 0}A_{4[\lambda_1,\lambda_2]}F^{224[\lambda_1,\lambda_2]}\\&+\left(\frac{g_{13}g_{24}}{g_{12}g_{34}}\right)^3\sum_{\lambda_1\geq \lambda_2\geq \lambda_3\geq 0}A_{6[\lambda_1,\lambda_2,\lambda_3]}F^{336[\lambda_1,\lambda_2,\lambda_3]}\Bigg),\end{align}
with coefficients
\begin{align}&\notag A_{2[\lambda]}=\frac{2 B (\lambda!)^2}{(2 \lambda)!}\text{ for }\lambda\in \mathbb{Z}_{\text{even}}\text{ zero otherwise},\\&\notag A_{4[\lambda_{1},\lambda_{2}]}=\frac{\lambda _1! \left(\lambda _1+1\right)! \left(\lambda _2!\right){}^2 \left(B \left(\lambda _1-\lambda _2+1\right) \left(\lambda _1+\lambda _2+2\right)+C (-1)^{\lambda_2}\right)}{\left(2 \lambda _2\right)! \left(2\lambda _1+1\right)!}\\&\text{    for $\lambda_{1}-\lambda_{2}\in \mathbb{Z}_\text{even}\geq 0$,  $\lambda_{2}\in \mathbb{Z}\geq 0$ and zero otherwise.} \end{align}
and exactly as is given in (\ref{eq:3333cof}), with colour factors
\begin{align}
\begin{tabular}{|l|l|l|l|}
\hline
Correlator type                                  & A      & B      & C    \\ \hline
$\left\langle A_3A_3A_3A_3\right\rangle$         & $\frac{9(N^2{-}4)^2(N^2{-}1)^2}{N^2}$ & $\frac{81 (N^2{-}4)^2(N^2{-}1)}{N^2}$ & $\frac{162(N^2{-}4)(N^2{-}1) \left(N^2{-}12\right)}{N^2}$    \\ \hline
\end{tabular}\end{align}

Here, the first manoeuver is to use the reducibility equation~\eqref{eq:51} to replace the short  Konishi and the succession of $\gamma=2$ semi-short operators by long operators as in the previous two cases. 

 However, now we need some additional information to help us with the twist four ($\gamma=4$) sector. In particular we need to know how many genuine short twist four operators there are in the theory (we already know from the $\langle2233\rangle$ correlator that it can not be zero). This can be answered by appealing to the classical interacting theory~\cite{Heslop:2001dr}. In analytic superspace the short twist four operators $\cO^{4[\lambda]}$ and  $\cO^{4[\lambda-1,1]}$ must be double trace operators of the form $A_2 \partial^\lambda A_2$ whereas those which combine to become long operators are single trace operators. Just as for the twist two operators, there is precisely one such operator for all even $\lambda$. The first few cases can also be checked with table 6 in the appendix of~\cite{Bianchi:2006ti}.

Armed with this knowledge that there is only one protected twist four operator for each case,  we can then use the considerations of section~\ref{sec:cons-checks-above} to {\em predict} the OPE coefficients, $\tilde A^{3333}_{4\ula}$,  after multiplet recombination, using  the corresponding coefficients from $\langle2222\rangle$ and $\langle2233\rangle$ via~\eqref{eq:46}. 

Namely we predict that
\begin{align}&\tilde{A}_{4[\lambda]}=\frac{\left(A_{4[\lambda]}^{ 2233}\right)^2}{A_{4[\lambda]}^{ 2222}}=\frac{1296 \left(N^2-4\right)^2 \left(N^2-1\right) \lambda ! (\lambda +1)!}{N^2 (2 \lambda +1)! \left(-\lambda  (\lambda +3)+(\lambda
   +1) (\lambda +2) N^2+2\right)},\\& \tilde{A}_{4[\lambda,1]}=\frac{\left(A_{4[\lambda,1]}^{'2233}\right)^2}{A_{4[\lambda,1]}^{' 2222}}=\frac{5184 \left(N^2-4\right)^2 \left(N^2-1\right) ((\lambda +1)!)^2}{N^2 (2 \lambda +2)! \left(\lambda  (\lambda +3)
   \left(N^2-1\right)-12\right)},\end{align}
where we may explicitly put in the colour factors.

We therefore deduce that we must use the reducibility equations to send part of the $\gamma=4$ superconformal partial waves to the $\gamma=6$ sectors, leaving the above coefficients. Moreover we find another consistency check in the fact that $\tilde{A}_{4[1,1]}=A'_{4[1,1]}$ corresponding to a protected quarter BPS operator which can not be combined with any higher weight operators to become long.

Altogether, this requires the use of the three reducibility equations, and the final equation comes from the redundancy of the Dynkin labels
\begin{align}\label{eq:recombination}&\notag F^{112[\lambda]}=\left(\frac{g_{13}g_{24}}{g_{12}g_{34}}\right)\left(F_{\text{long}}^{224[\lambda-1,1]}-F^{224[\lambda-1,1]}\right),\\&\notag F^{224[\lambda]}=\left(\frac{g_{13}g_{24}}{g_{12}g_{34}}\right)\left(F_{\text{long}}^{336[\lambda-1,1]}-F^{336[\lambda-1,1]}\right), \\&\notag F^{224[\lambda,1]}=\left(\frac{g_{13}g_{24}}{g_{12}g_{34}}\right)\left(F_{\text{long}}^{336[\lambda-1,1,1]}-F^{336[\lambda-1,1,1]}\right),\\& F^{224[\lambda_1,\lambda_2]}=\left(\frac{g_{13}g_{24}}{g_{12}g_{34}}\right)\left(F^{336[\lambda_1-1,\lambda_2-1,2]}\right).\end{align}

We thus obtain 
\begin{align}
&\frac{ \left\langle 3333\right\rangle}{g_{12}^3g_{34}^3}
  =\\
&\left.  \begin{array}{ll}
&A+\left(\frac{g_{13}g_{24}}{g_{12}g_{34}}\right)2BF^{112[0]}\\
&+\left(\frac{g_{13}g_{24}}{g_{12}g_{34}}\right)^2\Big[(2B+C)F^{224[0]}+\sum_{\lambda\geq 2}\tilde{A}_{4[\lambda]}F^{224[\lambda]}+\sum_{\lambda\geq 1}\tilde{A}_{4[\lambda,1]}F^{224[\lambda,1]}\Big]+\\[10pt]
&+\left(\frac{g_{13}g_{24}}{g_{12}g_{34}}\right)^3\Big[\sum_{\lambda\geq 0}A_{6[\lambda]}F^{336[\lambda]}+\frac{1}{10}(18A-14B-C)F^{336[1,1]}+\\
&\qquad \qquad \qquad \quad+\sum_{\lambda\geq 3}A'_{6[\lambda,1]}F^{336[\lambda,1]}+\sum_{\lambda\geq 2}A'_{6[\lambda,1,1]}F^{336[\lambda,1,1]}\Big]
  \end{array}\right\}\text{protected}\notag
\\
&\left.
  \begin{array}{ll}
&+
\left(\frac{g_{13}g_{24}}{g_{12}g_{34}}\right)^2\Big[\sum_{\lambda\geq 2}A_{2[\lambda,2]}F^{224[\lambda,2]} +\sum_{\lambda\geq 1}A_{2[\lambda+1]}F_{\text{long}}^{224[\lambda,1]}\Big]+\\[10pt]
& + \left(\frac{g_{13}g_{24}}{g_{12}g_{34}}\right)^3\Big[\sum_{\lambda_1\geq \lambda_2\geq 2}A_{6[\lambda_1,\lambda_2]}F^{336[\lambda_1,\lambda_2]}
+\sum_{\lambda_1\geq \lambda_2\geq 2}A_{6[\lambda_1,\lambda_2,1]}F^{336[\lambda_1,\lambda_2,1]}+\\
&\qquad \qquad \qquad \quad +\sum_{\lambda_1\geq \lambda_2\geq 2}A'_{6[\lambda_1,\lambda_2,2]}F^{336[\lambda_1,\lambda_2,2]}+ \sum_{\lambda\geq 2}A''_{6[\lambda,1,1]}F_{\text{long}}^{336[\lambda,1,1]}+\\
&\qquad \qquad \qquad \quad + \sum_{\lambda\geq 1}A'''_{6[\lambda+1]}F_{\text{long}}^{336[\lambda,1]}
  \Big]
  \end{array}\right\}\text{unprotected}
,\label{eq:44}\end{align}
where \begin{align}&\notag A'_{6[\lambda,1]}=A_{6[\lambda,1]}-A_{4[\lambda+1]}+\tilde{A}_{4[\lambda+1]}, \\&\notag A'_{6[\lambda,1,1]}=A_{6[\lambda,1,1]}-A_{4[\lambda+1,1]}+A_{2[\lambda+2]}+\tilde{A}_{4[\lambda+1,1]}, \\&\notag A'_{6[\lambda_1,\lambda_2,2]}=A_{6[\lambda_1,\lambda_2,2]}+A_{4[\lambda_1+1,\lambda_2+1]} ,\\&\notag A''_{6[\lambda,1,1]}=A_{4[\lambda+1,1]}-A_{2[\lambda+2]}-\tilde{A}_{4[\lambda+1,1]}, \\& A'''_{6[\lambda,1,1]}=A_{4[\lambda+1]}-\tilde{A}_{4[\lambda+1]}\end{align}
We have written~\eqref{eq:44} so that the first four lines correspond to the protected part whereas lines five to seven correspond to the unprotected piece.

The existence of a non-trivial protected twist four sector, $\tilde A$, differs from the assumption made in~\cite{Dolan:2244iy} that these should be absent and absorbed further into  long operators using~(\ref{eq:recombination}c). This question corresponds to the rather subtle point,  made in~\cite{Bianchi:2006ti}, that short operators which might combine to form long multiplets due to group theoretic considerations may in fact be protected dynamically.

Note that both the results here and the results of~\cite{Dolan:2244iy} are consistent with positivity of the OPE coefficients (we have checked and indeed all these coefficients remain  non-negative). Furthermore these results agree with~\cite{Dolan:2244iy}   in the large $N$ limit, since the coefficients $\tilde A$ are subleading.

\section{Conclusion}
\label{sec:conclusion}
In this paper we have provided the superconformal partial waves relevant for four-point functions of scalar operators in what we have called a super Grassmannian space $Gr(m|n,2m|2n)$. These are interesting mathematical objects in their own right, however they gain physical relevance for some selected values of the $(m,n)$ parameters, which yields $\cN=4$, $\cN=2$ and bosonic (super)conformal partial waves in four dimensions together with purely internal conformal partial waves. Critically, this all comes from the very same coefficient function $R^{\alpha\beta\gamma\ula}_{\umu}$ which does not depend on any particular group, but rather the Young tableaux $(\ula,\umu)$ only. The precise group only comes in via the (super) Schur polynomials. 

Further to this, we have re-summed the infinite expansion into a function. In particular, we made use of a determinant form of the super Schur polynomials to produce an analogous determinant like form for the superconformal partial wave in a re-summed form. Again, this is for completely arbitrary $(m,n)$ values. We expect that in the physically relevant cases, these forms will be useful for bootstrap applications.

We then considered $(m,n)=(2,2)$ which gives $\cN=4$ analytic superspace and initiated a detailed analysis of mixed charge half BPS four-point functions in the free theory.  We analysed the free theory OPE coefficients -- in both the $SU(N)$ and the $U(N)$ gauge theory -- of a number of correlators including the like-charge correlators $\left\langle 1111\right\rangle$, $\left\langle 2222\right\rangle$ and $\left\langle 3333\right\rangle$, along with the mixed charge cases $\left\langle 1122\right\rangle$, $\left\langle 2233\right\rangle$, $\left\langle 4233\right\rangle$ and finally $\left\langle 5344\right\rangle$, with the final two left for the appendix.  We finally considered the multiplet rearrangement due to the recombination of short operators into long operators for the $SU(N)$ theory. In particular the form of the $\left\langle 2233\right\rangle$ correlator in the $SU(N)$ gauge theory implies that there must be non-trivial twist four sector appearing in the $\langle3333\rangle$ correlator which remains protected.
Using the non-trivial information that can be extracted from $\left\langle 2233\right\rangle$ together with knowledge of the number of such protected operators only we are able to solve this degeneracy in this case. Thus we are able to fully determine the free-theory OPE coefficients of the $\left\langle 3333\right\rangle$ correlator in the interacting $SU(N)$ theory.

Looking forward, there are a number of directions to take. Computationally, in the $\cN=4$ SYM case there is much data --  anomalous dimensions and structure constants  --  to be extracted, which can then be compared to those computed via integrability.  Moreover, by understanding what the dimensionality of the vectors $C^{\cO}_{p_1p_2}$ are and using its inner product we could go ahead and work out the precise OPE coefficients for further correlators, in particular those which we have not studied all the way here.

On the bootstrap side it would be interesting to revisit and continue the work of~\cite{Beem:2013qxa,Alday:2014qfa}  analysing  the superconformal bootstrap in $\cN$=4 SYM for higher charge correlators.

Other supercofnromal theories not covered by the Grassmanian theories here the  mysterious six-dimensional  $(2,0)$ theory. 
A superconformal partial wave analysis of  the energy-momentum correlator in the $(2,0)$  theory was
performed in~\cite{Heslop:2004du} and superconformal partial waves were also considered in~\cite{Dolan:2004mu}. On the
bootstrap side there has been recent work analysing the restrictions on anomalous dimensions for this theory in~\cite{Beem:2015aoa}.
It would also be interesting to see if the method presented here can be modified to this and related theories.

\section*{Acknowledgements}

We would like to thank Burkhard Eden for collaboration in the early parts of this project and fruitful discussions. We would also like to thank Pedro Liendo, Dhritiman Nandan and Balt van Rees for interesting conversations. PH would also like to acknowledge the many conversations and previous work with Paul Howe out of  which the current work emerged.
RD acknowledges support from an STFC studentship,  PH from the STFC Consolidated Grant ST/L000407/1. We also both acknowledge support from the Marie Curie network GATIS (\href{http://gatis.desy.eu}{gatis.desy.eu}) of the European Union's Seventh Framework Programme FP7/2007-2013/ under REA Grant Agreement No 317089.

\appendix

\section{Proof of conformal partial wave for $GL(m)$}
\label{sec:proof-conf-part}

In this section we present a proof for the form of the conformal partial wave presented in (\ref{eq:29b}) and in particular the coefficients in (\ref{eq:27}). 
The proof follows a similar procedure to that of~\cite{Dolan:2243hv} for the conformal 4d case ($m=2,n=0$).
For conformal partial waves in $GL(m)$, the space-time coordinate $x^{\alpha\dot{\alpha}}$ is an $m$-dimensional matrix, where
\begin{align}x_{ij}^{2}:=\text{det}(x_{ij})=\frac{1}{m!}\left(x_{ij\dot{\alpha}_1}^{\alpha_1}\dots x_{ij\dot{\alpha}_m}^{\alpha_m}\right)\epsilon^{\dot{\alpha}_{1}\dots \dot{\alpha}_{m}}\epsilon_{\alpha_{1}\dots \alpha_{m}}.\end{align}
We may then consider some scalar operators $\Phi_{\Delta}(x)$ which take representation in ${SL}(m)$. The four-point function of these operators is given by
\begin{align}\label{eq:glm4pntfunction}\left\langle \Phi_{\Delta_1}(x_1)\Phi_{\Delta_2}(x_2)\Phi_{\Delta_3}(x_3)\Phi_{\Delta_4}(x_4)\right\rangle=\frac{1}{(x_{12}^2)^{\frac{\Delta_{1}+\Delta_{2}}{2}}(x_{34}^2)^{\frac{\Delta_{3}+\Delta_{4}}{2}}}\left(\frac{x_{14}^{2}}{x_{24}^{2}}\right)^{\frac12\Delta_{21}}\left(\frac{x_{13}^2}{x_{14}^{2}}\right)^{\frac12\Delta_{43}}F(x).\end{align}
Where as in the main text, $F(x)$ is a function of the $m$ many eigenvalues of $z=x_{12}x^{-1}_{24}x_{43}x^{-1}_{31}$ labeled $x_i$. We consider inverse variables in the first instance as it will be easier to apply the Casimir operator in this way, we call $\omega=z^{-1}$.
In fact, since we will be taking Schur polynomials of this matrix, we can diagonalise $\omega$ to be $\text{diag}(1/x_1,1/x_2,\dots,1/x_m)$,  and we call $w_{i}:=1/x_{i}$.

We are considering the Grassmannian $Gr(m,2m)$ which can be viewed as the space of $2m\times m$ matrices given by  $u^{A}_{\alpha}$. This is where the small Greek indices refer to the isotropy group whilst the big Latin indices refer to the global group. Explicitly, one can put coordinates on this by using the section
\begin{align}u^{A}_{\alpha}=\left(\delta^{\beta}_{\alpha},x^{\dot{\beta}}_{\alpha}\right)\text{ , }\bar{u}^{\dot{\alpha}}_{A}=\left(\begin{array}{c}-x^{\dot{\alpha}}_{\alpha}\\ \delta^{\dot{\alpha}}_{\dot{\beta}}\end{array}\right),\end{align}
So that we have $u^{A}_{i\alpha}\bar{u}^{\dot{\alpha}}_{jA}=x_{ij\alpha}^{\dot{\alpha}}$. In the $m=2$ case, we may view $u_{\alpha}^{A}$ as being a pair of twistors, as was used in a similar context in \cite{Fitzpatrick:2014oza}. The benefit of this is that the generators of $GL(m)$ are given by
\begin{align}D^{A}_{B}=u_{A}^{\alpha}\frac{\partial}{\partial u^{\alpha}_{B}},\end{align}
which satisfies the algebra:
\begin{align}\left[D^{A}_{B},D_{D}^{C}\right]=\delta^{C}_{B}D_{D}^{A}-\delta^{A}_{D}D^{C}_{B}.\end{align}
The conformal partial waves are 
eigenfunctions of the quadratic Casimir operator which will act on the four-point function (\ref{eq:glm4pntfunction}) at points $1$ and $2$. This is given by
\begin{align}\label{eq:cas}\frac{1}{2}D^{2}_{12}=\frac{1}{2}(D_{1B}^{A}+D_{1B}^{A})(D_{1A}^{B}+D_{1A}^{B}).\end{align}

In order to find the coefficients $r^{\alpha\beta\gamma\underline{\lambda}}_{\mu_1,\dots,\mu_m}$, in an expansion in Schur polynomials we will proceed by doing two things. Firstly we will reexpress (\ref{eq:cas}) in terms of the eigenvalues of $\omega$; namely $w_i$, by considering  its action on $GL(m)$ Schur polynomials of $\omega$. We can then trivially invert the eigenvalues, and then apply it to the correlation function (\ref{eq:glm4pntfunction}). This will lead to an action upon the conformal partial wave $F^{\underline{\lambda}}(x)=\sum_{\underline{\mu}\geq \underline{\lambda}}t_{\mu_1,\dots,\mu_m}^{\underline{\lambda}}s_{\underline{\mu}}(x)$, which in turn leads to a recursion relation on $t_{\mu_1,\dots,\mu_m}^{\underline{\lambda}}$. The derivation then concludes by finding that for the superconformal partial wave associated to this work a form of these coefficients is given by $r^{\alpha\beta\gamma\underline{\lambda}}_{\mu_1,\dots,\mu_m}$ given in  (\ref{eq:27}).

\subsection{Eigenvalue basis}

Now let us consider the entire correlator function in (\ref{eq:glm4pntfunction}), in which we take the function $F(w)$ to a be linear combination of Schur polynomials, a direct application of the Casimir gives
\begin{align}\label{eq:AA}&\notag \frac12 D_{12}^2 \left\langle \Phi(x_1)\Phi(x_2)\Phi(x_3)\Phi(x_4)\right\rangle=\frac{1}{(x_{12}^2)^{\frac{\Delta_{1}+\Delta_{2}}{2}}(x_{34}^2)^{\frac{\Delta_{3}+\Delta_{4}}{2}}}\left(\frac{x_{14}^{2}}{x_{24}^{2}}\right)^{\frac12\Delta_{21}}\left(\frac{x_{13}^2}{x_{14}^{2}}\right)^{\frac12\Delta_{43}}\\&\times\left[\left(\frac12\left(\Delta_{34}-\Delta_{12}\right)\frac{\partial}{\partial \tr(\omega)}-\frac14\Delta_{34}\Delta_{12}\sum_{i=1}^{m}\frac{1}{w_i}\right)F(w)+\frac12 D_{12}^{2}F(w)\right].\end{align}

Since $F(\omega)$ is a linear combination of Schur polynomials it is useful to consider the action of the Casimir upon these first.
We note that since $D_{12B}^{A}u_{iC}^{\alpha}=u_{iB}^{\alpha}\delta^{A}_{C}\text{ and }D_{12A}^{B}\bar{u}_{i\dot{\delta}}^{C}=-\delta^{C}_{A}\bar{u}_{i\dot{\delta}}^{B}$ for $i=1$ or $2$, it follows that
\begin{align}\label{eq:p1}&\notag D_{12}^{2}\omega^{\alpha}_{\beta}= 2(2m \omega^{\alpha}_\beta- m\delta^{\alpha}_{\beta}),\\&
D_{12J}^{I}\omega^{\alpha}_{\beta}D_{12I}^{J}\omega^{\gamma}_{\rho}=2\omega^{\alpha}_{\rho}\omega^{\gamma}_{\beta}-\omega^{\alpha}_{\rho}\delta^{\gamma}_{\beta}-\delta^{\gamma}_{\beta}\omega^{\alpha}_{\rho}.\end{align}
The $GL(m)$ Schur polynomial admits the following form in terms of the matrix $\omega$
\begin{align}\label{eq:schurmatrix}s_{\underline{\lambda}}(w)=\frac{1}{m!}\sum_{\sigma\in S_{m}}\chi_{\underline{\lambda}}(\sigma)\omega^{\alpha_{\sigma(1)}}_{\alpha_{1}}\omega^{\alpha_{\sigma(2)}}_{\alpha_{2}}\dots \omega^{\alpha_{\sigma(m)}}_{\alpha_{m}}=\frac{1}{m!}\sum_{a_{i}}\chi_{\underline{\lambda}}(\left\{a_{i}\right\})C(\left\{a_i\right\})\prod_{i=1}^{m}\tr(\omega^i)^{a_{i}},\end{align}
where $\sum_{i}\lambda_i=m$, and $\chi_{\underline{\lambda}}$ is the character of the corresponding $S_m$ representation in the first equality. In the second equality the set $\left\{a_i\right\}$ is the number of $i$-cycles (subject to the constraint $\sum_{i}a_{i}=m$), whilst $C(\left\{a_i\right\})$ is the number of terms in a given conjugacy class of $S_{m}$. By using this form of the Schur polynomial together with (\ref{eq:p1}), we find
\begin{align}\label{eq:operator}
\frac{1}{2}D_{12}^{2}s_{\underline{\lambda}}(w)=\left(2m \omega^{\alpha}_{\beta}-m\delta^{\alpha}_{\beta}\right)\frac{\partial s_{\underline{\lambda}}(w)}{\partial \omega^{\alpha}_{\beta}}+\omega^{\alpha}_{\rho}\left(\omega^{\gamma}_{\beta}-\delta^{\gamma}_{\beta}\right)\frac{\partial^2 s_{\underline{\lambda}}(w)}{\partial \omega^{\gamma}_{\rho}\partial \omega^{\alpha}_{\beta}}.\end{align}
In order to retrieve the usual form in terms of $m$ variables $w_{i}$, one simply diagonalises the $\omega$ matrices.

The first two terms of (\ref{eq:operator}) are linear in differential operators and are therefore trivial to diagonalise. The corresponding eigenvalue result will also be in terms of linear differential operators. The results are
\begin{align}&\notag  2m \omega^{\alpha}_{\beta}\frac{\partial s_{\underline{\lambda}}(w)}{\partial \omega^{\alpha}_{\beta}}=2m\left[\sum_{i=1}^{n} w_{i}\frac{\partial}{\partial w_i}\right]s_{\underline{\lambda}}(w)=2m\sum_{i=1}^{m}\lambda_i s_{\underline{\lambda}}(w),\\& m\delta^{\alpha}_{\beta}\frac{\partial s_{\underline{\lambda}}(w)}{\partial \omega^{\alpha}_{\beta}}=m\frac{\partial s_{\underline{\lambda}}(w)}{\partial \tr(\omega)}=m\left[\sum_{i=1}^{m}\frac{\partial}{\partial w_i}\right] s_{\underline{\lambda}}(w)=m\sum_{i=1}^{m}(\lambda_i-i+m)s_{(\lambda_1,\lambda_2,\dots,\lambda_i-1,\dots,\lambda_m)}(w).\end{align}
A proof of the RHS of the second expression can be found in appendix A of \cite{deMelloKoch:2007uu}. 

The last two terms of (\ref{eq:operator}) are slightly more non-trivial than the previous cases, since these are quadratic in differentials, however in the eigenvalue basis it may include quadratic as well as linear differentials. Instead, we can apply the matrix action of quadratic differential terms upon $\prod_{i=1}^{m}\tr(\omega^i)^{a_{i}}$, and consider as many different values of $m$ in which in it takes to find a consistent differential operator in terms of $w_i$. It is good enough to consider $\prod_{i=1}^{m}\tr(\omega^i)^{a_{i}}$ since this produces symmetric polynomials upon diagonalisation. 

We begin by defining the Vandermonde determinant:
\begin{align}\text{vdet}^{(m)}(w)=(-1)^{\scriptsize{\left(\begin{array}{c} m\\ 2\end{array}\right)}}\text{det}_{ij}(w_i^{j-1})=\text{det}_{ij}(w_i^{m-j})=\prod_{1\leq i<j\leq m}(w_i-w_j),\end{align}
one then finds that
\begin{align}&\notag \omega^{\alpha}_{\rho}\omega^{\gamma}_{\beta}\frac{\partial^2 }{\partial \omega^{\gamma}_{\rho}\partial \omega^{\alpha}_{\beta}}\prod_{i=1}^{m}\tr(\omega^{i})^{a_{i}}\\&=\left[-\sum _{j=1}^n \frac{j^2 a_j \tr\left(\omega ^{2 j}\right)}{\tr\left(\omega ^j\right)^2}+\sum _{j=1}^m \sum _{k=0}^{j-2} \frac{j a_j
   \tr\left(\omega ^{k+1}\right) \tr\left(\omega ^{j-k-1}\right)}{\tr\left(\omega ^j\right)}+\sum _{k=1}^m \sum _{j=1}^m \frac{j k a_j a_k
   \tr\left(\omega ^{j+k}\right)}{\tr\left(\omega ^j\right)\tr\left(\omega ^k\right)}\right]\prod_{i=1}^{m}\tr(\omega^{i})^{a_{i}},\end{align}
by putting in various examples for $m$, we find that the following operator always gives the correct result 
\begin{align}\omega^{\alpha}_{\rho}\omega^{\gamma}_{\beta}\frac{\partial^2 }{\partial \omega^{\gamma}_{\rho}\partial \omega^{\alpha}_{\beta}}=\frac{1}{\text{vdet}^{(m)}(w_i)}\sum_{i=1}^{n} w_{i}^{2}\frac{\partial}{\partial w_i^2}\text{vdet}^{(m)}(w_i)-2(m-1)\sum_{i=1}^{m}w_i \frac{\partial}{\partial w_i}-\frac{m}{3}(m-1)(m-2).\end{align}
Similarly we find
\begin{align}&\notag \omega^{\alpha}_{\rho}\delta^{\gamma}_{\beta}\frac{\partial^2 }{\partial \omega^{\gamma}_{\rho}\partial \omega^{\alpha}_{\beta}}\prod_{i=1}^{m}\tr(\omega^{i})^{a_{i}}\\&\left[-\sum _{j=1}^m \frac{j^2 a_j \tr\left(\omega ^{2 j-1}\right)}{\tr\left(\omega ^j\right)^2}+\sum _{j=1}^m \sum _{k=0}^{j-2} \frac{j a_j
   \tr\left(\omega ^k\right)\tr\left(\omega ^{j-k-1}\right)}{\tr\left(\omega ^j\right)}+\sum _{k=1}^m \sum _{j=1}^m \frac{j k a_j a_k
   \tr\left(\omega ^{j+k-1}\right)}{\tr\left(\omega ^j\right) \tr\left(\omega ^k\right)}\right]\prod_{i=1}^{m}\tr(\omega^{i})^{a_{i}},\end{align}
in which with various different values of $m$, always agrees with the operator:
\begin{align}\omega^{\alpha}_{\rho}\delta^{\gamma}_{\beta}\frac{\partial^2 }{\partial \omega^{\gamma}_{\rho}\partial \omega^{\alpha}_{\beta}}=\frac{1}{\text{vdet}^{(m)}(w)}\sum_{i=1}^{m}\frac{\partial}{\partial w_i}w_{i}\frac{\partial}{\partial w_i}\text{vdet}^{(m)}(w)-m\sum_{i=1}^{m}w_i \frac{\partial}{\partial w_i}.\end{align}

Putting this together with (\ref{eq:AA}), inverting the coordinates so that the Casimir is in terms of $x_i$ where $x_i=\frac{1}{w_i}$, namely with $D^{(m)}:=\frac12 D_{12}^{2}|_{w_i\rightarrow \frac{1}{x_i}}$, we find that
\begin{align}&\notag D^{(m)}=\frac{1}{\text{vdet}^{(m)}(x)}\Bigg[\sum_{i=1}^{m}\Bigg[x_i \left(-x_i \left(\frac12\left(\Delta_{34}-\Delta_{12}\right)-2 m+3\right)-2 m+2\right)\frac{\partial}{\partial x_i}+(1-x_i)x_i^2\frac{\partial^2}{\partial x_i^2}\\&\notag-\left(\frac12\Delta_{21}-m+1\right)\left(\frac12\Delta_{34}-m+1\right) x_i\Bigg]+\frac{m}{3}(m-1)(2m-1)\Bigg]\text{vdet}^{(m)}(x).\end{align}

\subsection{Recursion relation}

The action of the Casimir operator corresponding to the contribution of an operator in the OPE yields the eigenvalue equation on the four-point function 
\begin{align}\label{eq:6proof}D^{(m)}\left\langle \Phi(x_1)\Phi(x_2)\Phi(x_3)\Phi(x_4)\right\rangle= \sum_{i=1}^{m}\lambda_i(\lambda_i-(2i-1))\left\langle \Phi(x_1)\Phi(x_2)\Phi(x_3)\Phi(x_4)\right\rangle.\end{align}
This eigenvalue is simply  the value of the  Casimir for the corresponding representation of ${SL}(2m)$ (rather than the induced $SL(m)$ representation).

We define the $GL(m)$ conformal partial wave in (\ref{eq:glm4pntfunction}) to have the form of an expansion in Schur polynomials
\begin{align}F(x)=\sum_{\lambda_{i+1}\geq \lambda_i}F^{\underline{\lambda}}(x)\text{ where } F^{\underline{\lambda}}=\sum_{\underline{\mu}\geq \underline{\lambda}}t_{\mu_1,\dots,\mu_m}^{\underline{\lambda}}s_{\underline{\mu}}(x)\ .\end{align}
By noting the action of the Casimir upon the Schur polynomial
\begin{align}\label{eq:8proof}D^{(m)}s_{\underline{\mu}}(x)=\left(\sum_{i=1}^{m}\mu_{i}(\mu_{i}{-}(2i{-}1))\right)s_{\underline{\mu}}(x){-}\left(\sum_{i=1}^{m}(\mu_{i}{-}(i{-}1){-}\frac12\Delta_{12})(\mu_{i}{-}(i{-}1)+\frac12\Delta_{34})s_{(\dots,\mu_{i}+1,\dots)}(x)\right),\end{align}
and following (\ref{eq:6proof}), it follows that the action of the quadratic Casimir operator upon the four point function yields the recursion relation on $t_{\mu_1,\dots,\mu_m}^{\underline{\lambda}}$
\begin{align}\sum_{i=1}^{p} \left(\left(\mu_{i}-\lambda_i\right)\left(\lambda_{i}+\mu_{i}-(2i-1)\right)t_{\mu_1,\dots,\mu_m}^{\underline{\lambda}}-\left(\mu_{i}-i-\frac12\Delta_{12}\right)\left(\mu_{i}-i+\frac12\Delta_{34}\right) t_{\mu_1,\dots,\mu_i-1,\dots,\mu_m}^{\underline{\lambda}}\right)=0\end{align}
which is solved by:
\begin{align}t^{\underline{\lambda}}_{\mu_1\dots \mu_m}=\prod _{i=1}^m \frac{\left(\lambda_i+1-i+\frac12\Delta_{21}\right)^{\mu_i-\lambda_i} \left(\lambda_i+1-i+\frac12\Delta_{34}\right)^{\mu_i-\lambda_i}}{\left(\mu_i-\lambda_i\right)! \left(2 \lambda _i-2 i+2\right)^{\mu_i-\lambda_i}}\end{align}
where $(x)^{y}$ is the raising Pochhammer symbol. In taking $m=2$, we find agreement with \cite{Dolan:2243hv}. However, in the supersymmetric case the conformal partial wave is accompanied with the super-cross ratio
\begin{align}\left(\frac{g_{13}g_{24}}{g_{12}g_{34}}\right)^{\frac12\gamma}F^{\alpha\beta\gamma\underline{\lambda}}(Z)=\text{sdet}(Z)^{\frac12\gamma}F^{\alpha\beta\gamma\underline{\lambda}}(Z).\end{align}
In view of this we instead consider a shifted conformal partial wave
\begin{align}F^{\underline{\lambda}+m}=\sum_{\underline{\mu}\geq 0}t_{\mu_1,\dots,\mu_m}^{\underline{\lambda}+m}s_{\underline{\mu}+m}(x),\end{align}
where $\underline{\lambda}+m=[\lambda_1+m,\lambda_2+m,\dots,\lambda_m+m]$. Noting that $s_{\underline{\lambda}+m}=\left(\prod_{i=1}^{m}x_i\right)^m s_{\underline{\lambda}} =\text{det}(z)^m s_{\underline{\lambda}}$, we find that
\begin{align}F^{\underline{\lambda}+m}=\left(\prod_{i=1}^{m}x_i\right)^m\sum_{\underline{\mu}\geq \underline{\lambda}}t_{\mu_1,\dots,\mu_m}^{\underline{\lambda}+m}s_{\underline{\mu}}(x)\end{align}
where now we may now define the resulting coefficients by $r^{\alpha\beta\gamma\underline{\lambda}}_{\mu_1,\dots,\mu_m}$
\begin{align}r^{\alpha\beta\gamma\underline{\lambda}}_{\mu_1,\dots,\mu_m}:=t_{\mu_1,\dots,\mu_m}^{\underline{\lambda}+m}=\prod _{i=1}^m \frac{\left(\lambda_i+1-i+\alpha\right)^{\mu_i-\lambda_i} \left(\lambda_i+1-i+\beta\right)^{\mu_i-\lambda_i}}{\mu_i! \left(2 \lambda _i+2-2i+\gamma\right)^{\mu_i-\lambda_i}}\end{align}
Where here, $\alpha=\frac12\left(2m-\Delta_{12}\right)$, $\beta=\frac12\left(2m+\Delta_{34}\right)$ and $\gamma=2m$.

\section{Further results for the free theory}
\label{sec:further-results-free}
In this section, we give the free theory OPE coefficients of correlation functions $\left\langle 4233\right\rangle$ and $\left\langle 5344\right\rangle$. These cases distinguish themselves from the cases studied in the main text. Firstly, we now have $p_{12}=2\neq0$. Secondly, for the first time there can be more than one type of half BPS operator, even in the $\text{SU}(N)$ gauge theory (e.g. at charge four $\tr(W^4)$ as well as $\tr(W^2)^2$.)
\subsection*{\underline{$\mathbf{\left\langle 4233\right\rangle}$}}
The correlator is written as
\begin{align}&\notag\left\langle 4233\right\rangle=A \left(g_{14} g_{24}^2 g_{13}^3+g_{14}^3 g_{23}^2 g_{13}\right)+B g_{13}^2 g_{23} g_{24} g_{14}^2+C g_{12}^2 g_{13} g_{34}^2 g_{14}\\&\notag+D
   \left(g_{12} g_{14} g_{24} g_{34} g_{13}^2+g_{12} g_{14}^2 g_{23} g_{34} g_{13}\right)\\&=g_{12}^3g_{34}^3\frac{g_{14}}{g_{24}}\left(\left(\frac{g_{13}g_{24}}{g_{12}g_{34}}\right)f_{2}(C,0)+\left(\frac{g_{13}g_{24}}{g_{12}g_{34}}\right)^2f_{4}(D,D,0)+\left(\frac{g_{13}g_{24}}{g_{12}g_{34}}\right)^3f_{6}(A,B,A,0)\right)\end{align}
We can tabulate the SU$(N)$ colour factors
\begin{align}\begin{tabular}{|l|l|l|l|l|}
\hline
Correlator type                                  & A      & B      & C    & D\\ \hline
$\left\langle A_4A_2A_3A_3\right\rangle$         & $0$ & $\frac{72(N^2-1)(N^2-4)(N^2-6)}{N^2}$ & $\frac{72(N^2-1)(N^2-4)(2N^2-3)}{N^2}$ & $\frac{144(N^2-1)(N^2-4)(N^2-6)}{N^2}$ \\ 
$\left\langle (A_2A_2)A_2A_3A_3\right\rangle$         & $0$ & $\frac{144(N^2-1)(N^2-4)}{N}$ & $\frac{72(N^2-1)(N^2-4)(1+N^2)}{N}$ & $\frac{288(N^2-1)(N^2-4)}{N}$ \\ \hline
\end{tabular}\end{align}
There are many potential trace structures appropriate to the $\text{U}(N)$ theory, we tabulate some of the possible partitions
\begin{align}\hspace{-1cm}\begin{tabular}{|l|l|l|l|l|}
\hline
Correlator type                                  & A      & B      & C    & D\\ \hline
$\left\langle A_4A_2A_3A_3\right\rangle$         & $216 N^2 (1 + N^2)$ & $72 N^2 (5 + N^2)$ & $144 N^2 (2 + N^2)$ & $144 N^2 (5 + N^2)$ \\ 
$\left\langle (A_2A_2)A_2A_3A_3\right\rangle$         & $432 N^3$ & $144 N (1 + 2 N^2)$ & $72 N (2 + N) (1 + N^2)$ & $288 N (1 + 2 N^2)$ \\ 
$\left\langle (A_1A_3)A_2A_3A_3\right\rangle$         & $54 N^3 (7 + N^2)$ & $216 N (1 + N^2)$ & $108 N (1 + 3 N^2)$ & $432 N (1 + N^2)$ \\ 
$\left\langle (A_1^2A_2)A_2A_3A_3\right\rangle$         & $216N^2(1+N^2)$ & $432N^2$ & $36N^2(9+2N+N^2)$ & $864N^2$ \\
$\left\langle A_4(A_1)^2(A_1A_2)(A_1A_2)\right\rangle$         & $432N^3$ & $16N(12+13N^2+2N^4)$ & $48N(6+N+2N^2)$ & $96N(4+5N^2)$ \\
$\left\langle (A_1)^4(A_1)^2(A_1)^3(A_1)^3\right\rangle$         & $432N^6$ & $432N^6$ & $432N^6$ & $864N^6$ \\
$\left\langle (A_1^2A_2)(A_1)^2(A_1A_2)A_3\right\rangle$         & $72N^2(5+N)$ & $24N^2(14+N+3N^2)$ & $48N^2(5+4N^2)$ & $48N^2(15+N+2N^2)$ \\
$\left\langle (A_2A_2)(A_1)^2(A_1A_2)A_3\right\rangle$         & $144N(2+N)$ & $48N(4+4N^2+N^3)$ & $48N(4+5N^2)$ & $96N(6+N+2N^2)$ \\
\hline
\end{tabular}\end{align}

In comparing with the appropriate SCPW expansion one finds the result
\begin{align}\label{eq:OPEc4233}&\notag\left\langle 4233\right\rangle=g_{12}^3g_{34}^3\frac{g_{14}}{g_{24}}\left(\left(\frac{g_{13}g_{24}}{g_{12}g_{34}}\right)\sum_{\lambda\geq 0}A_{2[\lambda]}F^{012[\lambda]}+\Bigg(\frac{g_{13}g_{24}}{g_{12}g_{34}}\right)^2\sum_{\lambda_1\geq \lambda_2\geq 0}A_{4[\lambda_1,\lambda_2]}F^{124[\lambda_1,\lambda_2]}\\&+\left(\frac{g_{13}g_{24}}{g_{12}g_{34}}\right)^3\sum_{\lambda_1\geq \lambda_2\geq \lambda_3\geq 0}A_{6[\lambda_1,\lambda_2,\lambda_3]}F^{236[\lambda_1,\lambda_2,\lambda_3]}\Bigg),\end{align}
with the following coefficients
\begin{align}\label{eq:c4233}&\notag  A_{2[0]}=C\text{ all else 0},\\&\notag A_{4[\lambda_1]}=\frac{D \lambda_1 ! (\lambda_1 +2)!}{(2 \lambda_1 +1)!}\text{ for }\lambda_1\in \mathbb{Z}_{\text{even}}\text{ and all else }0,\\&\notag A_{6[\lambda_1,\lambda_2]}=\frac{4 (-1)^{\lambda _2} \left(\lambda _1+2\right) \left(\lambda _1+3\right) \left(\lambda _2+2\right) \left(\left(\lambda _1+2\right)!\right){}^2 \left(\left(\lambda _2+1\right)!\right){}^2}{\left(2 (-1)^{\lambda _2} \lambda_1+5 (-1)^{\lambda _1}-(-1)^{\lambda _2}\right) \left(2 \lambda _1+4\right)! \left(2 \lambda _2+2\right)!}\\&\notag \times \left(\frac{1}{24} A \left(12 \left(\lambda _1-3\right) \lambda _1+\left(96 \lambda _1-12 \lambda _2 \left(\lambda _2+3\right)+25\right)+23\right)+B (-1)^{\lambda _2}\right)\\& \text{ for $\lambda_{1}-\lambda_{2}\in \mathbb{Z}_{\text{even}}\geq 0, \lambda_{2}\geq 0$ }.\end{align}
All other coefficients are vanishing.

As a non-trivial check we can compute the OPE coefficients for the correlator $\left\langle 3342\right\rangle$. We find the the explicit ingredient of the SCPW expansion change, namely one uses $F^{122[\underline{\lambda}]}$, $F^{234[\underline{\lambda}]}$ and $F^{346[\underline{\lambda}]}$ instead of the SCPW's used in (\ref{eq:OPEc4233}). However, critically the result for the OPE coefficients give identically the same result as in (\ref{eq:OPEc4233}). Furthermore we also note that the results for $A_{6[\lambda_1,\lambda_2]}$ agree perfectly in the large $N$ limit with those obtained from free 3-point functions in~\cite{vw} (see the first row of table 5).

\subsection*{\underline{$\mathbf{\left\langle 5344\right\rangle}$}}
The correlator is given by
\begin{align}&\notag \left\langle 5344\right\rangle= A(g_{14} g_{24}^3 g_{13}^4+g_{14}^4 g_{23}^3 g_{13})+B(g_{14}^2 g_{23} g_{24}^2 g_{13}^3+g_{14}^3 g_{23}^2 g_{24} g_{13}^2)+C(g_{12} g_{14} g_{24}^2 g_{34} g_{13}^3+g_{12} g_{14}^3 g_{23}^2 g_{34} g_{13})\\&\notag+D(g_{12} g_{13}^2 g_{14}^2 g_{23} g_{24} g_{34})+E(g_{12}^2 g_{13} g_{14}^2 g_{23} g_{34}^2+g_{12}^2 g_{13}^2 g_{14} g_{24} g_{34}^2)+F(g_{12}^3 g_{13} g_{14} g_{34}^3)\\&\notag=g_{12}^4g_{34}^4\frac{g_{14}}{g_{24}}\Bigg(\left(\frac{g_{13}g_{24}}{g_{12}g_{34}}\right)f_{2}(F,0)+\left(\frac{g_{13}g_{24}}{g_{12}g_{34}}\right)^2f_{4}(E,E,0)+\left(\frac{g_{13}g_{24}}{g_{12}g_{34}}\right)^3f_{6}(C,D,C,0)\\&+\left(\frac{g_{13}g_{24}}{g_{12}g_{34}}\right)^4f_{8}(A,B,B,A,0)\Bigg)\end{align}
We have given some of the colour factors in the next page. The SCPW expansion is given by
\begin{sidewaystable}
\begin{tabular}{|l|l|l|l|l|l|}
\hline
Correlator type                                  & A      & B      & C    & D & E \\ \hline
$\left\langle A_5A_3A_4A_4\right\rangle$         & $0$ & $\frac{240M\left(N^2-6\right)\left(N^4-6 N^2+36\right)}{N^4}$ & $\frac{480M\left(N^2-6\right) \left(N^4-6 N^2+36\right)}{N^4}$ & $\frac{480M\left(N^6+3 N^4+72 N^2-864\right)}{N^4}$ & $\frac{480M\left(N^6-6N^4+99N^2-378\right)}{N^4}$ \\
$\left\langle (A_2A_3)A_3A_4A_4\right\rangle$         & $0$ & $\frac{480 M\left(N^2-6\right) \left(2 N^2-9\right)}{N^3}$ & $\frac{960M\left(N^2-6\right) \left(2 N^2-9\right)}{N^3}$ & $\frac{2880M\left(2 N^4-21 N^2+72\right)}{N^3}$ & $\frac{2160 M\left(N^4-10 N^2+42\right)}{N^3}$ \\
$\left\langle A_5A_3(A_2A_2)A_4\right\rangle$         & $0$ & $\frac{1440M \left(N^2-6\right) \left(N^2-2\right)}{N^3}$ & $\frac{2880 M \left(N^2-6\right) \left(N^2-2\right)}{N^3}$ & $\frac{5760 M \left(N^4-7 N^2+24\right)}{N^3}$ &$\frac{1440 M\left(3 N^4-13 N^2+42\right)}{N^3}$ \\
$\left\langle (A_2A_3)A_3(A_2A_2)A_4\right\rangle$         & $0$ & $\frac{160M\left(N^2-6\right) \left(N^2+9\right)}{N^2}$ & $\frac{320M\left(N^2-6\right) \left(N^2+9\right)}{N^2}$ & $\frac{960 M \left(N^3+16 N^2-6 N-78\right)}{N^2}$ & $\frac{4320 M \left(2 N^2-7\right)}{N^2}$ \\
\hline
\end{tabular}
\begin{tabular}{|l|l|}
\hline
Correlator type                                  &  F \\ \hline
$\left\langle A_5A_3A_4A_4\right\rangle$         & $\frac{480 M  \left(N^2-2\right) \left(N^4-6 N^2+18\right)}{N^4}$ \\
$\left\langle (A_2A_3)A_3A_4A_4\right\rangle$    & $\frac{480 M\left(N^4-6 N^2+18\right)}{N^3}$ \\
$\left\langle A_5A_3(A_2A_2)A_4\right\rangle$    &  $\frac{960 M\left(N^2-2\right) \left(2 N^2-3\right)}{N^3}$ \\
$\left\langle (A_2A_3)A_3(A_2A_2)A_4\right\rangle$  & $\frac{960 M \left(2 N^2-3\right)}{N^2}$ \\
\hline
\end{tabular}\label{tab:1}\caption{Colour factors for $\left\langle 5344\right\rangle$ in $\text{SU}(N)$ gauge theory and $M=(N^2-4) (N^2-1)$}

\begin{tabular}{|l|l|l|l|l|l|l|}
\hline
Correlator type                                  & A      & B      & C    & D  \\ \hline
$\left\langle A_5A_3A_4A_4\right\rangle$         & $34560 N^2 (1 + N^2) (5 + N^2)$ & $240 N^2 (2 + N^2) (23 + N^2)$ & $480 N^2 (47 + 24 N^2 + N^4)$ & $480 N^2 (158 + 57 N^2 + N^4)$ \\
$\left\langle (A_2A_3)A_3A_4A_4\right\rangle$         & $69120 N (1 + N^2) (1 + 2 N^2)$ & $480 N (9 + 23 N^2 + 4 N^4)$ & $480 N (19 + 46 N^2 + 7 N^4)$ & $480 N (65 + 139 N^2 + 12 N^4)$  \\
$\left\langle A_5A_3(A_2A_2)A_4\right\rangle$         & $69120 N^3 (5 + N^2)$ & $480 N (8 + N^2) (1 + 3 N^2)$ & $960 N (8 + N^2) (1 + 3 N^2)$ & $1920 N (16 + 35 N^2 + 3 N^4)$  \\
$\left\langle (A_2A_3)A_3(A_2A_2)A_4\right\rangle$         & $138240 N^2 (1 + 2 N^2)$ & $160 N^2 (63 + 38 N^2 + N^4)$ & $320 N^2 (64 + 37 N^2 + N^4)$ & $960 N^2 (77 + 5 N + 25 N^2 + N^3)$ \\
$\left\langle A_5(A_1A_2)A_4A_4\right\rangle$         & $69120 N^3 (5 + N^2)$ & $480 N (8 + N^2) (1 + 3 N^2)$ & $960 N (8 + N^2) (1 + 3 N^2)$ & $1920 N (16 + 35 N^2 + 3 N^4)$ \\
\hline
\end{tabular}

\begin{tabular}{|l|l|l|}
\hline
Correlator type                                & E  & F \\ \hline
$\left\langle A_5A_3A_4A_4\right\rangle$       & $480 N^2 (74 + 33 N^2 + N^4)$   &  $480 N^2 (13 + 10 N^2 + N^4)$ \\
$\left\langle (A_2A_3)A_3A_4A_4\right\rangle$   & $240 N (55 + 146 N^2 + 15 N^4)$      & $1440 N (1 + 6 N^2 + N^4)$ \\
$\left\langle A_5A_3(A_2A_2)A_4\right\rangle$       &$1440 N (10 + 23 N^2 + 3 N^4)$  & $1920 N (1 + 4 N^2 + N^4)$ \\
$\left\langle (A_2A_3)A_3(A_2A_2)A_4\right\rangle$     & $17280 N^2 (2 + N^2)$     &  $5760 N^2 (1 + N^2)$ \\
$\left\langle A_5(A_1A_2)A_4A_4\right\rangle$       &$1440 N (10 + 23 N^2 + 3 N^4)$   &  $640 N (3 + 13 N^2 + 2 N^4)$ \\
\hline
\end{tabular}\label{tab:2}\caption{Colour factors for $\left\langle 5344\right\rangle$ in $\text{U}(N)$ gauge theory}

\end{sidewaystable}
\begin{align}\label{eq:OPEc5344}&\notag\left\langle 5344\right\rangle=g_{12}^3g_{34}^3\frac{g_{14}}{g_{24}}\left(\left(\frac{g_{13}g_{24}}{g_{12}g_{34}}\right)\sum_{\lambda\geq 0}A_{2[\lambda]}F^{012[\lambda]}+\Bigg(\frac{g_{13}g_{24}}{g_{12}g_{34}}\right)^2\sum_{\lambda_1\geq \lambda_2\geq 0}A_{4[\lambda_1,\lambda_2]}F^{124[\lambda_1,\lambda_2]}\\&+\left(\frac{g_{13}g_{24}}{g_{12}g_{34}}\right)^3\sum_{\lambda_1\geq \lambda_2\geq \lambda_3\geq 0}A_{6[\lambda_1,\lambda_2,\lambda_3]}F^{236[\lambda_1,\lambda_2,\lambda_3]}+\left(\frac{g_{13}g_{24}}{g_{12}g_{34}}\right)^4\sum_{\lambda_1\geq \lambda_2\geq \lambda_3\geq \lambda_4\geq 0}A_{8[\lambda_1,\lambda_2,\lambda_3,\lambda_4]}F^{348[\lambda_1,\lambda_2,\lambda_3,\lambda_4]}\Bigg),\end{align}
whereby the result is structurally identical to (\ref{eq:c4233}) for the $\gamma=2,4\text{ and }6$ but for changes in the precise colour factors:
\begin{align}&\notag A_{2[0]}=F\text{ all else 0},\\&\notag A_{4[\lambda_1]}=\frac{E \lambda_1 ! (\lambda_1 +2)!}{(2 \lambda_1 +1)!}\text{ for }\lambda_1\in \mathbb{Z}_{\text{even}}\text{ and all else }0,\\&\notag A_{6[\lambda_1,\lambda_2]}=\frac{4 (-1)^{\lambda _2} \left(\lambda _1+2\right) \left(\lambda _1+3\right) \left(\lambda _2+2\right) \left(\left(\lambda _1+2\right)!\right){}^2 \left(\left(\lambda _2+1\right)!\right){}^2}{\left(2 (-1)^{\lambda _2} \lambda
   _1+5 (-1)^{\lambda _1}-(-1)^{\lambda _2}\right) \left(2 \lambda _1+4\right)! \left(2 \lambda _2+2\right)!}\\&\notag \times \left(\frac{1}{24} C \left(12 \left(\lambda _1-3\right) \lambda _1+\left(96 \lambda _1-12 \lambda _2 \left(\lambda _2+3\right)+25\right)+23\right)+D (-1)^{\lambda _2}\right)\\& \text{ for $\lambda_{1}-\lambda_{2}\in \mathbb{Z}_{\text{even}}\geq 0, \lambda_{2}\geq 0$ and all else zero}.\end{align}
For the $\gamma=8$ sector we get:
\begin{align}&\notag A_{8[\lambda_1,\lambda_2]}=n_{\lambda_1,\lambda_2} \frac{1}{6} \left(\lambda _1+4\right) \left(2 \lambda _2+5\right) \Big(A \left(\lambda _1+2\right) \left(\lambda _1+5\right) \left(\lambda _1-\lambda _2+1\right) \left(\lambda
   _2+1\right) \left(\lambda _2+4\right) \left(\lambda _1+\lambda _2+6\right)\\&\notag+12 B \left(\left((-1)^{\lambda _2}+1\right) \left(\lambda _1+2\right) \left(\lambda
   _1+5\right)+\left((-1)^{\lambda _2}-1\right) \left(\lambda _2+1\right) \left(\lambda _2+4\right)\right)\Big)\\&\notag \text{for $\lambda_{1}-\lambda_{2}\in \mathbb{Z}_{\text{even}}\geq 0, \lambda_{2}\geq 0$ and zero otherwise,}
	\\&\notag A_{8[\lambda_{1},\lambda_{2},1]}=n_{\lambda_1,\lambda_2}\frac{1}{12} \left(\lambda _1+4\right) \left(\lambda _1-\lambda _2+1\right) \left(\lambda _1+\lambda _2+6\right) \left(2 \lambda _2+5\right) \Big(A \left(\lambda _1+1\right)
   \left(\lambda _1+6\right) \lambda _2 \left(\lambda _2+5\right)\\&\notag+12 B \left((-1)^{\lambda _2}-1\right)\Big)\\&\notag \text{for $\lambda_{1}-\lambda_{2}\in \mathbb{Z}_{\text{odd}}\geq 1, \lambda_{2}\geq 1$ and zero otherwise,}
	\\&\notag A_{8[\lambda_1,\lambda_2,2]}=n_{\lambda_1,\lambda_2}\frac{1}{30} \left(\lambda _1+4\right) \left(2 \lambda _2+5\right) \Big(A \lambda _1 \left(\lambda _1+7\right) \left(\lambda _1-\lambda _2+1\right) \left(\lambda _2-1\right)
   \left(\lambda _2+6\right) \left(\lambda _1+\lambda _2+6\right)\\&\notag+12 B \left(\left((-1)^{\lambda _2}+1\right) \lambda _1^2+7 \left((-1)^{\lambda _2}+1\right) \lambda
   _1+\left((-1)^{\lambda _2}-1\right) \left(\lambda _2-1\right) \left(\lambda _2+6\right)\right)\Big)\\& \text{for $\lambda_{1}-\lambda_{2}\in \mathbb{Z}_{\text{even}}\geq 0, \lambda_{2}\geq 2$ and zero otherwise,}\end{align}
where 
\begin{align}n_{\lambda_{1},\lambda_{2}}=\frac{\left(\left(\lambda _1+3\right)!\right){}^2 \left(\left(\lambda _2+3\right)!\right){}^2}{\left(2 \lambda _1+6\right)! \left(2 \lambda _2+6\right)!}.\end{align}

\newpage

\section{Alternative form for $GL(m|n)$ characters}
\label{sec:altern-form-glmn}

In order to have a more direct link   between the determinantal formula for the conformal partial waves in (\ref{eq:2}), it will be useful to derive an alternative determinantal form for the super
Schur polynomial. It has a similar form to~\eqref{eq:5} but does not involve the
conjugate Young tableau and has a different dimension. The matrix
(whose determinant we take) has dimension $n+p$ where $p\geq 0 $ can be any
integer such that
\begin{align}
  \label{eq:19}
  p \geq m-n \quad \text{and} \quad p\geq \lambda^T_1\ .
\end{align}
Recall that $\lambda^T_1$ is the number of columns in the conjugate
Young tableau, i.e. the height of the Young tableau $\underline \lambda$.

The new formula is then given as 
 \begin{align}
   \label{eq:5b}
 s_{\underline \lambda}(x|y)=
(-1)^{\frac12(2 m + 2 p + n) (n - 1)}
D^{-1}
  \det \left(
     \begin{array}{cc}
\tilde X_{\underline\lambda}&R\\
       K_{\underline \lambda}   &   Y
         \end{array}
\right)\ ,
 \end{align}
where $D,R$ are just as defined in~\eqref{eq:6}, and $\tilde X_{\underline
  \lambda}$ is also very similar to $X_{\underline \lambda}$, just with a different range. However
the $Y$ matrix has no dependence on the representation and instead we
introduce a representation dependent matrix $K_{\underline \lambda}$ which only has zero's and
minus one's
\begin{align}
  \label{eq:6b}
\tilde X_{\underline \lambda}&=\Big([x_i^{\lambda_j+m-n-j}] \Big)_{
  \substack{
1\leq i
  \leq m\\
1\leq j\leq p
}
}
\nonumber\\[5pt]
 K_\ula&=\Big( -\delta_{i;-(\lambda_j+m-n-j)}\Big)_{\substack{1\leq i\leq p+n-m\\ 1\leq j  \leq p}}  \ 
&
Y&=\Big(y_j^{i-1} \Big)_{\substack{1\leq i\leq p+n-m\\1\leq j
  \leq n}}\ .
\end{align}
 Here we define
\begin{align}\label{eq:23b}
[x_i^{a}]:=\left\{
  \begin{array}{ll}
x_i^a \qquad &a\geq 0\\
0 \qquad &a<0\ ,
\end{array}
\right.
\end{align}
where the square brackets define the regular part, giving zero
if the power is negative.

Let us see this new form in the above example~\eqref{eq:16}
with  $GL(2|3)$ and
$\underline\lambda=(3,2,2,1)$.  We need $p\geq 4$ so we choose $p=4$,
then  this alternative formula~\eqref{eq:5b} gives 
\begin{align}\label{eq:20}
  s_{\underline\lambda}(x|y)= -D^{-1}\det
\left(
\begin{array}{ccccccc}
 x_1 & 0 & 0 & 0 & \frac{1}{x_1-y_1} & \frac{1}{x_1-y_2} & \frac{1}{x_1-y_3} \\
 x_2 & 0 & 0 & 0 & \frac{1}{x_2-y_1} & \frac{1}{x_2-y_2} & \frac{1}{x_2-y_3} \\
 0 & -1 & 0 & 0 & 1 & 1 & 1 \\
 0 & 0 & -1 & 0 & y_1 & y_2 & y_3 \\
 0 & 0 & 0 & 0 & y_1^2 & y_2^2 & y_3^2 \\
 0 & 0 & 0 & -1 & y_1^3 & y_2^3 & y_3^3 \\
 0 & 0 & 0 & 0 & y_1^4 & y_2^4 & y_3^4 \\
\end{array}
\right)
\ .
\end{align}
One can quickly see that~\eqref{eq:16} and~\eqref{eq:20} are
equal. Indeed
in~\eqref{eq:20} one can delete columns 2,3,4 (since they have only
one non-zero entry in) and the corresponding rows 3,4,6  
to arrive at the $4\times4$ matrix of~\eqref{eq:16} (up to a row swap).

The example illustrates the general proof that~\eqref{eq:5}
and~\eqref{eq:5b} are equal in general. Starting with~\eqref{eq:5b},
we first note that 
all non-zero entries of $K$ correspond to rows and columns
that can be trivially deleted to give the reduced matrix.  The
$K_{\underline \lambda}$ matrix has a non-zero entry in row
$j$ if and only if  $i=-(\lambda_j+m-n-j)$. This requires $\lambda_j-j-n+m < 0$ and so the corresponding
entries in column $j$ of $X_{\underline \lambda}$ vanish (since we take the regular
part~\eqref{eq:23b}). We conclude that any non-zero entry in the
$K_{\underline \lambda}$ matrix is the unique non-zero entry in its
column. We can therefore delete this
column and the corresponding row $i$ without changing the determinant
(up to a minus sign which we account for separately). On deleting the
columns $\tilde X_{\underline \lambda}$ reduces to $X_{\lambda}$ of~\eqref{eq:5} and
the matrix $K_{\underline \lambda}$ reduces to the zero matrix of~\eqref{eq:5}. 
We then just need to show that after all the corresponding rows  have been
deleted, $Y$ reduces to $Y_{\underline \lambda}$. 
The matrix $Y$ has powers $y_j^{i-1}$ for all $i=1\dots p+n-m$.
We delete (via $K_{\underline \lambda}$) rows $i=-(\lambda_j+m-n-j)$ for $j=k\dots p$. We wish to show
that we are left with $y_j^{\lambda_i^T+n-m-i}$ for $i=1\dots k'-1$. In
other words we need to show that the disjoint union of the two sets
\begin{align}
\cS_1=\Big\{\lambda^T_i+n-m-i:1\leq i\leq k'-1 \Big\}, \qquad \cS_2=\Big\{-(\lambda_j+m-n-j+1):k\leq
j\leq p \Big\}
\end{align}
form a partition of the set of integers from 0 to $p{+}n{-}m{-}1$:
\begin{align}
\cS_1+\cS_2 = \Big\{0,1,2,\,\dots,\,p{+}n{-}m{-}1\Big\}\ .
\end{align}

This is again most easily seen diagrammatically. The set $\cS_1$ is
represented by the number of boxes below
the shaded diagonal down to the bottom of the Young tableau. The
set $\cS_2$ is the number of boxes between the Young tableau
on the left  and the shaded boxes on the right. 
Together these sets count all numbers from $0$ to $p{+}n{-}m{-}1$ precisely once as we see in
the example below. Here we choose $p=9$ although one can easily check that it works for any $p\geq 9$. Recall that $m=7,n=10$ in this example. 
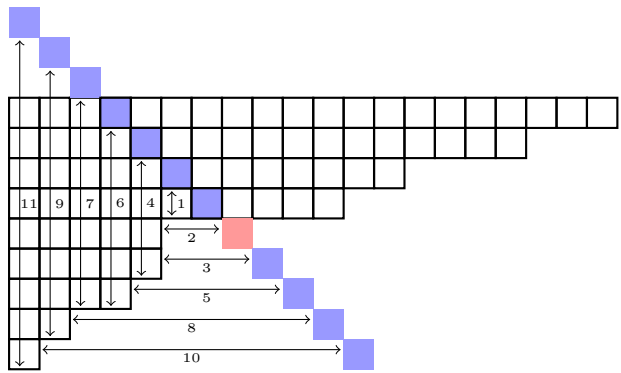
\begin{center}
  \begin{Tableau}{ {,,,,,,,,,,,,,,,,,,,}, {,,,,,,,,,,,,,,,,},
      {,,,,,,,,,,,,}, {,,,,,,,,,,}, {,,,,}, {,,,,}, {,,,}, {,}, {} }
\fill[blue!40!white] (0,2) rectangle (1,3);
\fill[blue!40!white] (1,1) rectangle (2,2);
\fill[blue!40!white] (2,0) rectangle (3,1);
\filldraw[draw=black, fill=blue!40!white] (3,-1) rectangle (4,0);
\filldraw[draw=black, fill=blue!40!white] (4,-2) rectangle (5,-1);
\filldraw[draw=black, fill=blue!40!white] (5,-3) rectangle (6,-2);
\filldraw[draw=black, fill=blue!40!white] (6,-4) rectangle (7,-3);
\fill[red!40!white ] (7,-5) rectangle (8,-4);
\fill[blue!40!white] (8,-6) rectangle (9,-5);
\fill[blue!40!white] (9,-7) rectangle (10,-6); 
\fill[blue!40!white] (10,-8) rectangle (11,-7);
\fill[blue!40!white] (11,-9) rectangle (12,-8);   
                    \draw[thin,black,<->]
    (5.35,-3.1)--node[anchor=west]{$\!\scriptscriptstyle 1$}(5.35,-3.9)  ;
\draw[thin,black,<->]
(4.35,-2.1)--node[above=.6em, anchor=west]{$\!\scriptscriptstyle 4$}(4.35,-5.9)  ;
\draw[thin,black,<->]
    (3.35,-1.1)--node[above=.6em, anchor=west]{$\!\scriptscriptstyle 6$}(3.35,-6.9)  ;
\draw[thin,black,<->]
    (2.35,-0.1)--node[anchor=west]{$\!\scriptscriptstyle      7$}(2.35,-6.9)  ;
\draw[thin,black,<->]
    (1.35,0.9)--node[anchor=west]{$\!\scriptscriptstyle      9$}(1.35,-7.9)  ;
\draw[thin,black,<->]
    (0.35,1.9)--node[anchor=west]{$\!\!\scriptscriptstyle      11$}(0.35,-8.9)  ;
\draw[thin,black,<->]
    (5.1,-4.35)--node[above=.2em , anchor=north]{$\scriptscriptstyle 2$}(6.9,-4.35)  ;
\draw[thin,black,<->]
    (5.1,-5.35)--node[above=.2em ,anchor=north]{$\scriptscriptstyle 3$}(7.9,-5.35)  ;
\draw[thin,black,<->]
    (4.1,-6.35)--node[above=.2em ,anchor=north]{$\scriptscriptstyle  5$}(8.9,-6.35)  ;
\draw[thin,black,<->]
    (2.1,-7.35)--node[above=.2em ,anchor=north]{$\scriptscriptstyle 8 $}(9.9,-7.35)  ;
\draw[thin,black,<->]
    (1.1,-8.35)--node[above=.2em ,anchor=north]{$\scriptscriptstyle 10$}(10.9,-8.35)  ;
 \end{Tableau}
\end{center}
In this example the set  $\{\lambda_i^T+n-m-i\}: i=1\dots k'-1\}=\{0,1,4,6,7,9,11\}$ corresponding to the vertical arrows, whereas the set  $\{-(\lambda_j+m-n-j):k\leq
j\leq p \}=\{2,3,5,8,10\}$, the horizontal arrows. Together they make the full set of numbers from 0 to $11=p-m+n-1$. 
To prove this in general, first convince oneself that a number cannot be in both $\cS_1$ and $\cS_2$ for a properly shape Young tableau, so the two sets are 
disjoint. Then
note that there are $(k'-1) + (p-k+1) = p-m+n$ elements in the two
sets. Finally, since all numbers are positive (or zero) and the highest value\footnote{If $p$ is the height of the Young tableau as in the example,
then $p-m+n$ is the height  of the first column extended up to the shaded box. If $p$ is
greater than the height of the Young tableau, then $p-m+n$ is the width of the
$p$th row to the left of the shaded box.}
 is
$p-m+n-1$ then they must correspond
precisely to all numbers from $0$ to $p-m+n$.

\end{document}